\documentclass[fleqn,usenatbib]{mnras}

\usepackage{newtxtext,newtxmath}

\usepackage[T1]{fontenc}

\let\VANthebibliography\thebibliography
\def\thebibliography{\DeclareRobustCommand{\VAN}[3]{##3}\VANthebibliography}

\usepackage{graphicx}	
\usepackage{amsmath}	
\usepackage{multicol}
\usepackage{subcaption}

\newcommand\HI{$\textrm{H}\scriptstyle\mathrm{I}$}

\newcommand{\ve}[1]{\mathbf{q1}}

\newcommand{\be}{\begin{equation}}      
\newcommand{\ee}{\end{equation}}      
      
\newcommand{\bef}{\begin{figure}}      
\newcommand{\eef}{\end{figure}}      
\newcommand{\bea}{\begin{eqnarray}}    
\newcommand{\eea}{\end{eqnarray}}

\newcommand{\av}[1]{\ensuremath{\left\langle q1 \right\rangle}}

\newcommand{\tve}[1]{\tilde{\boldsymbol{q1}}}

\def\bse{\begin{subequations}}
\def\ese{\end{subequations}}

\title[The Tully-Fisher relation and  the Bosma effect]{The Tully-Fisher relation and the Bosma effect}

\author[F. Sylos Labini et al.]{
Francesco Sylos Labini,$^{1,2}$\thanks{E-mail: sylos@cref.it (FSL)}
Giordano De Marzo$^{1,3}$
Matteo Straccamore,$^{1,3}$
and S\'ebastien Comer\'on$^{4,5}$
\\
$^{1}$Centro Ricerche Enrico Fermi,  I-00184, Roma, Italia\\
$^{2}$INFN Unit\'a Roma 1, Dipartimento di Fisica, Universit\'a di  Roma Sapienza, I-00185 Roma, Italia\\
$^{3}$Dipartimento di Fisica, Sapienza, Universit\'a di  Roma,  I-00185, Roma, Italia\\
$^{4}$Departamento de Astrof\'isica, Universidad de La Laguna, E-38200, La Laguna, Tenerife, Spain\\
$^{5}$Instituto de Astrof\'isica de Canarias E-38205, La Laguna, Tenerife, Spain
}

\date{Accepted XXX. Received YYY; in original form ZZZ}

\pubyear{2023}

\begin{document}
\label{firstpage}
\pagerange{\pageref{firstpage}--\pageref{lastpage}}
\maketitle

\begin{abstract}
We show that the rotation curves of 16 nearby   disc galaxies in the THINGS sample and the Milky Way can be described by the NFW halo model and by the  Bosma effect at approximately the same level of accuracy. The latter effect suggests that the behavior of the rotation curve at large radii is determined by the rescaled gas component and thus that dark matter and gas distributions are tightly correlated. By focusing on galaxies with exponential decay in their gas surface density, we can normalize their rotation curves to match the exponential thin   disc model at large  enough  radii. This normalization assumes that the galaxy mass is estimated consistently within this model, assuming a thin disc structure.  We show that this rescaling allows us to derive a new version of the Tully-Fisher (TF) relation, the Bosma TF relation that nicely fit the data.    In the framework of this model, the connection between the Bosma Tully-Fisher (TF) relation and the baryonic TF relation can be established by considering an additional empirical relation between the baryonic mass and the total mass of the disc, as measured in the data.
\end{abstract}


\section{Introduction}
\label{intro}

The rotation curves of spiral galaxies appear to be roughly flat out to the outermost observed points, which is in contrast to the nearly Keplerian decline expected from the luminous   disc (see, e.g., \cite{Rubin_Ford_Thonnard_1980,Sancisi_1999} and references therein). This discrepancy has led to the conclusion that spiral galaxies are dominated by a massive dark halo which has a different nature from the luminous matter. The dark halo is thought to have an almost spherical { mass} distribution with an isotropic velocity dispersion, while the luminous matter forms a  rotationally supported   disc. This difference in kinematic and geometrical structure of the   disc and the halo is due to the physical properties of the matter of which they are made of:  dissipational baryonic matter the former,  dissipation-less and collision-less non-baryonic matter the latter. 

The halo model is the prevailing physical framework used to interpret observations of galactic kinematics. This model is able to explain the variety of rotation curves observed: many galaxies show a flat rotation curve, others a rising one without a clear tendency toward a flattening, while in some cases the rotation curve  shows a declining behavior. 
{ The ability to fit different rotation curve shapes in  the standard halo model  \citep{Navarro_etal_1997} is related to the shape of the dark matter (DM) halo, which decays with a logarithmic slope ranging from -1 as the radius $r$ approaches zero to -3 for $r \rightarrow \infty$. In cases where a central cusp is not present, a generalized version of the halo model can be utilized \citep{DiCintio_etal_2014,Read_etal_2016,Ou_etal_2023}. This generalized model, allows for the logarithmic slope to approach 0 as $r$ tends to zero. This modification provides additional flexibility in capturing the rotation curve behavior, particularly in the central regions of galaxies where deviations from the NFW profile may be observed.  These characteristics allow  the halo  model to successfully fit rotation curves that exhibit various shapes.  The standard halo profile depends on two free parameters, while the generalized halo profile requires three free parameters. It is worth noting that these parameters have been found to exhibit strong correlations in numerical $N$-body calculations. (see, e.g., \cite{Maccio_etal_2008,Dutton_etal_2014}).  }

A different theoretical mass model to interpret   disc galaxy kinematics,  proposed  long ago \citep{Milgrom_1983},  is based on the Modified Newtonian Dynamics (MOND) and, interestingly,  has passed a number of observational tests (see, e.g., \cite{McGaugh_etal_2016_PRL}). This model does not require the introduction of an additional dark component and only considers the mass of the visible stellar and gas components. The key feature is that the gravitational force is assumed to decline more slowly than in Newtonian dynamics, which allows for the rotation velocity to remain  constant without the need for additional mass beyond the baryonic components. 

In this paper, beyond the  halo model, we consider a mass model based on standard Newtonian dynamics. The key difference with the halo model is that we place DM into the galactic   disc. The roots of this model lie in the observed correlation between DM  and the distribution of   \HI{}  gas, first noted by \cite{Bosma_1978,Bosma_1981}. This correlation refers to the observation, in a sample of nearby   disc galaxies, that the ratio of the total disc surface density (derived from rotation curve measurements) and the gas surface density (measured from   \HI{}  observations) is roughly constant beyond about one-third of the optical radius. The correlation between DM and gas in   disc galaxies implies that rotation curves at large radii can be rescaled versions of those derived from the   \HI{}  gas distribution \citep{Sancisi_1999,Hoekstra_etal_2001,Hessman+Ziebart_2011,Swaters_etal_2012,SylosLabini_etal_2023a}. { In this framework, which we refer to as the Bosma model,} at small radii, the total disc surface density is dominated, completely or partially, by stars, while at large radii it is dominated by a mass component obtained by rescaling the surface density of the gas. Since the mass of the gas component is a negligible fraction of the galaxy mass, the Bosma effect implies that the gas surface density is a proxy for DM. This means that the distribution of DM follows that of the gas and is confined to a   disc, but is not still observed. 
{ Note that the Bosma effect does not inherently imply that DM is exclusively confined to a disc. It is crucial to acknowledge that, based on a given radial acceleration profile, such as the one predicted by the Bosma effect, it is always possible to construct a spherical distribution of matter that produces the observed acceleration. However, for the sake of our discussion, we will proceed with the assumption that DM is restricted to a disc.
This assumption allows us to simplify the modeling and analysis by focusing on a specific configuration where DM is concentrated within a disc-like structure. While it may not capture all possible scenarios, this hypothesis provides a useful starting point for understanding the observed rotational velocities and their relation to the distribution of DM in galaxies.}

In this respect, it is worth recalling that \cite{Pfenniger_etal_1994} proposed  the idea that  DM could reside in the galactic   disc in the form of very cold molecular hydrogen (i.e., H$_2$).  Extensive vast clouds of cold (i.e., $T\approx$10K-20K) dust and dark gas, invisible in \HI{}  and CO but detected in $\gamma$-rays, were  detected, in the solar neighborhood,  by \cite{Grenier_etal_2005}  who concluded  that dark gas mass in the Milky Way was comparable to the molecular one. Later, dark gas was found also by the Planck mission \citep{Planck_2011} --- see \cite{Casandjian_etal_2022} for an recent discussion of cold dark gas clouds measurements. 
H$_2$ in  less excited and colder regions than those seen with CO cannot be inferred yet: CO freezes on grains below 20K, so molecules { do not rotate anymore thus emitting }rotational lines.   Regions colder than 20K are not rare and for this reason there is room for more undetected H$_2$ in outer gas   discs where  excitation from stellar radiation  decays fast: unfortunately, a direct detection such clouds in the outer   disc of galaxies appears to be a very hard task from an observational point of view  \citep{Combes+Pfenniger_1997}.  

Irrespective of the type of DM in the galactic   disc, it has been shown that the Bosma effect can provide highly accurate rotation curve fits for several   disc galaxies, by using both the observed stellar disc and  \HI{}  gas as proxies for DM  \citep{Hessman+Ziebart_2011,Swaters_etal_2012}. Recent work by \cite{SylosLabini_etal_2023a}  has demonstrated that the rotation curve of the Milky Way can also be accurately fit using the Bosma effect. Although the presence of correlation does not necessarily imply causality, its detection may be related to a fundamental property of   disc galaxies --- that DM is distributed on the   disc, rather than in a spherical halo around the galaxy, as currently assumed. Therefore, we believe it is worthwhile to investigate this effect in more detail than has been previously done.

In this paper, we use 16 galaxies from the The  \HI{}  Nearby Galaxy Survey (THINGS)  data-set \citep{Walter_etal_2008} and the Milky Way to further examine the Bosma effect. This sub-sample of the THINGS galaxies is the same used by \cite{Hessman+Ziebart_2011} (with the exception of NGC 3031 that shows signs of strong tidal interactions). However, unlike previous studies, for a large fraction of the galaxies in the THINGS sample we use the full gas surface density and mass, not just the  \HI{}  component. In particular, we use the measurements of CO inferred H$_2$ by \cite{Schruba_etal_2011}, which include 11 galaxies in our sample. In addition, for   12 galaxies we use the more accurate stellar surface density profiles and masses from the stellar density maps based on the recipes in  \cite{Querejeta_etal_2015}  applied to the   Spitzer Survey of Stellar Structure in Galaxies  (S$^4$G ---   \cite{Sheth_etal_2010,Munoz-Mateos_etal_2015})  Pipeline 5  dust-corrected 3.5 $\mu$m  surface brightness maps, instead of previous measurements based on the 2MASS survey \citep{deBlok_etal_2008,Hessman+Ziebart_2011}. To quantitatively evaluate our results, we fit the Bosma mass model and compare it to { both  the Navarro-Frenk-White (NFW) halo mass model \citep{Navarro_etal_1997} and its generalized version \citep{Ou_etal_2023}}

{ By assuming, consistently with the Bosma model, that the toal mass is confined on the galactic disk, we shown that rotational curves shown a universal shape when properly normalized. In particular,  } when the radius is expressed in units of the exponential scale-length $R_{\text{d}}$  of the gas (i.e.,  \HI{}) density profile and the velocity is in units of the typical system velocity that depends on the total mass and on $R_{\text{d}}$, we find that the rotation curves of the different galaxies approximately  follow  the exponential thin   disc approximation \citep{Freeman_1970}. Given this result, we are able to show that  a simple analytical explanation of the  Tully-Fisher relation can be formulated. This is then compared to observations and related to the baryonic Tully-Fisher { (BTF)} relation \citep{McGaugh_etal_2000,McGaugh_2005b}.

The structure of the paper is as follows: in  Sect.\ref{sample}, the properties of the sample galaxies are discussed, including their rotation curves, star and gas surface brightness, and masses.   Then in Sect.\ref{fits}, we discuss the { fits of the Bosma and  NFW models, and their modifications}  to the rotation curves (the results of fitting the individual galaxies are presented in the Appendix), whereas in  Sect.\ref{univ}, the { computation of the average rotation curve among the galaxies in our sample and its comparison to the} the exponential thin disc model is { discussed. Furthermore, we show that it is possible to derive a modified version of the Tully-Fisher (TF) relation  in the framework of the Bosma model and we discusss its relation to the  classical TF and BTF  relations (Sect.\ref{sec:TF}). Furthermore we examine the correlation between the observed acceleration and the acceleration predicted by baryonic matter. We discuss that this correlation can be naturally understood within the framework of the Bosma model. } Finally in Sect.\ref{conclusions}, the main conclusions are drawn.


\section{The sample} 
\label{sample} 

The THINGS survey \citep{Walter_etal_2008} is a high spectral and spatial resolution survey of the \HI{}  emission of nearby galaxies that was obtained using the NRAO Very Large Array. The sample has a velocity resolution of 5.2 km/s or better and an angular resolution of 6" corresponding to a linear resolution of 0.1/0.5 kpc for galaxies at a distance of 2/15 Mpc \citep{Walter_etal_2008}. This high resolution makes the THINGS survey a unique sample for the study of galaxy kinematics. The astrophysical parameters of the galaxies studied in this work can be found in \cite{Walter_etal_2008}. We considered the same sample of \cite{Hessman+Ziebart_2011}  but with the exception of NGC3031 because its  \HI{}  distribution is affected by strong tidal interactions with neighboring galaxies and its surface density  does not show a simple exponential decay: all other galaxies show a gas surface brightness exponentially decaying.

The stellar profiles and masses from the stellar density maps were produced from the dust-corrected 3.6$\mu$m maps in the Pipeline 5 of the S$^4$G \citep{Sheth_etal_2010,Munoz-Mateos_etal_2015}  using the recipe provided in \cite{Querejeta_etal_2015}. We obtained the stellar surface density profile and mass (see Tab. \ref{table:results1}) for galaxies included in this sample. For galaxies not included in this sample, we used the mass from \cite{Hessman+Ziebart_2011}  and the stellar surface density profile from the 2MASS data \citep{deBlok_etal_2008}. Note that in some cases, there are differences between the values of the stellar mass we used and those reported in \cite{Hessman+Ziebart_2011}.

The  \HI{}  mass and surface density profile were estimated from the THINGS data set \citep{Walter_etal_2008}. Note that the values of  \HI{}  masses we report in Tab.\ref{table:results1} are not multiplied by 1.39 to correct for He and heavy elements, and thus coincide with those listed in Tab. 5 of  \cite{Walter_etal_2008}. The H$_2$  mass and surface density profile are available only for the galaxies included in the sample of  \cite{Schruba_etal_2011}. For all galaxies except NGC 2366, NGC 3621, NGC 7793, DDO 154, and IC 2574 for which there are no data, we use the gas surface density profile defined as the sum of the  \HI{}  and H$_2$ profiles. Note that the H$_2$ distribution may affect the fitting with models at small radii, but it does not change the behavior at large radii, which is the regime that matters for the Bosma effect.

The fluxes derived from Pipeline 5 have uncertainties of the order of 10\% (see  Appendix A of  \cite{Querejeta_etal_2015}). As mentioned above, we  convert  the flux to mass  following the recipe in \cite{Querejeta_etal_2015}, which has an accuracy of 0.1 dex. 
For this reason, quadratically summing these two uncertainties, we find an error of  30\% for the stellar mass. We consider the same error for the mass of the gas. 

In general, the stellar surface brightness of a galaxy may show a simple or double exponential decay. The latter occurs when the galaxy has a central bulge. However, the consideration of the  bulge implies the inclusion of a new free parameter in the model. As the bulge is relevant only at very small radii, we exclude the very inner   disc from the fit and thus we avoid adding another free parameter.  { The rationale for this choice is that the Bosma effect primarily pertains to the outer regions of a galaxy. Therefore, considering the scope of the present work, it is unnecessary to intricately model the behavior of the inner disc.

Concerning the gas surface brightness profile in the inner   disc, this is generally not exponential. This is true especially for the}  \HI{}  profile, which is roughly flat at small radii and then has an exponential tail with a characteristic scale-length, $R_{\text{g}}$, reported in Tab.\ref{table:results1}. In many cases, the  \HI{}+H$_2$ profile displays exponential decay from the very inner regions of the   disc. A detailed description of the stellar and gas surface densities of individual galaxies is reported in the Appendix.  
 

\section{Mass models} 
\label{fits} 

{ In Section \ref{dmd}, we provide a detailed description of the Bosma effect, focusing on its relevance to the outer regions of a galaxy. Following that, in Section \ref{nfw}, we provide a brief overview of the key elements of the halo model. We then proceed to discuss the results of the fits in Section \ref{results}, specifically considering the four models that we have tested, which include two halo models and two disc models. For a more comprehensive analysis of each galaxy, a detailed discussion is provided in the Appendix.}

\subsection{The Bosma effect and the dark matter   disc model}
\label{dmd} 

The correlation referred to as the "Bosma effect," first noticed by \cite{Bosma_1978,Bosma_1981}, consists in the fact that the ratio of atomic hydrogen (\HI{}) gas to total matter surface density as inferred from the rotation curves appears to be relatively constant in   disc galaxies. This ratio, the Bosma factor, can be defined as 
the total matter to  \HI{}  surface densities averaged over the   disc, i.e., 
\be
\label{bosma} 
\frac{\Sigma_\text{b}(R)}{\Sigma_\text{g}(R)} \sim \frac{v_\text{c}^2(R)}{v_\text{g}^2(R)} \sim \mbox{const.}  \gg 1 \;
\ee
where $\Sigma_\text{b}$ is the total mass surface density, obtained from the observed rotation curve $v_\text{c}$,  $v_{\text{g}}$ is  the  rotation curve expected from the observed  gas surface density $\Sigma_{\text{g}}$  and $R$ is the galacto-centric distance. The Bosma factor in Eq. \ref{bosma}, being generally much larger than unity,  can  be interpreted as  the constant of proportionality between the centripetal effects of  gas and DM, or equivalently  as an effective mass-to-light ratio contribution above that normally attributable to standard gas components. 

This  factor has been measured in about hundred galaxies of different Hubble types from massive early spirals to late dwarf irregulars 
\citep{Sancisi_1983,Carignan_1985,Carignan+Beaulieu_1985,Carignan+Puche_1985,Jobin+Carignan_1990,Carignan_etal_1990,Puche_etal_1990,Sancisi_1999,Hoekstra_etal_2001}. The value of the Bosma factor spans about one order of magnitude, i.e.,  ranges between 5 and 50 and more massive galaxies tended to have larger values. 

\cite{Hoekstra_etal_2001}  analyzed a sample of 24 spiral galaxies with high-quality rotation curves. They  found that, with a few exceptions, the rotation curves generated by scaling up the centripetal contribution of the  \HI{}  gas by a constant factor of approximately 10 and not including a spherical DM halo, were comparable in accuracy to those generated by the NFW halo model. However, they also claimed that it was not possible to confirm that a real coupling between  \HI{}  and DM exists due to several imperfections in their correlation.  

Then, \cite{Hessman+Ziebart_2011} made a quantitative comparison between the results of fits with the Bosma effect and of the NFW models. They have pointed out  that the rotation curves of   disc galaxies obtained by relying solely on  \HI{}  scaling are unsatisfactory because the concentration of  \HI{}  decreases in the central regions of galaxies. This is because the high-density regions of stars consume the  \HI{}  gas and convert it into stars or molecules. To account for this physical effect, the authors suggest using the stellar discs as proxies, or including both neutral and ionized gas, instead of only the neutral  \HI{}  component, as the total gas component is believed to have a closer correlation with the distribution of DM. \cite{Hessman+Ziebart_2011} confirmed the correlation between DM and  \HI{}  distribution using 17 galaxies from the THINGS dataset \citep{Walter_etal_2008}  and rebutted several arguments against the effect by  \cite{Hoekstra_etal_2001}. They found fits of similar or even better quality than those obtained by the standard NFW halo model. This is indeed the case especially when the additional correlation between the two parameters of the NFW halo model found in $N$-body simulations \citep{Maccio_etal_2008} in taken into account.

\cite{Swaters_etal_2012} further studied a sample of 43   disc galaxies, ranging from late-type dwarf galaxies to early-type spirals, by fitting the rotation curves with mass models based on scaling up the stellar and  \HI{}    discs. They found that such scaling models fit the observed rotation curves well in the vast majority of cases, even though the models have only two or three free parameters (depending on whether the galaxy has a bulge or not). They also found that these models reproduce some of the detailed small-scale features of rotation curves such as the "bumps" and "wiggles", a feature that point toward a real coupling between  \HI{}  and DM.

The dark matter   disc (DMD) model assumes, consistently with the Bosma effect, that the DM is constrained on the   disc and has the same 
distribution as the gas, {\it thus the same scale-length  of the exponential decay}.  The model can be parametrized as  
\be
\label{dmd2} 
v_{\text{c}}^2 =\Upsilon_{\text{s}} v_{\text{s}}^2  + \Upsilon_{\text{g}} v_{\text{g}}^2 
\ee
where $\Upsilon_{\text{s}}$ and $\Upsilon_{\text{g}}$\footnote{Note that we used the pure   \HI{}  mass not multiplied by 1.39: this factor  is generally considered to correct for He and heavy elements. In our fit this corresponds to a simple rescaling of $\Upsilon_{\text{g}}$.}  are  two free parameters, while 
$v_{\text{s}}$ ($v_{\text{g}}$) is the rotation velocity that the stellar (gas)   mass component would induce on a test particle in the plane
of the galaxy if it were placed in isolation without any external influences. 
Note that Eq. \ref{dmd2}  corresponds to assume that the proportionality  between  gas and DM  is independent of the galactocentric radius:  in view of the very different physical conditions in the inner   disc with respect to those of the outermost regions of the galaxy we believe this is a strong approximation. The key feature of the Bosma effect is the rescaling of the  \HI{}  surface brightness at very large radii, where the stellar surface density becomes negligible. In addition, as mentioned above, in Eq. \ref{dmd2} we have neglected the bulge in the fits (both in the case of the Bosma effect and the NFW halo --- see below)  
{ so to avoid introducing}  an additional free parameter in the models. Indeed, the very inner   disc of the galaxy does not represent the main region where the Bosma effect was found. For this reason, the range of radii where the bulge is present can be constrained from observations of the stellar surface density and excluded from fits.

From  Eq. \ref{dmd2} it follows that the galaxy mass is 
\be
\label{dmd3} 
M_{\text{dmd}} =\Upsilon_{\text{s}} M_{\text{s}} + \Upsilon_{\text{g}} M_{\text{g}}  \;,
\ee
where $M_{\text{s}}$ is the mass of the stellar component and $M_{\text{g}}$ that of the gas component. The mass of the bulge is typically 10\% that of the   disc and it is thus inside the our error bars which are of $\sim 30\%$ (see discussion in Sect.\ref{sample}).
In what follows we compute $v_{\text{s}}$  (or $v_{\text{g}}$) from simulations as an analytical computation of the centrifugal force for a   disc is not straightforward \citep{Feng+Gallo_2011}: we thus generate a thin   disc with the observed surface brightness of the stellar (or gas) component and same mass and we numerically compute the centripetal acceleration.


\subsection{The halo  model} 
\label{nfw} 

The Navarro Frenk and White  (NFW)  mass model  can be written as 
\be
v_{\text{c}}^2 = v_{\text{s}}^2 + v_{\text{g}}^2+ v_{h}^2
\ee
where $v_{h}$ 
corresponds to the equilibrium velocity in a NFW halo profile described by
  \citep{Navarro_etal_1997}
\be
\label{nfw_profile} 
\rho(r)= \frac{\rho_0}{\frac{r}{R_{\text{s}}} \left( 1 + \frac{r}{R_{\text{s}}} \right)^2} \;,
\ee
where $R_{\text{s}}$ and $\rho_0$ are two free parameters. Results can be expressed in terms of 
the virial radius 
$R_{\text{vir}} = c R_{\text{s}} $
and  halo virial mass  $M_{\text{vir}} = M(R_{\text{vir}})$,
where $c$ is the  concentration parameter   \citep{Navarro_etal_1997}.

Normally,  NFW profiles
are characterized by two model parameters ($\rho_0, R_{\text{s}}$ in Eq. \ref{nfw_profile}), and mass model 
fits usually consider both parameters as free (see, e.g., \cite{deBlok_etal_2008,Eilers_etal_2019}). 
However, it was  shown by \cite{Maccio_etal_2008} that the $N$-body calculations
which resulted in the NFW profile model clearly show that the concentration parameter is
not an independent parameter but is in fact strongly correlated
with $V_{\text{vir}}$, defined as the rotational velocity at $R_{\text{vir}}$. 
Quantitatively they found { the following empirical relation}   
\be
\label{cnfw} 
c_{\text{vir}} \approx 7.80 
\left(
\frac{V_{\text{vir}}} {100 \; \; \mbox{km s}^{-1}}  
\right)^{-0.294}
\ee
(derived from  Eq. (10) of \cite{Maccio_etal_2008} --- see  \cite{Dutton_etal_2014} for a slightly different phenomenological fit). 
Including this intrinsic correlation reduces the number of fit parameters by one. In what follows, we consider the two parameter  fit and report the value 
of one of the two parameters (we choose the concentration parameter) if the additional correlation of Eq. \ref{cnfw} is taken into account. This is because a  potential systematic effect of adiabatic compression was not taken into account in the mass models  \citep{Gnedin_etal_2004,Sellwood+McGaugh_2005}  and this process may help reconcile the parameters found with the Eq. \ref{cnfw} as it has the effect of raising the effective concentration of the resulting halo above that of the primordial initial condition to which the mass-concentration relation applies \citep{McGaugh_2016}.  It is worth noticing that when the additional correlation given by Eq. \ref{cnfw} is considered, the quality of the fit generally worsen as it was noticed by \cite{Hessman+Ziebart_2011}.

{ 
To establish a benchmark for our analysis, we have adopted a core DM  model that has a compact core thus avoiding the cusp for $r \rightarrow 0$. This is known to provide a better fit to observed rotation curves compared to cusp profiles like the NFW  profile. In the literature, several functional behaviors have been considered to account for the interactions between baryons and DM, such as the models proposed by \cite{DiCintio_etal_2014,Read_etal_2016}. In this study, we utilize the generalized NFW (gNFW) profile, which has been recently employed for the Milky Way  by  \cite{Ou_etal_2023}. Unlike the standard NFW profile, which exhibits a divergence towards smaller radii, the gNFW profile introduces an additional free parameter that modulates the inner and outer asymptotic power law slopes of the NFW profile. This additional parameter allows for a fully cored density profile, enabling the gNFW profile to possess a power law slope down to $-3$ at radii larger than the scale radius.
By introducing this additional free parameter, the gNFW profile offers a more straightforward and manageable framework while still capturing important features of the density distribution in DM halos. For our analysis, we compute the circular velocity curve based on the density profile described by the following equation:
\be
\label{gnfw_profile} 
\rho(r)= \frac{\rho_0}{\left( \frac{r}{R_{\text{s}}} \right)^\beta  \left( 1 + \frac{r}{R_{\text{s}}} \right)^{3-\beta}} \;,
\ee
where $\beta$ is the additional free parameter. Note that   $\beta =1$ gives the usual NFW profile of Eq. \ref{nfw_profile}.
}


\subsection{Results of the fits}
\label{results} 

Rotation curves of the galaxies in our sample were published by \cite{deBlok_etal_2008}; these same rotation curves were measured also by \cite{SylosLabini_etal_2023b} by using a different method.
They  have three different behaviors:
NGC 925, NGC 2976 and  IC 2574 show a growing rotation curve without a clear tendency for a flattening;    
NGC 2366, NGC 2403, NGC 3198, NGC 3521, NGC 3621,  NGC 6946, NGC 7331, DDO 154 show a well defined flattening after the 
inner velocity  increase; finally  
NGC 2841, NGC 2903, NGC 4736, NGC 5055 and  NGC 7793   show a decreasing rotation curve. 
For this last group of galaxies the decrease between the maximum velocity and the outermost point 
corresponds to a change between 15\% (NGC 2841, NGC 2903) and 40\% (NGC 4736).

{ Both the numerical fits of the Bosma model, where we consider $\Upsilon_{\text{s}}$ and $\Upsilon_{\text{g}}$ as free parameters (see Eq. \ref{dmd2}), and the fits  with the standard NFW mass model, are similar to those of \cite{Hessman+Ziebart_2011}. However, there are some discrepancies deriving } from the high sensitivity to to the details of the stellar and gas mass and surface density profiles, which, as mentioned above, are generally different. In addition, in cases where the stellar surface density displays a double exponential decay corresponding to the presence of a bulge in addition to the   disc, we have excluded the very inner   disc from the fit. Given the sensitivity of the fit to the details of the range of radii considered, this is another reason for obtaining results that are different from \cite{Hessman+Ziebart_2011} for the precise values of the parameters, but, of course, not for the general trend.

{ 
An intriguing test for the DMD model involves treating the characteristic  length-scale of the exponential decay of the DM disk ($R_{\text{d}}$) as a free parameter in the fitting process, rather then being equal to that of the gas surface density profile. This approach allows for an exploration of whether the value of $R_{\text{d}}$ obtained from the gaseous disk provides the best match to the data. By doing so, it relaxes the strong assumption that gas and dark matter are tightly correlated in their distribution.
To this end, we conducted a DMD analysis (that we named modified DMD model, or mDMD) by considering $\Upsilon_{\text{g}}$ and $R_{\text{d}}$ as the free parameters, while $\Upsilon_{\text{s}}$ is fixed and equal to one. The results of this analysis are  included in Table \ref{table:results1a}. Upon examining the new DMD fits, we observe that, in general, the quality of the fit is relatively poorer compared to the fits where the free parameters were $\Upsilon_{\text{g}}$ and $\Upsilon_{\text{s}}$. This is evident from the higher values of the $\chi^2$ statistic obtained in the new DMD fits. 
This result is attributed to the poor fit in the inner regions, where the stellar component dominates, rather than the outer disc.

However, there are a few cases where the fits are comparable, or in some instances, even better, particularly when $\Upsilon_{\text{s}}$ is close to unity. Furthermore, we have determined that there is a linear correlation between $\Upsilon_{\text{g}}$ and $R_d$, with a correlation coefficient of approximately $r \approx 0.7$. This correlation implies that there is some degree of interdependence between the gas mass-to-light ratio $\Upsilon_{\text{g}}$ and the scale length of the disc $R_d$.

Results of fit to the  DMD, mDMD, NFW and gNFW models for individual galaxies are reported in the Appendix (see, respectively,  Tab.\ref{table:results1}-Tab.\ref{table:results2a}).
In summary, our findings suggest that the DMD model fits generally result in reduced $\chi^2$ values comparable to those obtained by the NFW fits. As expected, the gNFW profiles tend to provide improved fits compared to the NFW profiles due to the increased flexibility of the density profile shape. This improvement is primarily attributed to the additional free parameter in the gNFW model, enabling better alignment with the observed data. However, it is important to note that the primary distinction between the two profiles emerges at small radii, which are often excluded from the fitting procedure to avoid complications arising from the inclusion of a bulge component. Consequently, the exclusion of the inner disc region, combined with the avoidance of an additional free parameter associated with the bulge, may contribute to the comparable performance of the NFW fits in many cases. 

Furthermore, it is worth noting that the suboptimal reduced $\chi^2$ values observed in several cases are often a consequence of irregular features, such as bumps and wiggles, present in the rotation curves. These non-smooth characteristics pose challenges to the fitting with a smooth model and can contribute to the overall discrepancies observed between the model predictions and the observed data. 

Of course, this same problem affect the fits with the DMD model. The Bosma effect corresponds to the fact that the distribution of   \HI{}  is a proxy of the global matter surface density in galaxies.  \cite{Hessman+Ziebart_2011} have shown for this same sample of galaxies that it can be a noisy tracer. A close comparison of the gas and total rotation curves has revealed that in some galaxies, the observed "bumps" and "wiggles" in the rotation curve can be explained by the Bosma model, where the rotation curve is more closely tied to the gas density. However, in other galaxies, there is not such a close correlation. This is because the wiggles can be created by spiral arm structures, which can convert  \HI{}  to other molecules, rendering it an inadequate tracer.

Regarding the DMD model, our analysis suggests that the mass-to-light factor for the stellar component is peaked at approximately $\Upsilon_{\text{s}} \approx 1$, indicating a close correspondence between the stellar mass and the observed light. In a few cases, the mass-to-light factor exceeds unity, suggesting a higher stellar mass relative to the observed light. This variation in $\Upsilon_{\text{s}}$ highlights the diversity in the stellar populations and properties among the analyzed galaxies. Overall, these observations provide valuable insights into the fitting performance of the DMD and gNFW models, as well as the characteristics of the stellar mass-to-light ratios in the analyzed galaxies.

For the NFW and gNFW cases, we report (see Tabs.\ref{table:results2}-\ref{table:results2a}) } the values of the best-fit concentration parameter $c$, obtained from the fit with two free parameters. In addition, we also report the value of $c_{\text{vir}}$ computed from Eq. \ref{cnfw} by inserting the value of the rotational velocity at $R_{\text{vir}}$ that is obtained from fit with to free parameters. It is noticed that in all cases these are not consistent with the predictions of $\Lambda$CDM, which provides the well-defined mass-concentration relation Eq. \ref{cnfw}. We thus find that the Galactic halo has a concentration parameter which is different, and generally higher, than theoretical expectations based on cosmological simulations \citep{Maccio_etal_2008}. High values of the concentration parameter have also been found by other studies \citep{Bovy_etal_2012,Deason_etal_2012,Kafle_etal_2014,McMillan_2017,Monari_etal_2018,Lin+Li_2019,Eilers_etal_2019}: these are in tension with theoretical expectations based on cosmological simulations.


\section{The exponential thin disc  model approximation for nearby   disc galaxies}
\label{univ}

{ The rotation curves of different galaxies can vary significantly, not only due to differences in their intrinsic masses and sizes but also because of variations in their shapes.  Here, we present a normalization method that enables us to calculate their average behavior that can be further compared with a theoretical model. The  two main ingredients are the following.

(i) The first is that the mass surface density profile  can be approximated  by an exponential function. Indeed, the Bosma surface density profile is} 
\be
\Sigma_{\text{b}}(R)  =  \Upsilon_{\text{g}} \Sigma_{\text{g}} + \Upsilon_{\text{s}} \Sigma_{\text{s}} \;. 
\ee
At sufficiently large radii $\Sigma_{\text{b}}(R) $ is proportional to  the gas surface brightness, i.e., $\Sigma_{\text{b}}(r) \approx \Upsilon_{\text{g}} \Sigma_{\text{g}}(r)$. 
Thus if $\Sigma_{\text{g}}(r) \sim \exp(-R/R_{\text{g}})$, the full mass density surface brightness also decreases exponentially with a characteristic scale of $R_{\text{g}}$ and at large enough radii we have 
\be
\label{SBB}  
\Sigma_{\text{b}}(R)  \sim \exp \left( - \frac{R}{R_{\text{d}}}\right) 
\ee
where, in general if the rotation curve is extended enough beyond the optical radius, we have $R_{\text{d}} \approx R_{\text{g}}$.  
\begin{figure}
\centering 
\includegraphics[width=8cm,angle=0]{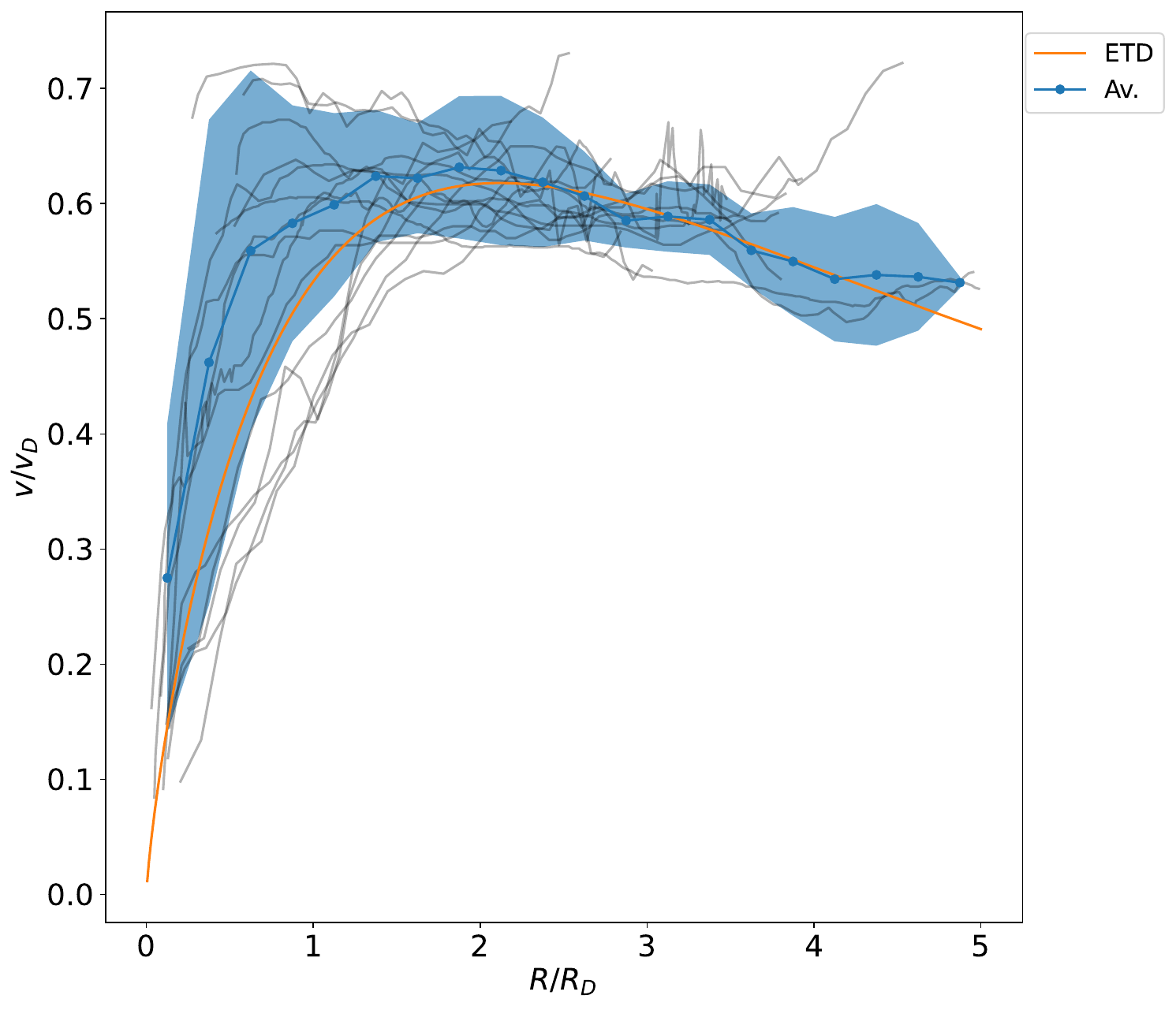}
\caption{The { average  rotation curve  (Av.) of the galaxies in our sample is obtained by normalizing, for each galaxy, the scale radius and the rotation curve respectively to the characteristic scale of $R_{\text{g}}$ and to the 
characteristic velocity Eq. \ref{eq:td} and then by making the average over all galaxies in the sample. }  Grey lines are the rotation curves of the  galaxies reported in the labels. The analytical 
behavior of the exponential thin   disc  model (ETD)   is reported as comparison.} 
\label{URC} 
\end{figure}

(ii) The { second ingredient is that a galaxy} rotation curve $v_{\text{c}}(r)$ can be well approximated, at large enough radii where  Eq. \ref{SBB} is satisfied,  
by the Exponential Thin   disc (ETD) behavior  \citep{Freeman_1970}:
\begin{equation}
\label{eq:td}
v_{\text{c}}^2(R)=4 \pi G \Sigma_{\text{b}}^0 R_{\text{d}} y^2[I_0(y)K_0(y)-I_1(y)K_1(y)],
\end{equation}
where $y=R/R_{\text{d}}$ and $I_n$ and $K_n$ are the modified Bessel functions
and 
\be
\Sigma_{\text{b}}^0  = \Upsilon_{\text{s}} \Sigma_{\text{s}}^0 + \Upsilon_{\text{g}} \Sigma_{\text{g}}^0 \approx \Upsilon_{\text{g}} \Sigma_{\text{g}}^0 \;, 
\ee
is the amplitude of the total mass surface density (see Eq. \ref{dmd3}).

{ 
Given the points mentioned above, we derive the average rotation curve of the galaxies in our sample using the following procedure. First for each galaxy's rotation curve, we normalize the scale radius to characteristic lenght-scale  $R_{\text{d}}$ and the observed velocity to the characteristic velocity defined as
\be
\label{vd} 
v_{\text{d}} = \sqrt{\frac{GM_{\text{dmd}}}{R_{\text{d}}}} \;, 
\ee
where $M_{\text{dmd}}$ is the mass of the   disc given by Eq. \ref{dmd2}. Then, we determine the average  of the so-normalized rotation curves.

Note that at small radii, Eq. \ref{SBB} is not well satisfied because the stellar component, which dominates the total mass distribution, has a different scale-length for its exponential decay (or even exhibits two different exponential behaviors in cases where there is a bulge) compared to that of the gas. This variability in scale-lengths makes the re-scaling of the rotation curves of different galaxies highly variable. Conversely, at larger radii, the agreement is much stronger. The results are depicted in Fig. \ref{URC}, where the average rotation curve aligns well with the ETD model. }
%


\section{The Tully Fisher relation and the Dark Matter   disc}
\label{sec:TF} 

 \cite{Tully+Fisher_1975} discovered an intriguing observational relationship between the luminosity of a   disc galaxy and its maximum rotational velocity, denoted by $v_{\text{max}}$. This relationship, known as the Tully-Fisher (TF) relation, has  been widely employed in estimating distances beyond our own galaxy and has become a cornerstone of observational astrophysics. However, the underlying theoretical explanation for the TF relation remains elusive, as the connection between the production of stellar light and rotational velocity is not straightforward from a physical point of view. 
  
 \cite{McGaugh_etal_2000} extended the TF relation by correlating the rotation velocity with the detected baryon mass over five orders of magnitude in stellar mass and one order of magnitude in velocity, resulting in what is known as the { baryonic TF (BTF)} relation. This relation is believed to have a more direct physical causality link between mass and rotational velocity through gravitational dynamics than the observational TF relation.  The baryonic mass considered by  \cite{McGaugh_etal_2000} includes the stellar component, whose mass is estimated using a plausible stellar mass-to-light ratio, as well as neutral hydrogen mass measured through 21 cm observations. By combining the  \HI{}   mass with the stellar mass, the scatter plot of velocity versus mass for a sample of 243 galaxies was found to be straighter. This inclusion of  \HI{}   mass was necessary because low surface brightness galaxies in the sample can have a very low stellar mass content, which can even be smaller than the gas mass, as shown studies focusing on the BTF at the smallest scale (see, e.g., \cite{Iorio_etal_2017}).  The  BTF relation has the form 
\be
\label{BTF} 
M_{\text{bar}} = (M_{\text{s}} + M_{\text{g}}) \propto v_{\text{max}}^\beta \;, 
\ee
where $v_{\text{max}}$ is the peak value of the rotation curve and the exponent was empirically found to be  $\beta \approx 4$. As noticed by \cite{McGaugh_2005b} the use of a stellar mass-to-light ratio estimated from stellar population models causes a significant systematic uncertainty in the stellar mass, which is particularly sensitive to the initial mass function. Depending on the choice of model, one can derive BTF relations with slopes anywhere in the range of  between 3 and 4. Although the formulation of the BFT relation expressed by Eq. \ref{BTF} represents an important step towards understanding its physical basis, the problem is still unresolved. 
\begin{figure*}
\centering 
\includegraphics[width=8cm,angle=0]{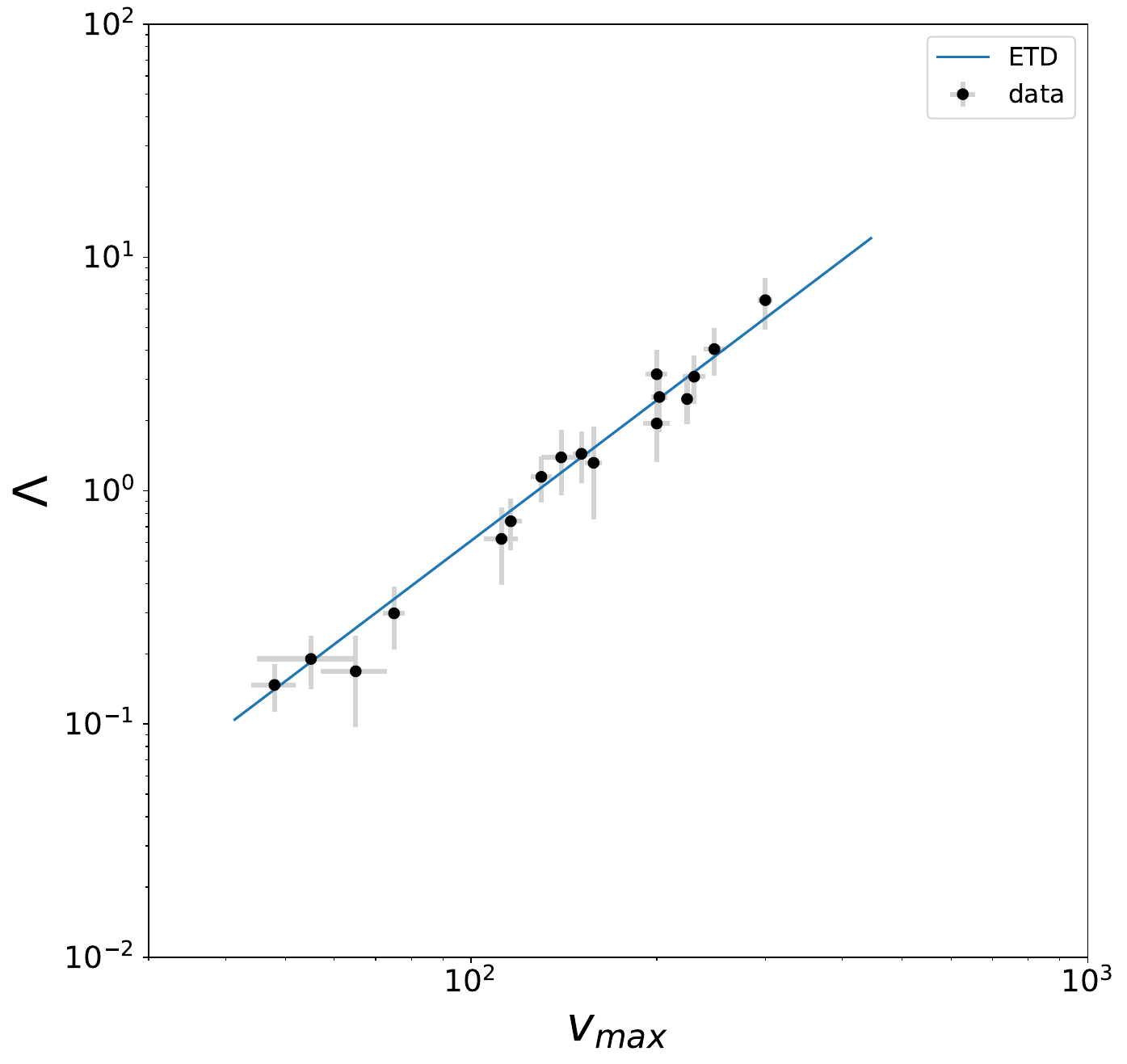}  
\includegraphics[width=8cm,angle=0]{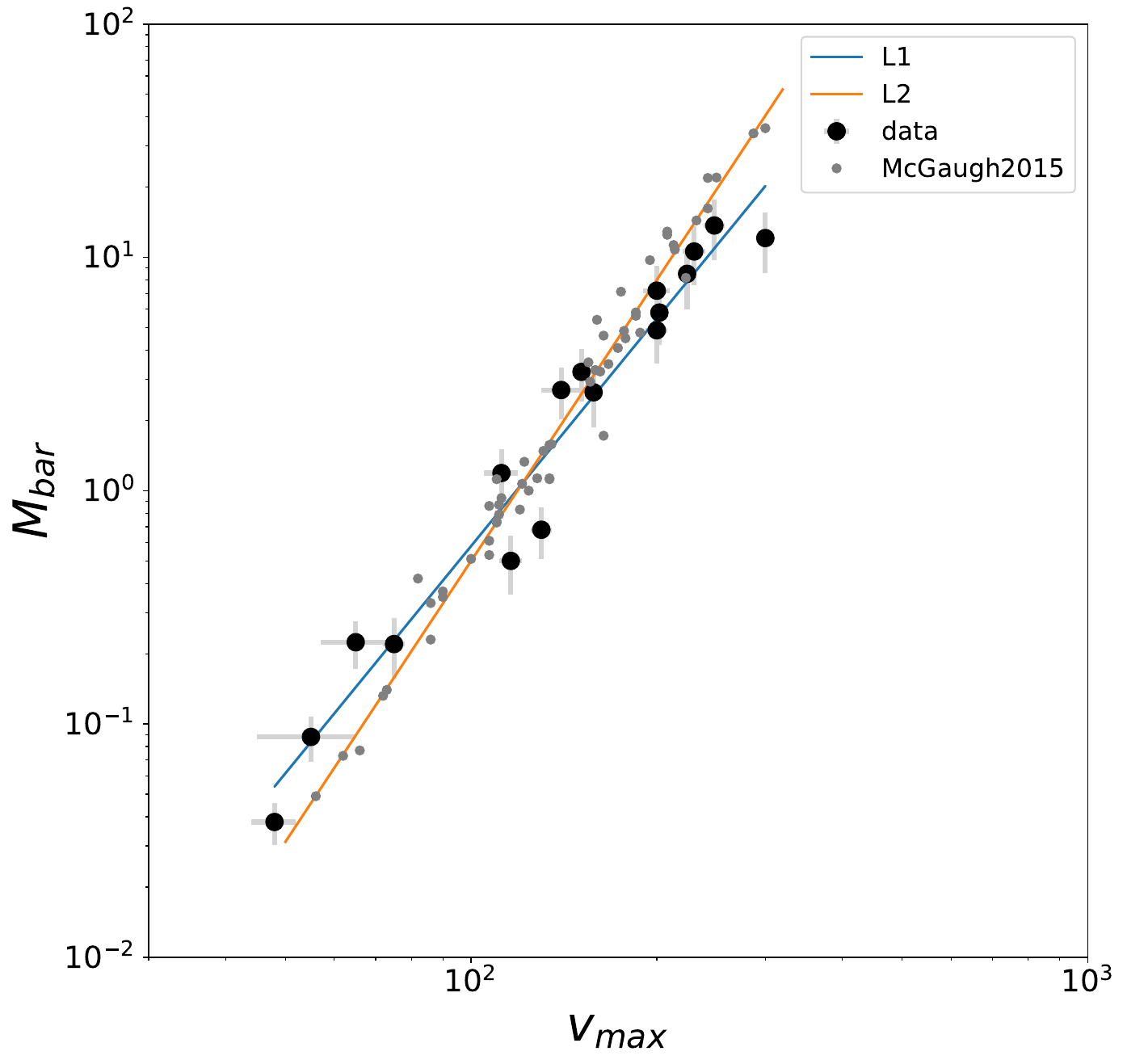}  
\caption{
Left panel: the Bosma Tully Fisher relation (see Eq. \ref{tf2}): $\Lambda$  (in $10^{10} M_\odot$  kpc$^{-1}$) vs $v_{\text{max}}$ (in km s$^{-1}$); the  
exponential thin   disc (ETD) model  of Eq. \ref{tf3} is reported: note that this is not a fit.
Right Panel: the baryonic Tully Fisher relation $M_{\text{bar}}$ (in units of  $10^{10} M_\odot$) vs $v_{\text{max}}$ (in km sec$^{-1}$). 
The best fit line L1 has slope $3.2$; for reference we report a line (L2) with slope 4.  For comparison  the data 
by 
McGaugh (2005)
are also shown.
}
\label{TF} 
\end{figure*}

In order to consider the connection between the Bosma effect and the TF relation,   \cite{Pfenniger+Revaz_2005} made a fit of the type 
\be
\label{BTF2} 
\log(M_{\text{s}} + c M_{\text{g}} ) = a + b \log (v_{\text{max}})
\ee
where $a,b,c$ are considered as free parameters so that the usual BFT relation corresponds to $c = 1$. A value $c>1$ corresponds to a re-scaling of the  \HI{}  mass in agreement with the Bosma model. They found  at the optimal value $c = 2.98$, where the
linear fit parameters are $a = 3.11$ and $b = 3.36$. Eq. \ref{BTF2} means that the mass of the gas (specifically, only hydrogen) was considered as proxy of a more massive DM component in agreement with the prescription of the Bosma effect.  The conclusion of \cite{Pfenniger+Revaz_2005}  was that the  BFT relation is  improved when the  \HI{}   mass is multiplied by a factor of about 3, but  they also  noticed that larger factors up to 11-16 still improve the fit over the original one using only the detected baryons. This result suggests that a substantial amount of baryons would remain to be found in spirals.

{ 
We can now advance beyond the results presented in \cite{Pfenniger+Revaz_2005} by connecting the  TF relation to the average rotation curve we have obtained, taking the Bosma effect into account. Since the average rotation curve of the galaxies in our sample aligns well with the ETD model, we can establish a simple relationship between the observed peak velocity $v_{\text{max}}$ in the rotation curve and the characteristic velocity $v_{\text{d}}$ as given by Eq. \ref{vd}. Indeed, for the ETD model, we find from Eq. \ref{eq:td} that \citep{Freeman_1970}.
}
\be
\label{tf1}
\frac{v_{\text{max}}}{v_{\text{d}}} = \alpha \approx 0.62 \;. 
\ee 
{ Note that, for the case of the ETD model,   the peak of the rotation curve $v_{\text{max}}$  is reached for $R=2.2 R_{\text{d}}$.}  
In our sample the estimation of the Bosma surface density gives $R_{\text{d}}$ and thus we can determine $v_{\text{max}}=v_{\text{c}}(R=2.2R_{\text{d}})$ from the rotation curve. 
In this respect, it should be noted that while the rotation curve is flat in many cases, there are instances where it may increase or decrease: 
for this reason our definition of $v_{\text{max}}$ is unambiguous and can be applied for all rotation curve shapes.

Considering Eq. \ref{vd} and Eq. \ref{tf1} we find 
\be
\label{tf2}
\Lambda \equiv \frac{M_{\text{dmd}}}{R_{\text{d}}} =  \frac{v_{\text{max}} ^2}{\alpha^2 G} \;,
\ee  
that we name the Bosma TF ({ BoTF}) relation. This can be rewritten as 
\be
\label{tf3}
\Lambda \approx 6.1\times 10^{-5} v_{\text{max}}^2 
\ee 
where  $\Lambda$ has units $10^{10} M_\odot$  kpc$^{-1}$ and  $v_{\text{max}}$ is in  km s$^{-1}$.  

{ In  the left panel of  Fig. \ref{TF} we plot  the  BoTF relation (Eq. \ref{tf2}): 
 we note that the data points are very nicely fitted by Eq. \ref{tf3}, i.e., 
 $\Lambda$ scales as $v_{\text{max}}^2$ with very small scattering. Moreover, the constant of proportionality between them
 can be analytically determined and it is simply $(\alpha^2 G)^{-1}$. }

{ 
We emphasize  that if the average rotation curve approximately matches with the ETD model then the relation between $\Lambda$ and $v_{\text{max}}$, i.e. the BoTF, is entirely justified. However, the BoTF does not contain any information other than the fact that we have fitted the DMD model to the rotation curves. Indeed, when the DMD mass is divided by the scale radius, the only dependence remaining is from $v^2(R_{\text{d}})$ (and thus $v_{\text{max}}$). As a result, $\Lambda$ is a parameter that does not provide information about the location of the maximum, and hence on the rotation curve shape. It merely depends on the possibility of the models to match the 'average magnitude' of the rotational velocity. This is achieved by construction when we fit the DMD to the observed rotation curves. So, the BoTF relation should always be present, even if the ETD does not provide a good overall match to the rotation curve shape. 

In brief, since the BoTF is inherent of the ETD model and given that the ETD model well describes the rotation curves of the galaxies we considered, not surprisingly we observe the BoTF relation also in the real data.
}

{ 

The right panel of Fig. \ref{TF} illustrates the BTF relation (Eq. \ref{BTF}). It is evident that for the galaxies within our sample, the best-fit exponent of Eq. \ref{BTF} is approximately $\beta \approx 3.3$, and a fit with $\beta=4$ remains acceptable.
It's worth noting that \cite{Lelli_etal_2019} demonstrated that the slope of the BTFR can depend on the choice of the velocity axis. Specifically, for $v_{max}$, they reported a slope of approximately $\approx 3.52$ (as shown in their Figure 2 and Table 1), which aligns closely with our findings.

The BoTF relation (Eq. \ref{tf3}) is indirectly connected to the BTF (Eq. \ref{BTF}): if there is a relationship between $M_{\text{dmd}}$ and $M_{\text{bar}}$, then it is possible to establish a connection between the BTF and the BoTF. Let's delve further into this issue.

}
\begin{figure*}
\centering 
\includegraphics[width=8cm,angle=0]{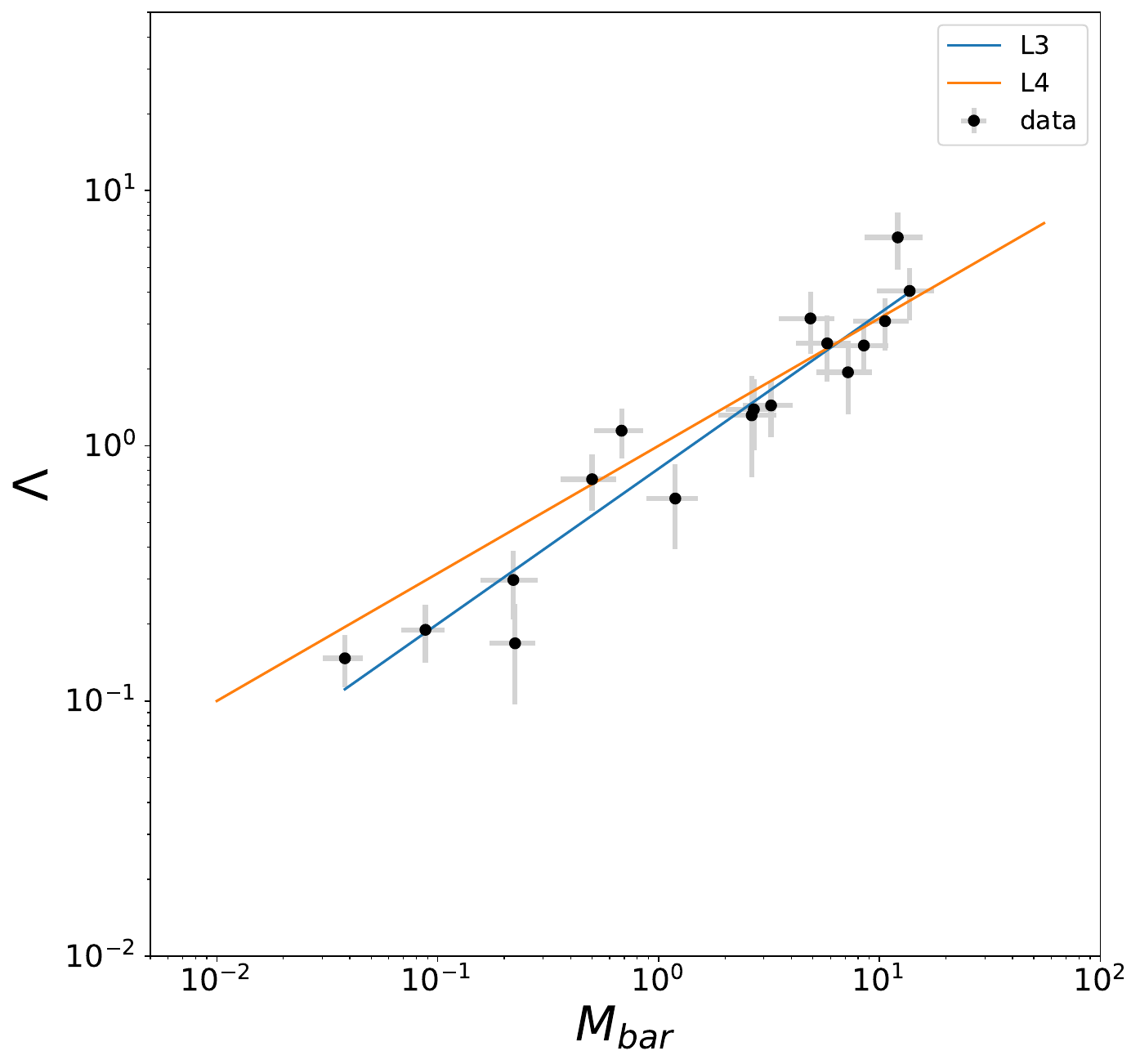}  
\includegraphics[width=8cm,angle=0]{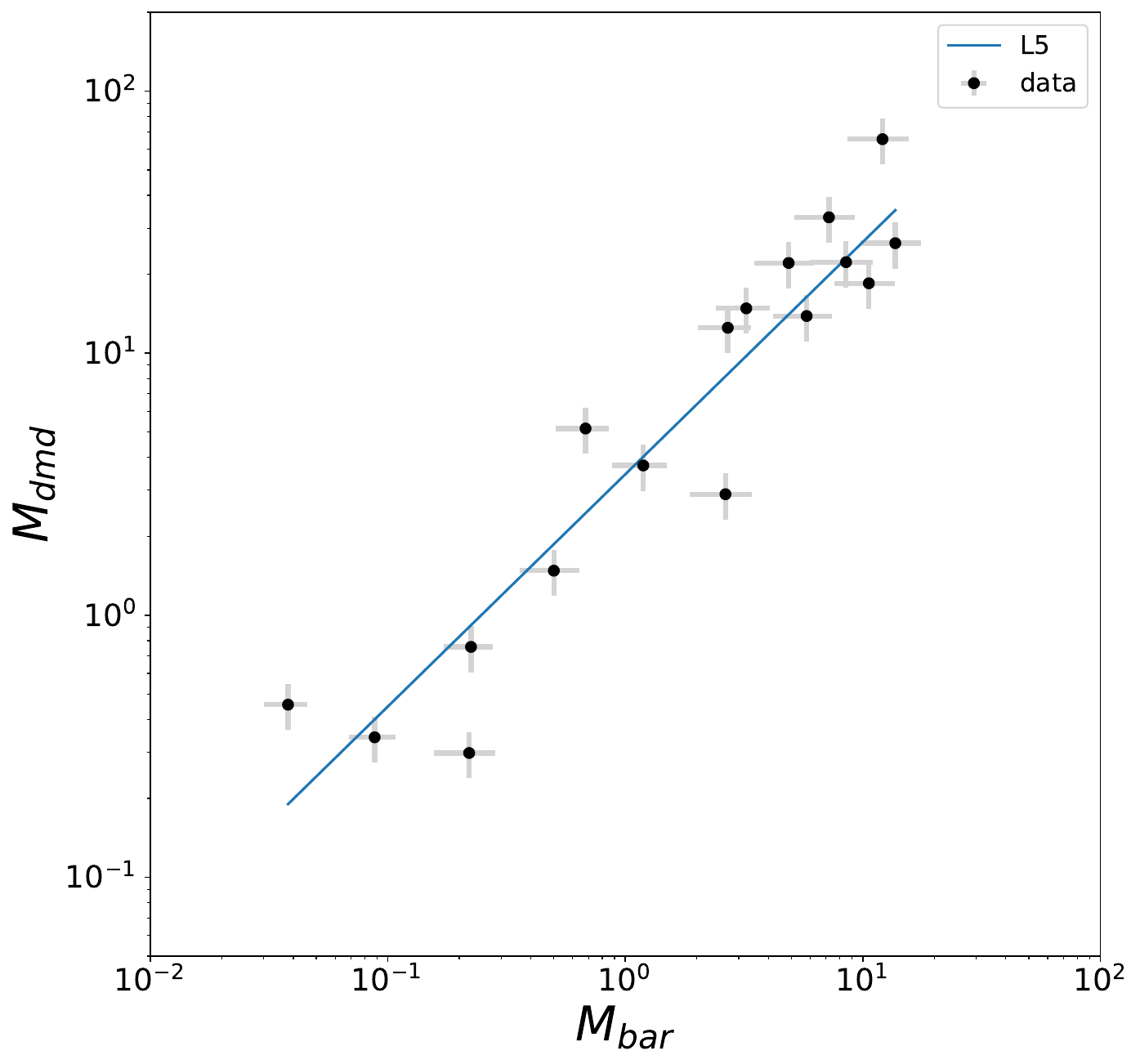}  
\caption{
Left panel: $\Lambda$  (in $10^{10} M_\odot$  kpc$^{-1}$)  in function of the baryonic mass (in $10^{10} M_\odot$), i.e. sum of mass of the stellar and gas components $M_{\text{bar}}$.
The best fit line L3  ($\Lambda= 0.81\times M_{\text{bar}}^{0.61}$) and the  line L4 with slope $\Lambda \propto M_{\text{bar}}^{0.5}$ are reported.
Right panel:  the DMD mass $M_{\text{dmd}}$ (in $10^{10} M_\odot$) as a function of the baryonic mass $M_{\text{bar}}$ (in $10^{10} M_\odot$). The best fit line L5  ($M_{\text{dmd}}=3.5 \times M_{\text{bar}}^{0.9}$) is reported.} 
\label{mass} 
\end{figure*}

{ 
Fig.\ref{mass} (left panel) shows that $\Lambda$ and $M_{\text{bar}}$ are tightly correlated: this correlation originates from the fact  that { the DM mass in the Bosma model}  $M_{\text{dmd}}$, { the} baryonic mass $M_{\text{bar}}$, { the stellar mass} $M_{\text{s}}$ and { the mass of the gaseous component} $M_{\text{g}}$ are linked by Eq. \ref{dmd3}. 
However, we cannot a priori determine the functional relation between $\Lambda$ and $M_{\text{bar}}$. 
 Empirically, in our sample, we find that  
 \be
 \label{Lambda_Mbar} 
 \Lambda  \propto M_{\text{bar}}^{\gamma}
 \ee where $\gamma\approx 0.6$;
in addition we also find that 
\[M_{\text{bar}} \propto M_{\text{dmd}}^{0.86} \; ,
\]
i.e., the baryonic and DMD mass are almost linearly correlated  (see the right panel of Fig.\ref{mass}) .

 Using  that $\Lambda  \propto M_{\text{bar}}^{\gamma}$ and Eq. \ref{tf2}, we find that the  BTF relation can be written as}  
\be
\label{eq:ML} 
M_{\text{bar}} \propto \Lambda^{\frac{1}{\gamma}} \approx \left(\frac{v_{\text{max}} ^2}{\alpha^2 G}\right)^{\frac{1}{\gamma}} \propto v_{\text{max}}^\frac{2}{\gamma} \;. 
\ee
{ If we take $\gamma=0.6$, as measured in our sample, the exponent in Eq. \ref{eq:ML} is 3.3 as, indeed, we found  in our data 
(see the right panel of Fig.\ref{TF}). However, as shown in  the right panel of Fig. \ref{TF}, 
 the value $\gamma=0.5$, that is still a good fit,  gives the exponent 4 in Eq. \ref{eq:ML}   equal to the value reported by  \cite{McGaugh_2005b}. 
Note that in our sample the maximum velocity spans about one decade, i.e. from 50 km s$^{-1}$ to $\approx 300$ km s$^{-1}$, which is about the same range in the sample considered by \cite{McGaugh_2005b} (also reported in the right panel of Fig.\ref{TF}): the main difference is that in the latter case the scatter is larger. 
}

 \subsection{Discussion}

{
In summary, we can conclude that the BoTF  relation observed in our sample of galaxies can be attributed to the following factors:

i) The surface density of the gas component follows an exponential decay for sufficiently large radii.

ii) By rescaling the rotation curve of the gas component by a constant factor, we obtain a rotation curve that matches the observed data. This phenomenon, known as the Bosma effect, provides reasonable fits to the rotation curves at large radii.

As a result of these empirical behaviors, the observed rotation curve can be well approximated by an ETD model, which establishes a relationship between $\Lambda$, the ratio  between the total mass of the DMD disc and its characteristic length-scale,  and the maximum rotational velocity ($v_{\text{max}}$) as expressed in Eq. \ref{tf3}. Additionally, considering the empirical correlation between $\Lambda$ and the baryonic mass ($M_{\text{bar}}$) we can establish a scaling argument that connects the BoTF relation (Eq. \ref{tf3}) to the BTF relation (Eq. \ref{BTF}). While the presence of such a correlation is expected based on the construction of the model (Eq. \ref{dmd2}), the precise value of the exponent needs to be determined from the data.
}


\begin{figure*}
\centering 
\includegraphics[width=8cm,angle=0]{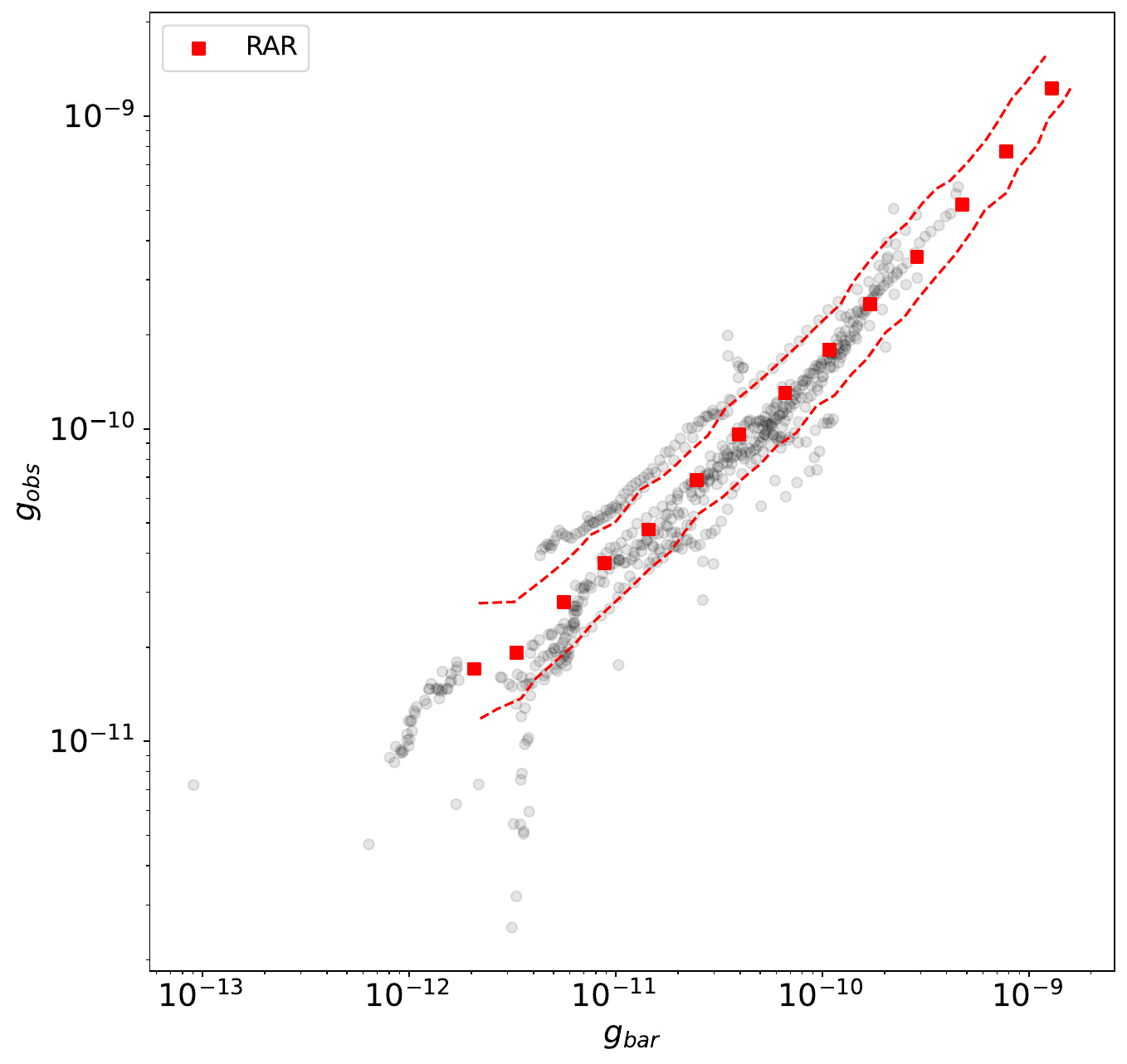} 
\includegraphics[width=8cm,angle=0]{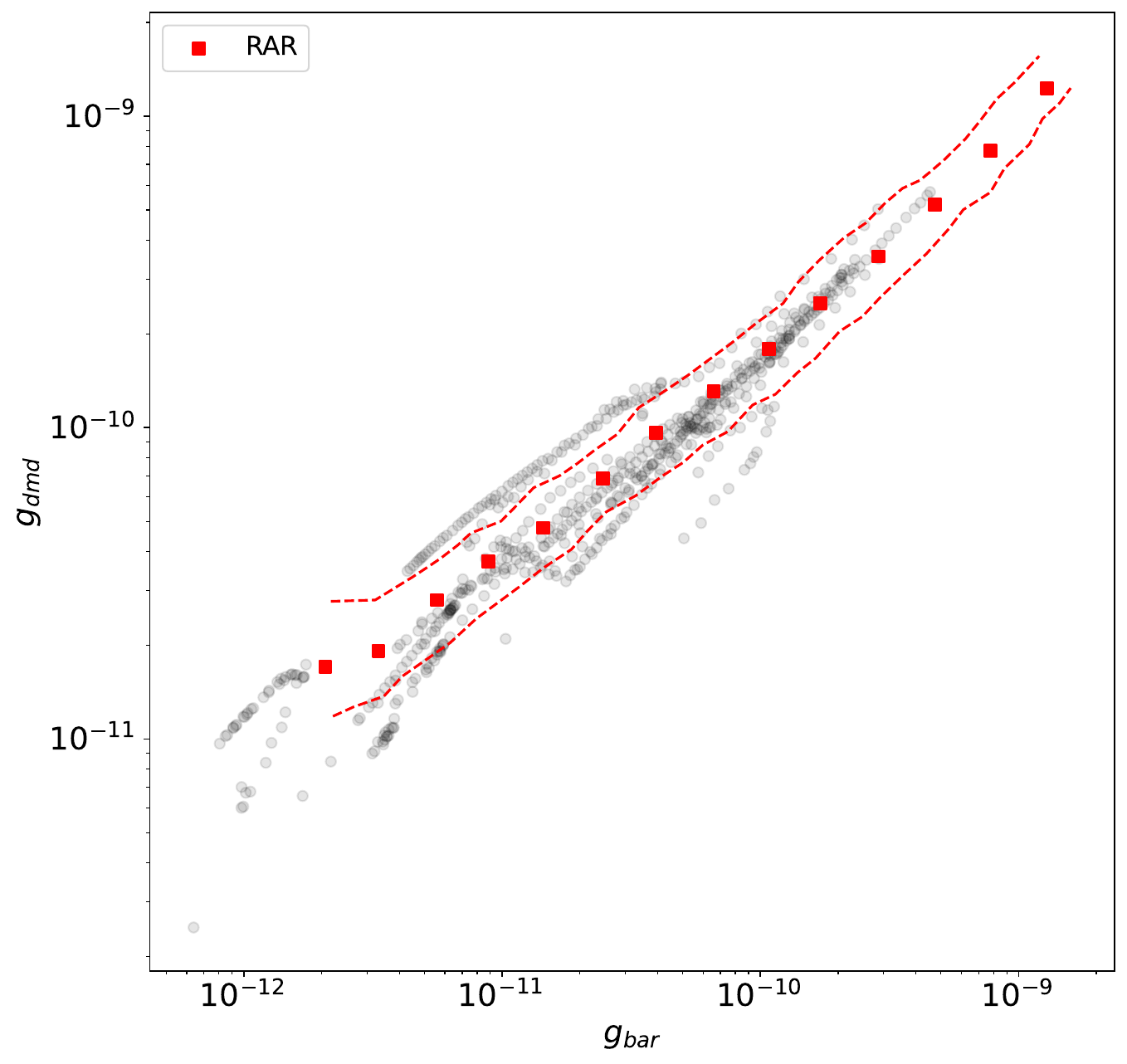} 
\caption{
 Left panel: the radial acceleration relation, $g_{\text{obs}}$ vs $g_{\text{bar}}$,   for the galaxies of our sample. The data of McGaugh et al. (2016) (average values and dispersion) are reported as reference (RAR).
Right panel: the radial acceleration relation in the DMD model, i.e., $g_{\text{dmd}}$ vs $g_{\text{bar}}$ (see text for details).  }
\label{fig:TF3} 
\end{figure*}

\subsection{The Bosma effect and the radial acceleration relation} 
{ 
The radial acceleration relation (RAR) is an empirical relationship that has been observed to apply to a broad spectrum of galaxies \citep{McGaugh_etal_2016_PRL,Lelli_etal_2017}. It suggests a connection between the visible matter and the overall gravitational potential within these galaxies. More precisely, the RAR establishes a connection between the radial acceleration $g_{\text{obs}}$, which is inferred from the rotation curves of disc galaxies and is attributable to the total mass of the galaxy, including both visible and dark matter components 
\[
g_{\text{obs}}  = \frac{V_{\text{c}}(R)^2}{R}  = \frac{\partial \Phi_{\text{tot}}(R)}{\partial R}  \;,
\] 
where $V_{\text{c}}(R)$ is the rotation curve and where $ \Phi_{\text{tot}}$ is the gravitational potential due to the total mass of the galaxy, and  the radial acceleration predicted by the observed distribution of baryons $g_{\text{bar}}$.
Specifically, by observing the distribution of stars and gas within a galaxy, one can create baryonic mass models. Subsequently, by considering that galactic discs possess a small but non-zero thickness, we numerically solve the Poisson equation, allowing the computation of $\Phi_{\text{bar}}(R)$, which represents the gravitational potential attributed to the baryonic components. This potential results in the acceleration experienced due to the cumulative effect of these baryonic components
\[
g_{\text{bar}}  = \frac{\partial \Phi_{\text{bar}} (R)}{\partial R}  \;.
\] 
Even though the mass models for individual galaxies exhibit considerable diversity, as we have previously discussed for the case of our sample,  there is a robust and consistent correlation between the observed acceleration $g_{\text{obs}}$ and the acceleration predicted by the baryonic matter $g_{\text{bar}}$ for all galaxies. It's important to note that in principle, there is no inherent guarantee that $g_{\text{obs}}$ should exhibit such a strong correlation with $g_{\text{bar}}$ when DM dominates the gravitational dynamics of galaxies \citep{McGaugh_etal_2016_PRL,Lelli_etal_2017}.
The distribution of DM  follows directly from the relation, and can be written entirely in terms of the
baryons:
$g_{\text{dm}} =  g_{\text{obs}} - g_{\text{bar}}  \;.$
The observed correlation between 
$g_{\text{obs}}$
and
$g_{\text{bar}}$ 
implies that dark and baryonic mass are strongly coupled  and  suggests that the baryons are the source of the gravitational potential.
 
The RAR finds a natural interpretation in the framework of the Bosma effect and  DMD model, as in this case  DM follows the distribution of baryons as traced by the gas component. 
To show that this is the case, first we have computed $g_{\text{obs}}$  and $g_{\text{bar}}$ from our data and we have found that galaxies in our sample 
 follow the RAR  (see the left panel of Fig.\ref{fig:TF3}) as measured by  \cite{McGaugh_etal_2016_PRL}. Then, we have  computed the correlation between  $g_{\text{bar}}$ and  the  radial acceleration    from  the best DMD model $ g_{\text{dmd}} $ finding that they are also tightly correlated  (see the right panel of Fig.\ref{fig:TF3}).   Even in this case,  it is not possible to predict in advance the specific exponent that relates these two radial accelerations: however,  we can infer that such a correlation is expected to naturally arise in the DMD model, where the distribution of baryonic matter traces that of DM.  }

 \subsection{The Tully Fisher relation in the halo model}

\begin{figure*}
\centering 
\includegraphics[width=8cm,angle=0]{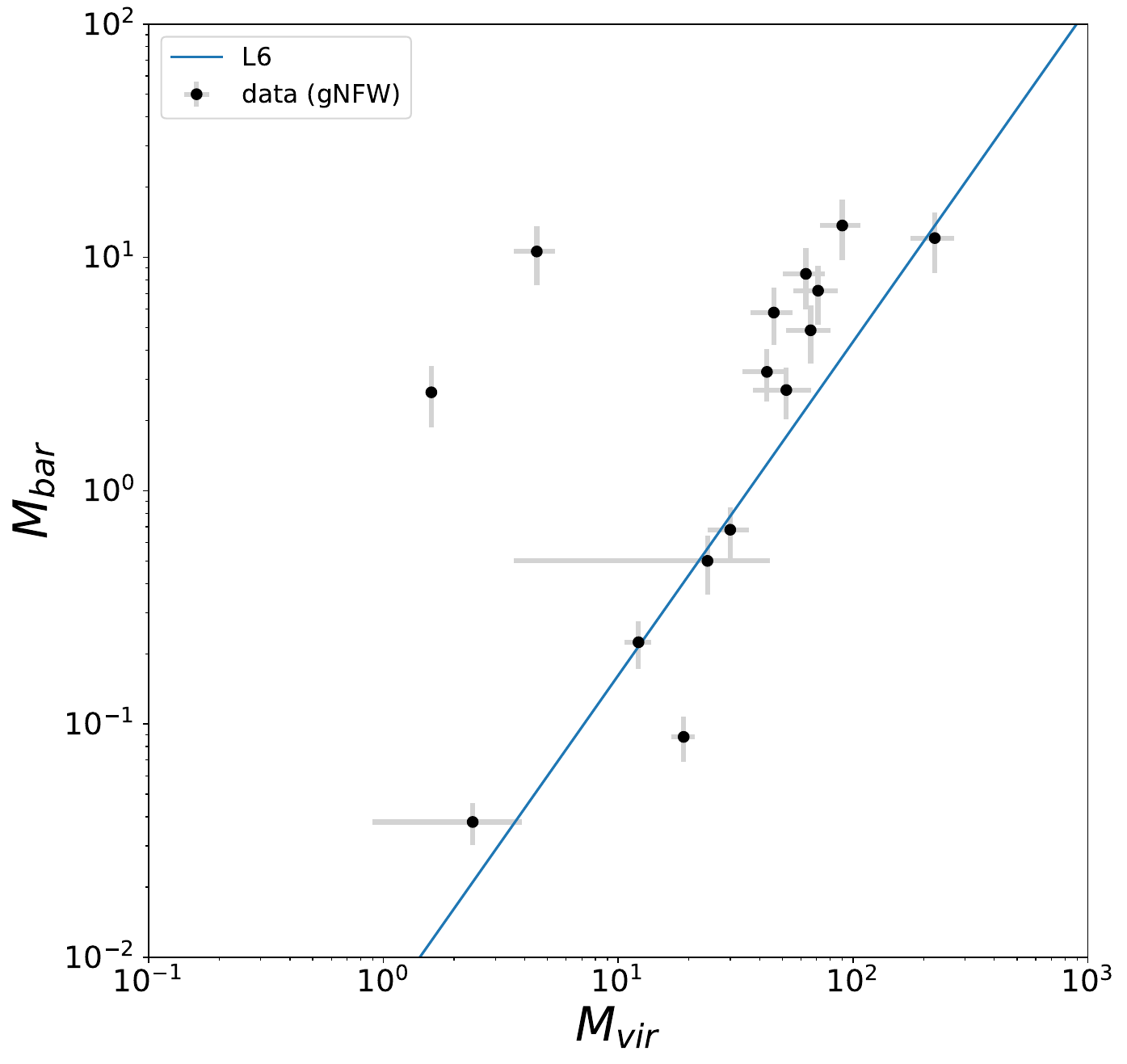} 
\includegraphics[width=8cm,angle=0]{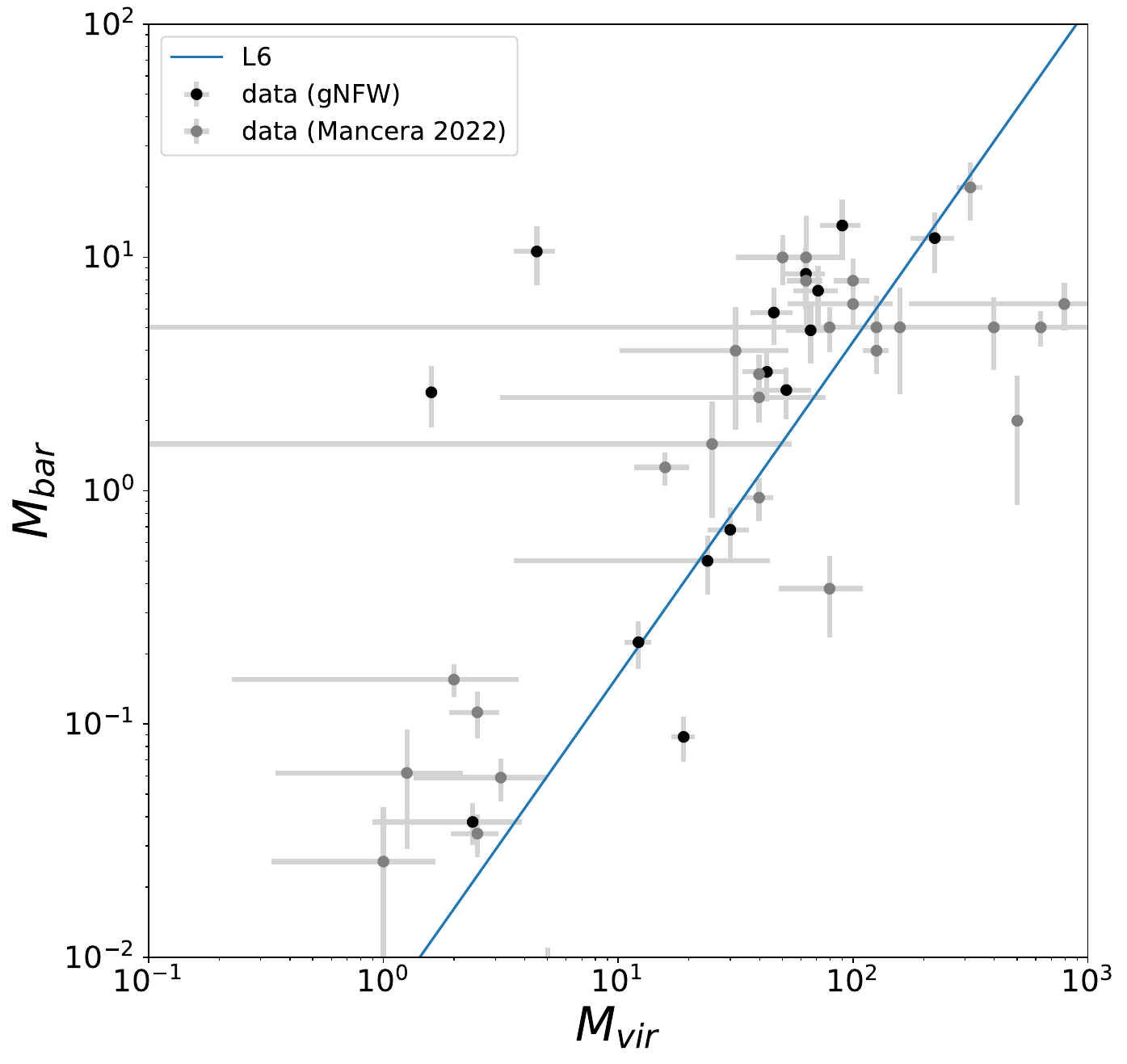} 
\caption{
 Left panel: baryonic mass $M_{\text{bar}}$ (in $10^{10} M_\odot$) vs NFW virial halo mass $M_{\text{vir}}$ (in $10^{10} M_\odot$). The reference line L6 has slope 1.43.
Right panel: The same but for the  gNFW virial halo mass $M_{\text{vir}}$ (in $10^{10} M_\odot$). 
The gray dots are the data from Mancera Pi\~{n}a et al. (2022).}
\label{fig:TF2} 
\end{figure*}

It is worth noticing that, in the framework of the NFW halo model,  the interpretation of the BFT relation  depends on how  dark and baryonic mass are related.  For an halo at virial equilibrium one finds (see, e.g., \cite{McGaugh_2012}) 
\be
\label{vhalo}	
M_{\text{vir}} = A V_{\text{vir}}^3  
\ee
where $A=4.6 \times 10^5 M_\odot$ km$^{-3}$ s$^3$ so that DM halos  obey a scaling relation reminiscent of the TF  relation with some important differences from the  observed one: indeed, both mass  and velocity refer to the virial radius $R_{\text{vir}}$ that is much larger than the typical range of observed radii. In order to map these theoretical quantities into the observed baryonic mass  and circular velocity, we need to introduce some factors which crucially depend on the fact that $M_{\text{vir}} $ is correlated with $M_{\text{bar}}$. 
{ 
In our sample, we have observed a significant scatter in the relation between the baryonic mass ($M_{\text{bar}}$) and the halo mass ($M_{\text{vir}}$). This finding is consistent with the results obtained by  \cite{Mancera_etal_2022} (see Fig. \ref{fig:TF2})). However, upon excluding two galaxies from the fit, we have identified a correlation between the baryonic mass and the halo mass, given by $M_{\text{bar}} \sim M_{\text{vir}}^\xi$, where $\xi \approx 1.4$ (see the left panel of Fig. \ref{fig:TF2}). This correlation can be considered as analogous to Eq. \ref{Lambda_Mbar} for the case of the BoTF relation. From Eq. \ref{vhalo} we thus find 
\be
M_{\text{bar}} \sim  V_{\text{vir}}^{3 \xi} \sim  V_{\text{vir}}^{4.3} \;.
\ee
This behavior is in proximity to the observed trend (see Fig. \ref{TF}).
The right panel of Figure \ref{fig:TF2} illustrates the same plot, but in this case, the halo mass was computed using the gNFW model. The trend observed in this plot is very similar to the previous one: most of the galaxies exhibit a consistent relationship between the baryonic mass and the halo mass. However, there are only two galaxies for which the halo mass differs significantly. These two galaxies, NGC 925 and NGC 2976, are characterized by a growing rotation curve.
The similarity in trends between the two plots, obtained using different halo mass calculation methods (NFW and gNFW models), suggests that the overall relationship between baryonic mass and halo mass remains robust. The slight discrepancies observed in the halo mass for specific galaxies can be attributed to the unique characteristics of their rotation curves. }

 \subsection{The Tully Fisher relation in Modified Newton Dynamics model}

As a final remark, we note that   MOND \citep{Milgrom_1983} makes strong a priori predictions about the BFT relation, which is a consequence of the form of the force law in MOND. This, for small enough  accelerations,  deviates from purely the Newtonian  behavior: in the deep modified regime the effective acceleration $a  = \sqrt{a_0 g_N}$ where  $g_N$ is the Newtonian acceleration calculated for the observed baryonic mass in the usual way and $a_0$ is the characteristic acceleration at which  the MOND regime becomes effective. For circular motion around a point source, by equating the centripetal acceleration to this effective force, one obtains 
\be 
a_0 G M_{\text{bar}} = v_{\text{max}}^4 \;. 
\ee
Note that the normalization is set by the constants of nature $a_0,G$. As discussed above in our sample we find a value of the exponent 
of the BFT relation that is definitively smaller than 4. 


\section{Conclusion} 
\label{conclusions} 

We have shown that the rotation curves of a group of nearby   disc galaxies can be accurately modeled using the Bosma effect and that it is possible to rescale the rotation curves of different   disc galaxies to the simple exponential thin disc behavior. This model, referred to as the Dark Matter Disc model (DMD), posits that DM is confined to a thin  disc. The   disc's mass surface density is dominated by that of the stars  at small radii and by that of the gas rescaled at larger ones. Both the star and gas surface brightness show approximately  exponential decays, but with different characteristic scale-lengths  where the stars have a smaller scale-length  than the gas. At small radii, the circular velocity is proportional to that caused by the stellar component, with the proportionality constant close to 1 in most cases, i.e.,  the rotation curve in the inner   disc can be almost entirely explained by solely luminous matter.  Therefore, the inner shape of rotation curves is well predicted by the distribution of observed baryons \citep{vanAlbada+Sancisi_1986,Palunas+Williams_2000}, implying that galactic   discs are maximal in the sense that they contribute the bulk of the mass at small radii. 

At large radii, the circular velocity is proportional to that caused by the gas, where the proportionality constant can be as large as 10 or more, indicating that the observed  \HI{}  and gas is just a proxy for a more massive DM component as assumed by the Bosma effect. 

{ 
Due to the exponential decay of the gas surface brightness in the galaxies under consideration, and assuming that the DM is confined to a thin disc due to the Bosma effect, the rotation curves at large radii for the selected sample of disc galaxies can be effectively approximated by the behavior of a thin disc \citep{Freeman_1970}.
To achieve this renormalization, the galaxy mass $M_{\text{dmd}}$ needs to be estimated consistently within the framework of the exponential thin disc model. 
It is important to note that this renormalization procedure is applicable primarily to galaxies with a surface brightness that decays exponentially. For galaxies whose surface brightness does not follow an exponential decay, the simple renormalization described above may not be valid, and the rotation curves of such galaxies are expected to deviate from the behavior expected for an exponential thin disc.
}

{ A global link between the baryonic mass of a galaxy and the amplitude of its rotation curve has already been established in the baryonic Tully-Fisher (BTF) relation  \citep{McGaugh_etal_2000,McGaugh_2005b}. The success of the baryonic scaling model indicates there is a more local coupling between the distribution of the baryons and the rotation curve shape.  However, the reason why the  BTF relation should be a power law relation between the   disc baryonic mass and the   disc rotational velocity remains unclear, because the improvement found on the TF relation still does not explain its physical origin; it just points toward an explanation based on the   disc baryonic content instead of either only the stellar or only the detected baryonic content.

In our study, we have demonstrated that the BTF relation can be formulated within the framework of the Bosma model, which we refer to as the Bosma TF (BoTF) relation. The BoTF relation arises from the observed exponential thin disc behavior exhibited by the disc galaxies in our sample. This behavior manifests as a strong correlation between the total mass $M_{\text{dmd}}$, the characteristic scale-length of the exponential decay $R_{\text{d}}$, and the rotation velocity $v_{\text{c}}$. To establish a connection between the BoTF relation and the more widely known BTF relation, an additional relationship between the disc DM mass $M_{\text{dmd}}$ and the baryonic mass $M_{\text{bar}}$ is required. Within the DMD model, the specific functional form of this relation cannot be determined a priori. However, our analysis of the data has revealed a tight power-law relationship between these two quantities.

Given this situation, it is indeed not surprising that the Radial Acceleration Relation (RAR) finds a natural framework within the context of the Bosma model. The RAR is an empirical relationship that has been observed to hold for a wide range of galaxies, as demonstrated by studies such as \citep{McGaugh_etal_2016_PRL,Lelli_etal_2017} . This relation suggests a connection between the visible matter and the overall gravitational potential in these galaxies. In the Bosma model, where DM follows the distribution of gas, the RAR can be readily incorporated. This model naturally accommodates the observed empirical relationship, as the distribution of DM is tied to the distribution of baryonic matter, particularly the gas component.
}

{

The hypothesis that DM is confined to a disc presents several theoretical challenges, including the dynamical stability of massive discs and the origin of their exponential mass distribution. In the standard Cold Dark Matter (CDM) scenario, the process of structure formation follows a bottom-up hierarchical pattern, where non-collisional and non-dissipational non-baryonic DM forms a nearly spherical halo structure with an almost isotropic velocity dispersion. In this framework, baryonic matter undergoes dissipative collapse in the gravitational field of the halo structure, leading to the formation of a disc dominated by rotational motions. This disc is embedded in a DM component, which is instead dominated by isotropic motions and is nearly spherical in shape. The prevalence of rotational motion in the disc is a consequence of the conservation of angular momentum during the collapse process.

The confinement of DM to a disc challenges the conventional framework and raises questions about the stability and origin of such disc structures. In response to this challenge, it has been proposed that heavy DM discs can form through non-collisional and non-dissipational processes if gravitational collapse follows a top-down scenario. This  approach, as discussed in studies by \cite{Benhaiem+SylosLabini+Joyce_2019,SylosLabini_etal_2020,SylosLabini_CapuzzoDolcetta_2020}, suggests that in a top-down gravitational collapse model,  as initial deviations from spherical symmetry are amplified during the violent collapse phase, isolated  out-of-equilibrium perturbations  may give rise to massive discs in quasi-stable configurations.   This  formation mechanism broadens the range of possibilities for the origin of disc structures and offers an alternative perspective to the bottom-up hierarchical scenario. Notably, recent work by \cite{Peebles_2020} advocates for  this kind of monolithic top-down scenario for galaxy formation. This approach aims to address the challenges faced by standard CDM-like models in explaining crucial observational aspects of galaxies. A monolithic top-down collapse could potentially be facilitated by a sharp cutoff in the matter power spectrum at small scales, such as the one associated with a warm dark matter initial mass fluctuation power spectrum.

From an observational standpoint, as mentioned in the introduction, a direct detection such cold clouds in the outer   disc of galaxies appears to be a very hard task \citep{Combes+Pfenniger_1997}. However, we note that disc galaxies offer valuable opportunities to impose stringent constraints on the symmetries of their mass distribution through the study of strong gravitational lensing. Strong lensing provides a distinct advantage compared to methods relying on stellar or gas kinematics, as it does not depend on the detection of luminous tracers within the galaxy under investigation. Instead, lensing directly probes the mass distribution in regions predominantly influenced by DM  (see, e.g., \cite{Sonnenfeld_eta_2020} and references therein).
In this regard, future samples of strong lensing events in disc galaxies, coupled with high-resolution imaging, will enable the measurement of the relative mass ratios between the bulge, disc, and DM halo of these galaxies. Furthermore, it will facilitate the determination of the distribution of the total mass flattening parameter, which characterizes the overall shape of the mass distribution. These observations hold the potential to refine our understanding of the mass distribution within disc galaxies and the differences and/or similarities between the distributions of  the baryonic and DM  components.

}

\section*{Acknowledgments}

FSL thanks for useful  comments and discussions Roberto Capuzzo-Dolcetta, \v{Z}ofia Chrob\'akov\'a, Frederic Hessman, Michael Joyce, 
Martin Lopez-Corredoira, Daniel Pfenniger,  Alessandro Sonnefeld and Hai-Feng Wang. 
{ We also  thank an anonymous referee for a number of  interesting comments and suggestions that have allowed us to improve the presentation of our results.}
This work has made use of data from the  The  \HI{}  Nearby Galaxy Survey (THINGS)  data-set available at the web address\\ 
{\tt https://www2.mpia-hd.mpg.de/THINGS/}

\section*{ Data availability} 
The data used in this paper are from the THINGS survey \citep{Walter_etal_2008}.

\bibliographystyle{mnras}

\section{Appendix: Results for individual galaxies}


\begin{table*}
\centering 
\begin{tabular}{c c c c c c c c c c  } 
\hline
\hline 
    Galaxy                           &$R_{\text{d}}$ & $M_{\text{s}}$ & $M_{\text{g}}$   & $\Upsilon_{\text{s}}$      &    $\Upsilon_{\text{g}}$  & $M_{\text{bar}}$ &   $M_{\text{dmd}}$    &  $\chi^2$  & $v_{\text{max}}$ \\ 
\hline 
NGC 925$^{2,3}$              &     6.0$\pm$1      & 0.72     & 0.5         & 1.0                   &6.4                         &  1.22     &4.0    (0.73,3.2)         & 2.7           &   117$\pm$7   \\
NGC 2366$^3$                  &    1.8$\pm$0.1   & 0.018   & 0.07       & 5.0                   &3.6                         &   0.088  &0.35  (0.1,0.25)         & 0.3           &   55$\pm$10  \\
NGC 2403$^{1,2}$            &     4.5$\pm$0.1   &0.33      & 0.35       & 2.9                  &12.0                        &  0.68     & 5.2   (1.0,4.2)           & 1.1           &  130$\pm$5   \\
NGC 2841$^{1,2}$            &    10.0$\pm$0.5  & 11.0     & 1.1         & 1.3                   &46.6                       &  12.1     & 65.6 (14.3,51.3)       & 1.9           &  300$\pm$8   \\
NGC 2903$^{1,2}$            &    7.0$\pm$0.5    & 4.0       & 0.9        & 1.5                   &18.5                        &  4.9       & 16.6 (6.0,10.6)         & 2.6           &  200$\pm$8   \\
NGC 2976$^{1,2}$            &    1.0$\pm$0.1    & 0.2       & 0.02       & 0.8                   &7.3                         &  0.22     & 0.30 (0.15,0.15)       & 1.4           &  75$\pm$3     \\
NGC 3198$^{1,2}$            &    10.3$\pm$0.5  & 1.8       & 1.4         & 1.8                   &8.1                         &  3.2       & 14.5 (3.2, 11.3)         & 0.7           &   151$\pm$5  \\
NGC 3521$^{1,2}$            &    6.0$\pm$0.5    & 9.0       & 1.6         & 1.2                   &4.8                         & 10.6      & 18.7  (11,7.7)            & 0.9           &   230$\pm$10 \\
NGC 3621$^3$                 &     9.0$\pm$1      & 1.4       &  1.3        &  1.5                  &8.0                          &  2.7       & 12.5   (2.1,10.4)        & 0.6          &   140$\pm$10 \\
NGC 4736$^{1,2}$            &    2.5$\pm$0.5    & 2.5       & 0.14       &  0.24                  &16.4                       &  2.6        & 2.9    (0.6, 2.3)        & 0.9         &   158$\pm$5   \\
NGC 5055$^{1,2}$            &    15.0$\pm$2     & 6.0        & 1.2         &  1.1                  &22                          &  7.3        & 33   (6.6, 26.4)          & 1.2         &   200$\pm$10 \\
NGC 6946$^{2}$               &    5.5$\pm$0.5    & 4.4        & 1.4         &  1.4                  &5.5                         &  5.8        & 13.9   (6.2, 7.7)         & 1.7         &  202$\pm$5    \\
NGC 7331$^{2}$               &    6.5$\pm$0.2    & 12.0      & 1.7         &  1.1                  &7.7                         & 13.7     & 26.3   (13.2, 13.1)       & 0.2         &   248$\pm$10 \\
NGC 7793$^{1}$               &    2.0$\pm$0.1    & 0.41      & 0.09       &  1.5                  & 9.6                         & 0.5       & 1.5     (0.6, 0.9)           & 0.3        & 116$\pm$5     \\
DDO 154$^3$                   &    3.1$\pm$0.1    & 0.002     & 0.036    &   1.0                  &12.6                        &  0.038  &  0.46   (0.002,0,46)     & 1.2        &  48$\pm$4      \\
IC 2574$^3$                     &    4.5$\pm$1.0     & 0.074    & 0.15       &   2.7                  & 3.7                        &  0.22    & 0.75    (0.2,0.55)           & 0.6         &  65$\pm$8      \\
MW$^2$                           &    9.0$\pm$0.2     &  8.0     &  0.5          &  1.2                   & 26.1                      &   8.5      & 22.3    (9.2,13.1)          & 1.1       & 229$\pm$5   \\
\hline
\end{tabular}
\caption{The index 
$^1$ is for the galaxies for which the data from the Pipeline 5 of S$^4$G are available providing both the stellar surface brightness profile and the stellar mass;  the index  
$^2$ when the H$_2$ surface density profile  from  
Schruba et al. (2011)
is known ;  the index    
$^3$ when the stellar surface brightness is from the 2MASS survey  
deBlok et al. (2008)
and the stellar mass from  
Hessman \& Ziebart (2011).
The best-fit values  of $\Upsilon_{\text{s}}$,  $\Upsilon_{\text{g}}$  and  $M_{\text{s}},M_{\text{g}},M_{\text{bar}}, M_{\text{dmd}}$ (in units of $10^{10} M_\odot$) are reported;  the contributions related to the stellar  $\Upsilon_{\text{s}} M_{\text{s}}$ and
gas $\Upsilon_{\text{g}}  M_{\text{g}}$ components are shown in parenthesis.
For the case of the Milky Way (MW) we refer the reader to Sylos Labini et al. (2023b)
for further details.}
\label{table:results1} 
\end{table*}

\begin{table}
\centering 
\begin{tabular}{c c c c   } 
\hline
\hline 
    Galaxy             &$R_{\text{d}}$ &    $\Upsilon_{\text{g}}$    &  $\chi^2$  \\
\hline  
NGC 925           &   25.5     & 90   & 1.9          \\
NGC 2366         &   3.0       & 7.9  & 0.2      \\
NGC 2403         &   5.1       & 14.9& 2.2         \\
NGC 2841         &   13.0     & 52.5& 2.5          \\
NGC 2903         &    8.7      & 16.7 & 0.7        \\
NGC 2976         &    16.2    &1080  & 1.5      \\
NGC 3198         &    10.4    & 9.3    & 0.9        \\
NGC 3521         &    5.2      & 5.4    & 3.2         \\
NGC 3621         &     10.0   & 8.7    &  0.5       \\
NGC 4736         &    18.6    & 56     & 0.7     \\
NGC 5055         &    42.6    & 69     & 5.3         \\
NGC 6946         &   5.3       & 5.4     & 2.2           \\
NGC 7331         &    8.7      & 9.8     & 0.3          \\
NGC 7793         &    3.6      & 24.4   & 1.4           \\
DDO 154           &    3.3      & 12.4   & 0.1         \\
IC 2574             &    24.7    & 178    & 0.6          \\
MW                  &    7.4       &  24     &  2.0             \\
\hline
\end{tabular}
\caption{ Dark Matter disc fits obtained by considering free parameters the gas length-scale $R_d$ (in kpc) and the DMD mass, i.e. $M_{\text{DMD}} = \Upsilon_{\text{g}} M_g + M_s $ where $M_g, M_s$ are reported in Tab.\ref{table:results1}. In the last column it is reported the value of reduced $\chi^2$ obtained from the fit.}
\label{table:results1a} 
\end{table}

\begin{table}
\centering 
\begin{tabular}{c c c c c c } 
\hline
\hline 
    Galaxy              &  $R_{\text{vir}}$ &  $M_{\text{vir}}$              &  $c$     & $c_{\text{vir}}$  &     $\chi^2$\\ 
\hline 
NGC 925              &    240        &     155                   & 2.4      & 7.1       &      4.8         \\
NGC 2366            &    99          &     11                     & 5.5      & 9.0       &      0.72         \\
NGC 2403            &    138        &     30                     & 15.0    & 8.0       &      0.43         \\
NGC 2841            &    275        &     233                   & 16.0    &  6.6      &      0.4         \\
NGC 2903            &   185         &     71                     & 13.3    & 7.5       &      0.3         \\
NGC 2976            &    118        &     18                     & 2.0      & 10.2     &      1.7         \\
NGC 3198            &    160        &     46                     & 10.1    & 8.0      &       0.3          \\
NGC 3521            &    9.6         &     4.5                    & 2.0      & 7.9      &       0.5         \\
NGC 3621            &    186        &     72                     & 6.8      & 7.8      &       0.1        \\
NGC 4736           &      25         &     0.17                  & 19.0    & 13.2    &       1.1        \\
NGC 5055            &    211       &     107                    &  6.5     & 7.5       &      0.5        \\
NGC 6946            &   161         &     47                     &  15       & 7.7      &        0.8        \\
NGC 7331            &   200         &     90                     &  11.8    &7.4      &       0.1         \\
NGC 7793            &   208         &     102                   &  5.6     & 7.6      &       2.6                \\
DDO154               &     87         &     7.5                    &  4.5     & 9.9      &       0.5               \\
IC2574                  &     89        &     8.0                    &  4.5      & 9.9      &       2.0               \\
MW                       &     178       &     63                     &  14.3    & 7.5      &       1.7               \\
\hline
\end{tabular}
\caption{ The best-fit values of the virial radius  $R_{\text{vir}}$ (in kpc) and the mass $M_{\text{vir}}$ (in units of $10^{10} M_\odot$) of the NFW halo, together  with the best-fit concentration parameter $c$ 
and the expected concentration parameter $c_{\text{vir}}$ from Eq. \ref{cnfw} are reported.
}\label{table:results2} 
\end{table}

\begin{table}
\centering 
\begin{tabular}{c c c c c c c} 
\hline
\hline 
    Galaxy                             &  $R_{\text{vir}}$ &  $M_{\text{vir}}$    &$\beta$           &  $c$     & $c_{\text{vir}}$  &     $\chi^2$\\ 
\hline 
NGC 925              &    484        &     1274            & 0.4       & 4.9      & 5.6       &     2.2         \\
NGC 2366            &    120        &     19                & 0.9       & 5.4      & 8.5       &     0.71         \\
NGC 2403            &    139        &     30                & 0.9       & 16.8    & 8.0       &     0.40         \\
NGC 2841            &    270        &     223              & 0.9       & 18.0    & 6.5       &     0.4         \\
NGC 2903            &   180         &     66                & 0.7       & 18.0    & 7.4       &     0.3         \\
NGC 2976            &    486        &    1300             & 0.5       & 4.9      & 5.8       &     1.7         \\
NGC 3198            &    156        &     43                & 0.8       & 13.0    & 7.9       &      0.3          \\
NGC 3521            &    9.6         &     4.5               & 1.0       & 2.0      & 7.9       &      0.5         \\
NGC 3621            &    167        &     52                & 0.7       & 11.3    & 7.8       &       0.1        \\
NGC 4736           &     53          &     1.6              & 0.1       & 10.5    & 10.9     &      1.0        \\
NGC 5055            &    184        &     71                & 0.2       &  16.8    & 7.3      &       0.5        \\
NGC 6946            &   160         &     46                & 1.0       &  15.0    & 7.7      &       0.8        \\
NGC 7331            &   200         &     90                & 1.0       &  11.8    &7.4       &       0.1         \\
NGC 7793            &   129         &     24                &  0.4      &  17.8    & 5.6      &       2.0         \\
DDO154               &     60         &     2.4               &  0.2      &  17.1    & 10.2     &       0.2         \\
IC2574                  &   103        &     12.2             &   0.1      &  10.5      & 9.0     &       0.8               \\
MW                       &   175         &     60                 &  0.8       &  17.8    & 7.4      &       0.8               \\
\hline
\end{tabular}
\caption{
The best-fit values of the virial radius  $R_{\text{vir}}$ (in kpc) and the mass $M_{\text{vir}}$ (in units of $10^{10} M_\odot$) of the gNFW halo (see Eq. \ref{gnfw_profile}), 
together  with the best-fit concentration parameter $c$ 
and the expected concentration parameter $c_{\text{vir}}$ from Eq. \ref{cnfw} are reported.}
\label{table:results2a} 
\end{table}


\begin{figure*}
\centering
\begin{subfigure}{0.35\textwidth}
    \includegraphics[width=\textwidth]{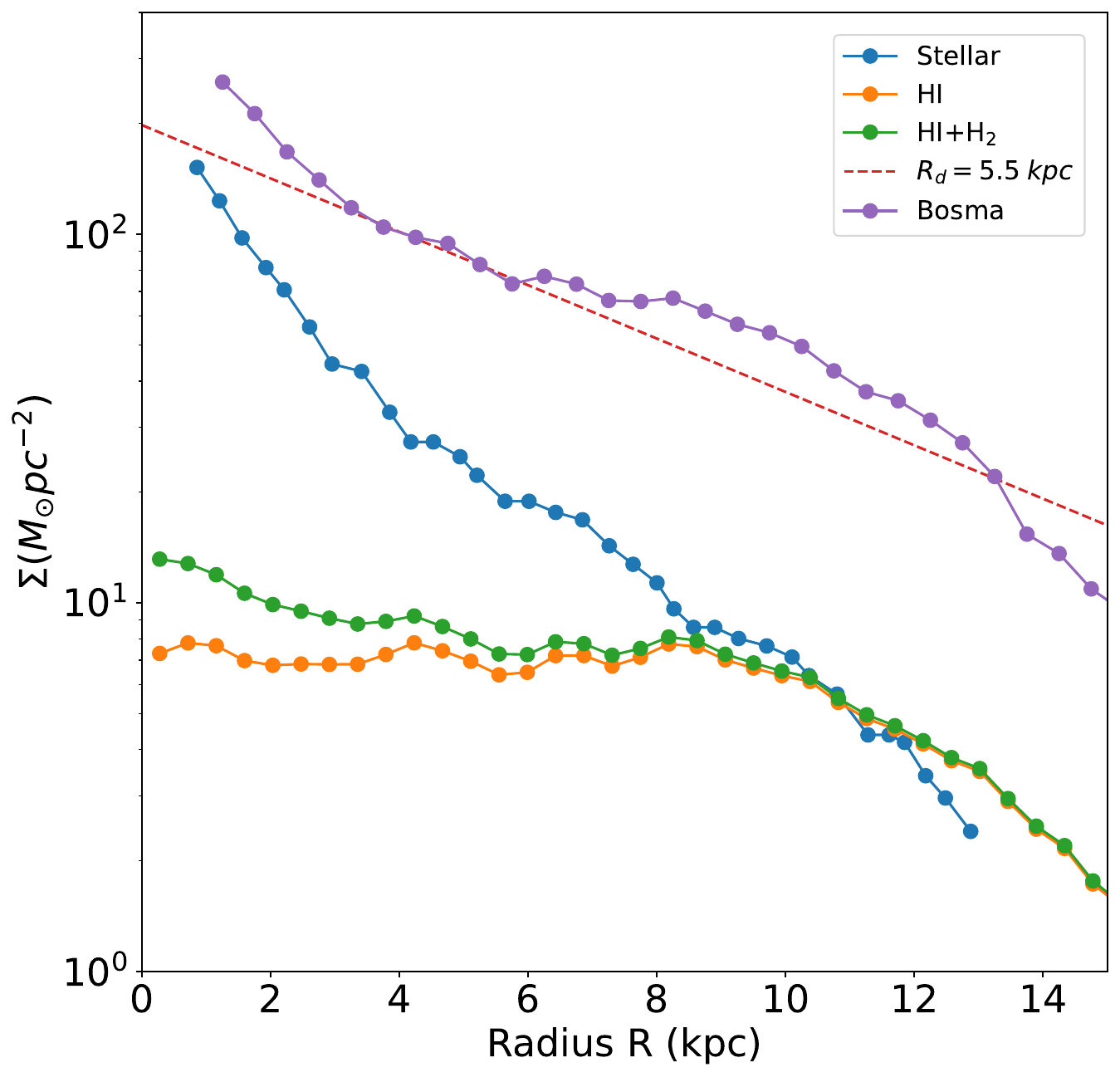}
    \caption{NGC 925}
    \label{fig:first}
\end{subfigure}
\begin{subfigure}{0.35\textwidth}
    \includegraphics[width=\textwidth]{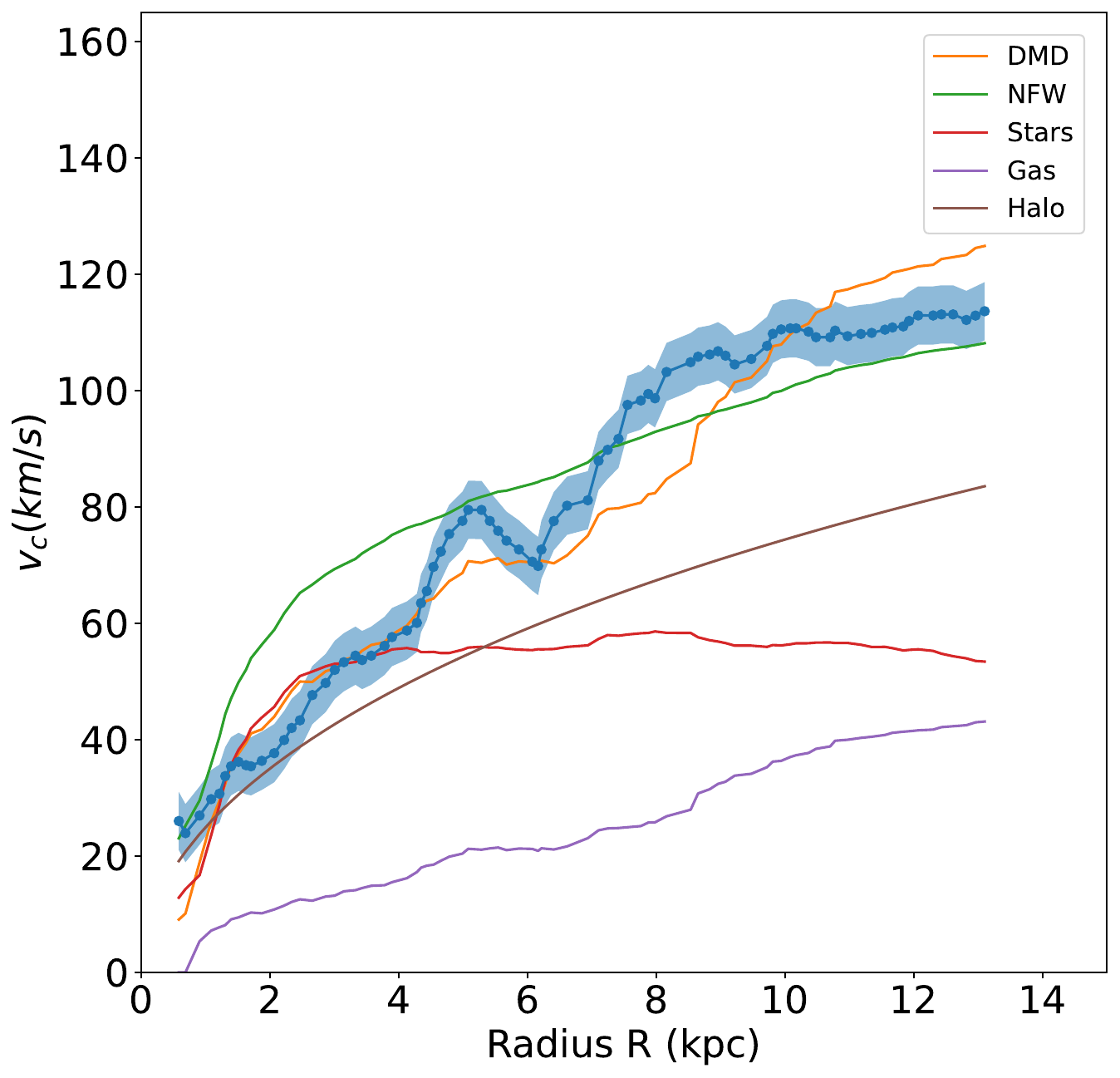}
    \caption{NGC 925}
    \label{fig:second}
\end{subfigure}
\begin{subfigure}{0.35\textwidth}
    \includegraphics[width=\textwidth]{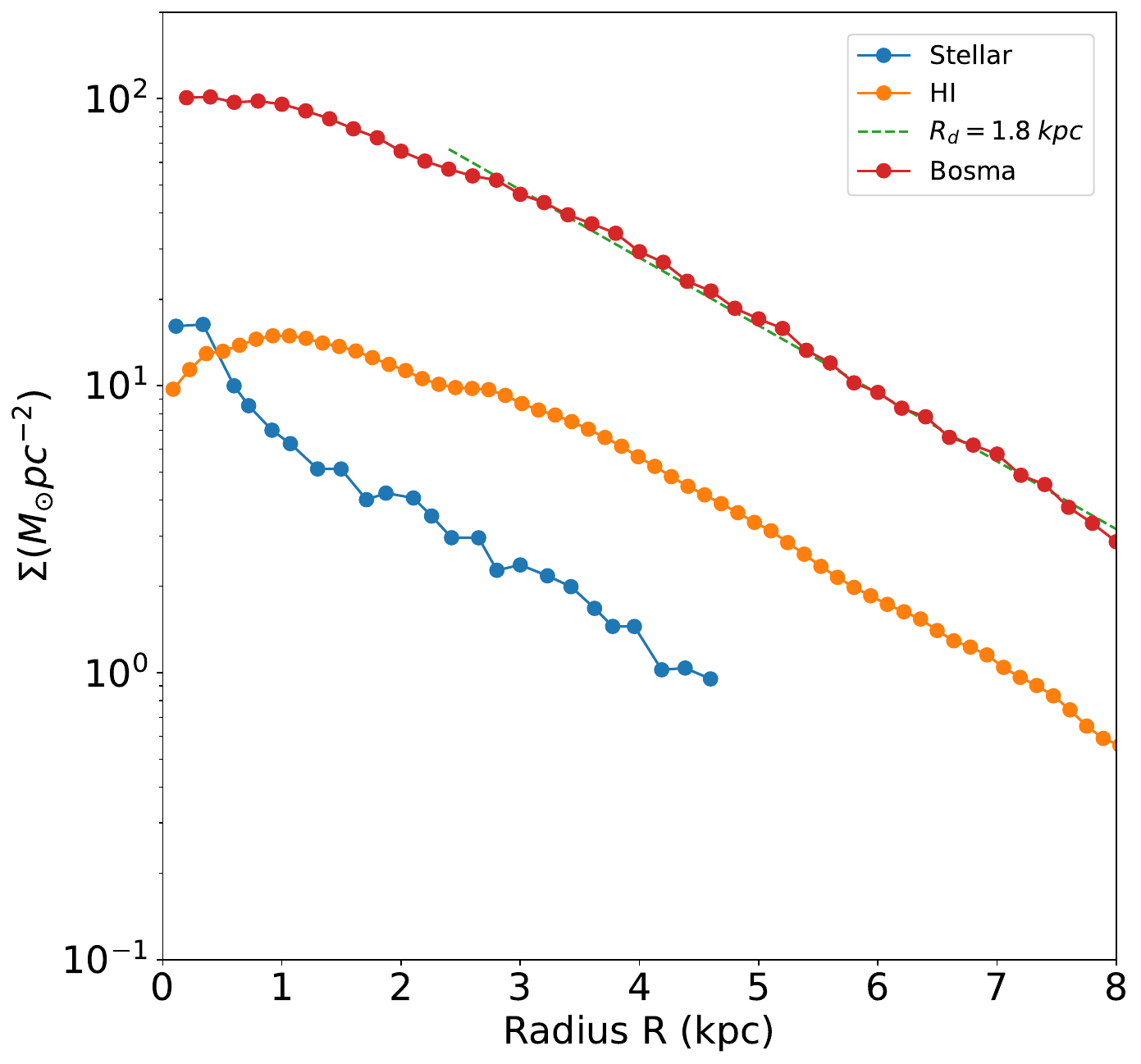}
        \caption{NGC 2366}
    \label{fig:third}
\end{subfigure}
\begin{subfigure}{0.35\textwidth}
    \includegraphics[width=\textwidth]{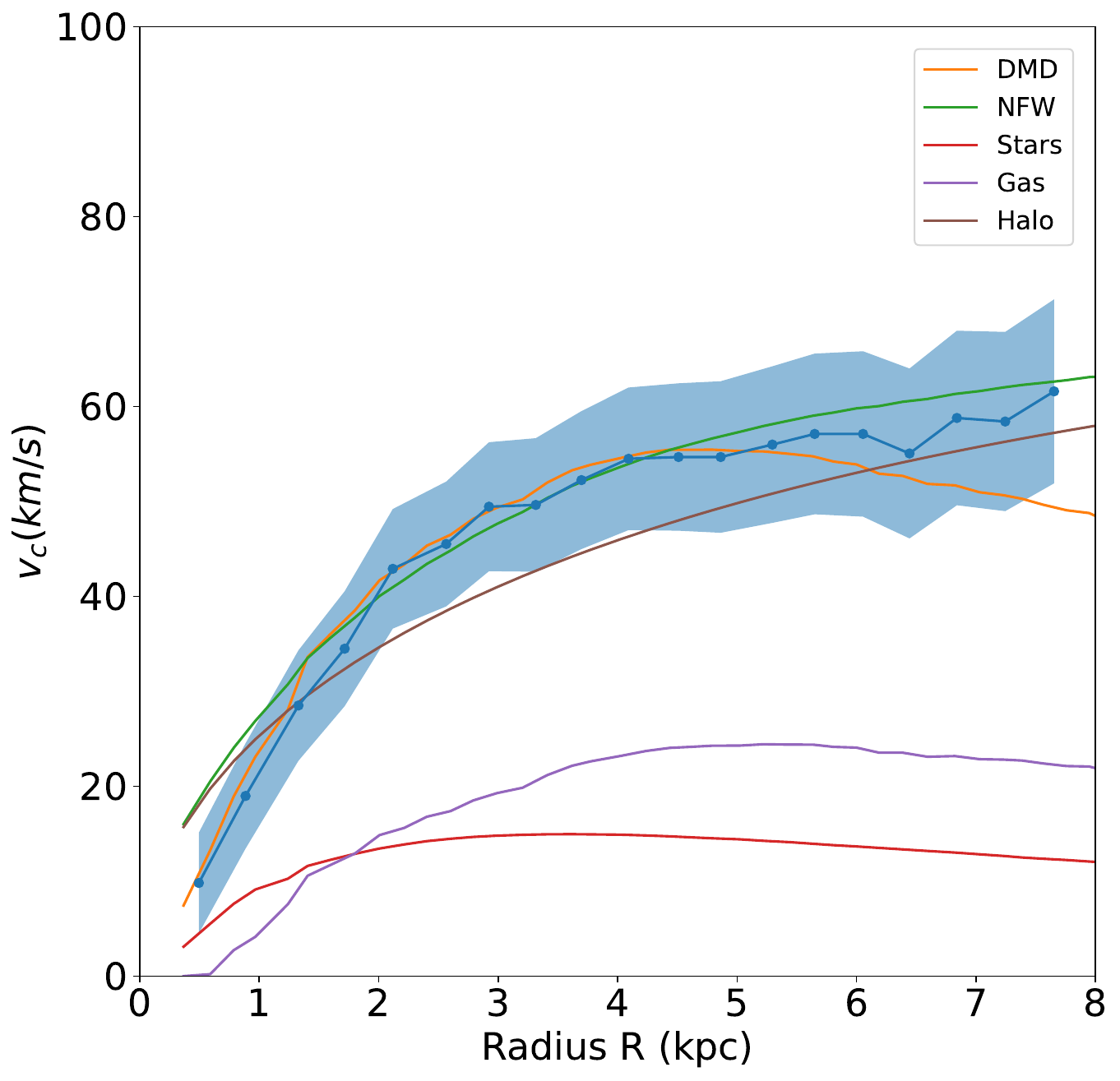}
            \caption{NGC 2366}
    \label{fig:third}
\end{subfigure}
\begin{subfigure}{0.35\textwidth}
    \includegraphics[width=\textwidth]{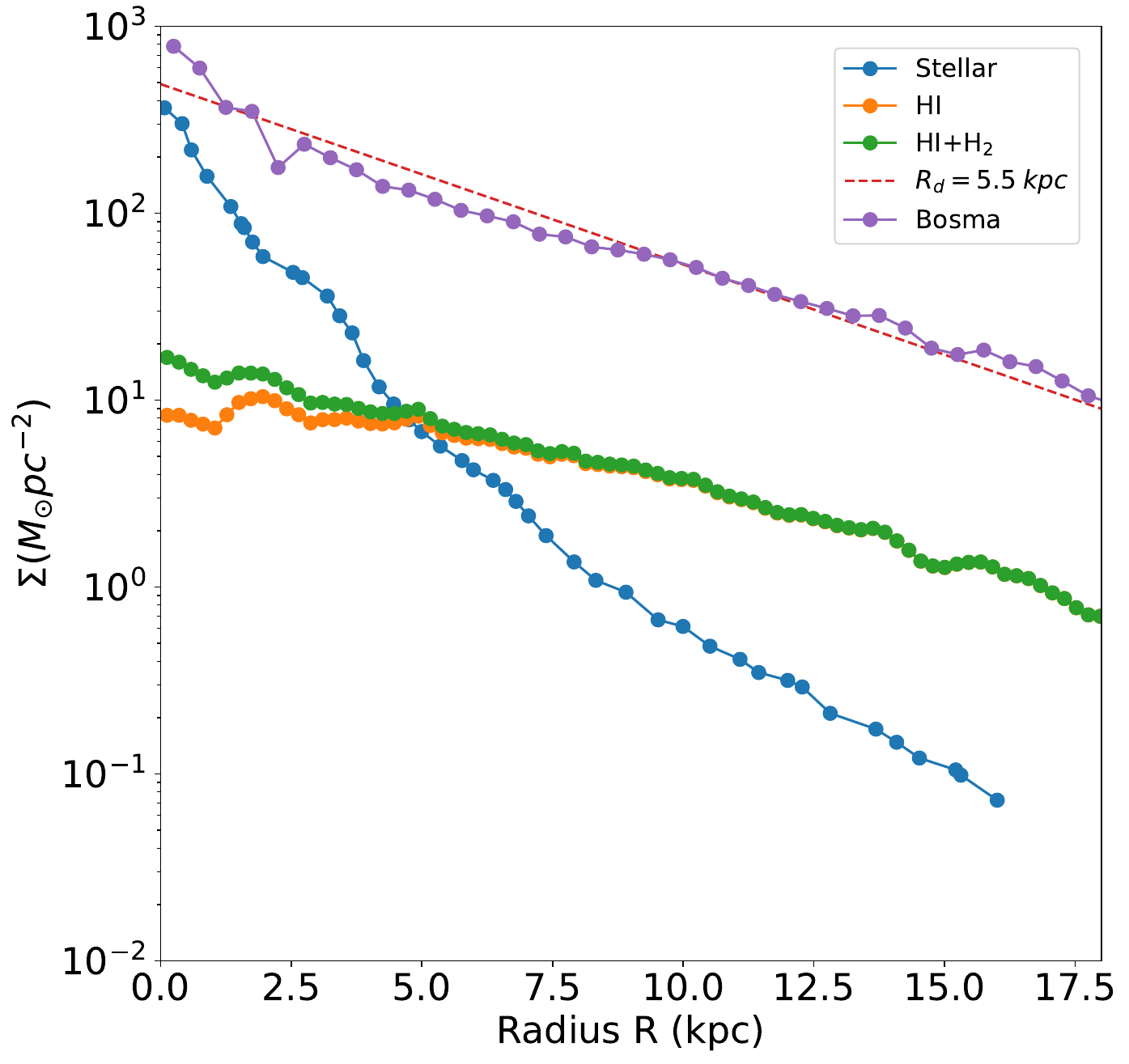}
    \caption{NGC 2403}
    \label{fig:first}
\end{subfigure}
\begin{subfigure}{0.35\textwidth}
    \includegraphics[width=\textwidth]{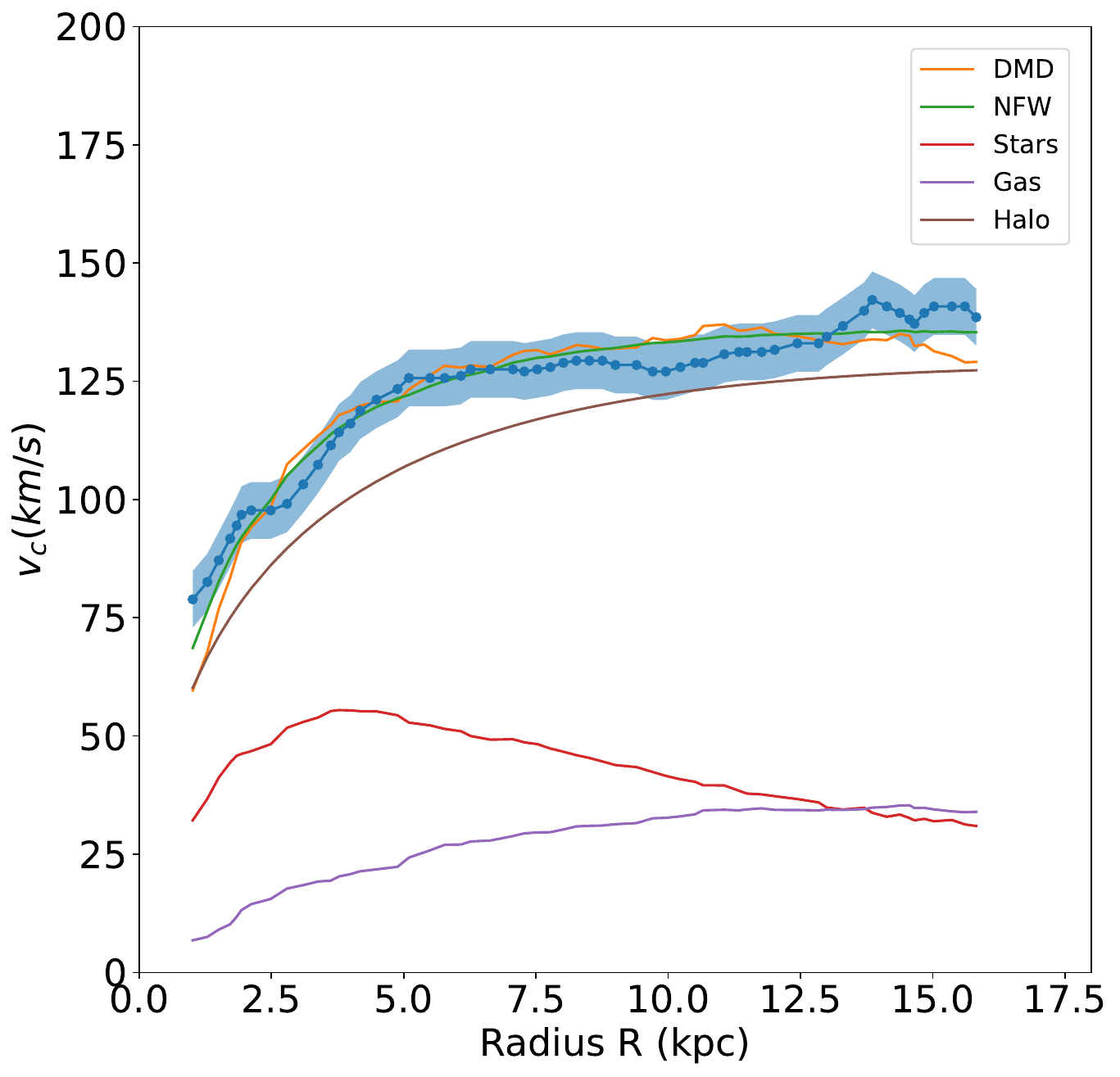}
    \caption{NGC 2403}
    \label{fig:second}
\end{subfigure}
\caption{For each galaxy we show the surface brightness profile of the stellar and gas component and the rotation curve}
\label{fig1}
\end{figure*}


\begin{figure*}
\begin{subfigure}{0.35\textwidth}
    \includegraphics[width=\textwidth]{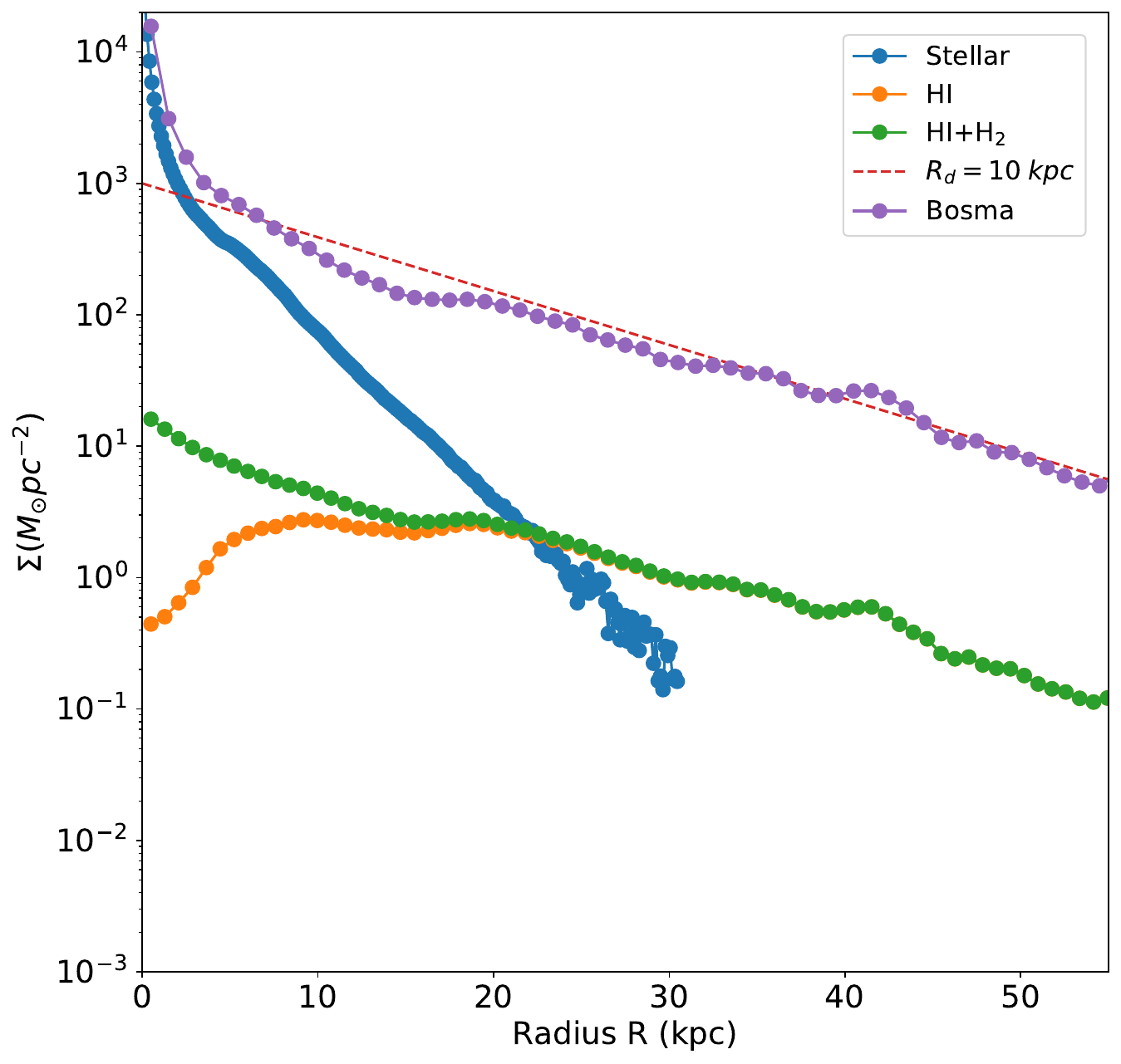}
        \caption{NGC 2841}
    \label{fig:third}
\end{subfigure}
\begin{subfigure}{0.35\textwidth}
    \includegraphics[width=\textwidth]{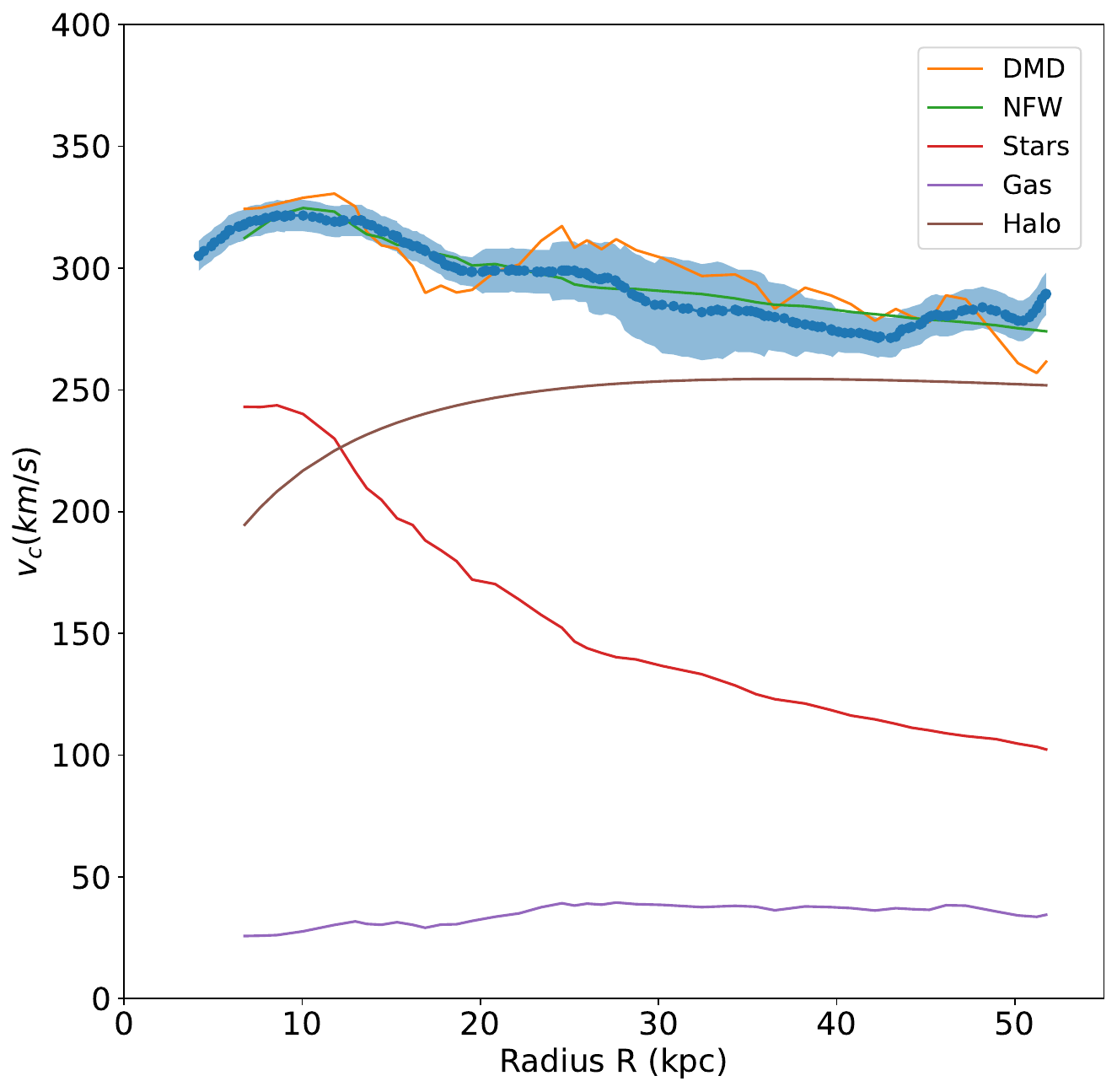}
            \caption{NGC 2841}
    \label{fig:third}
\end{subfigure}

\begin{subfigure}{0.35\textwidth}
    \includegraphics[width=\textwidth]{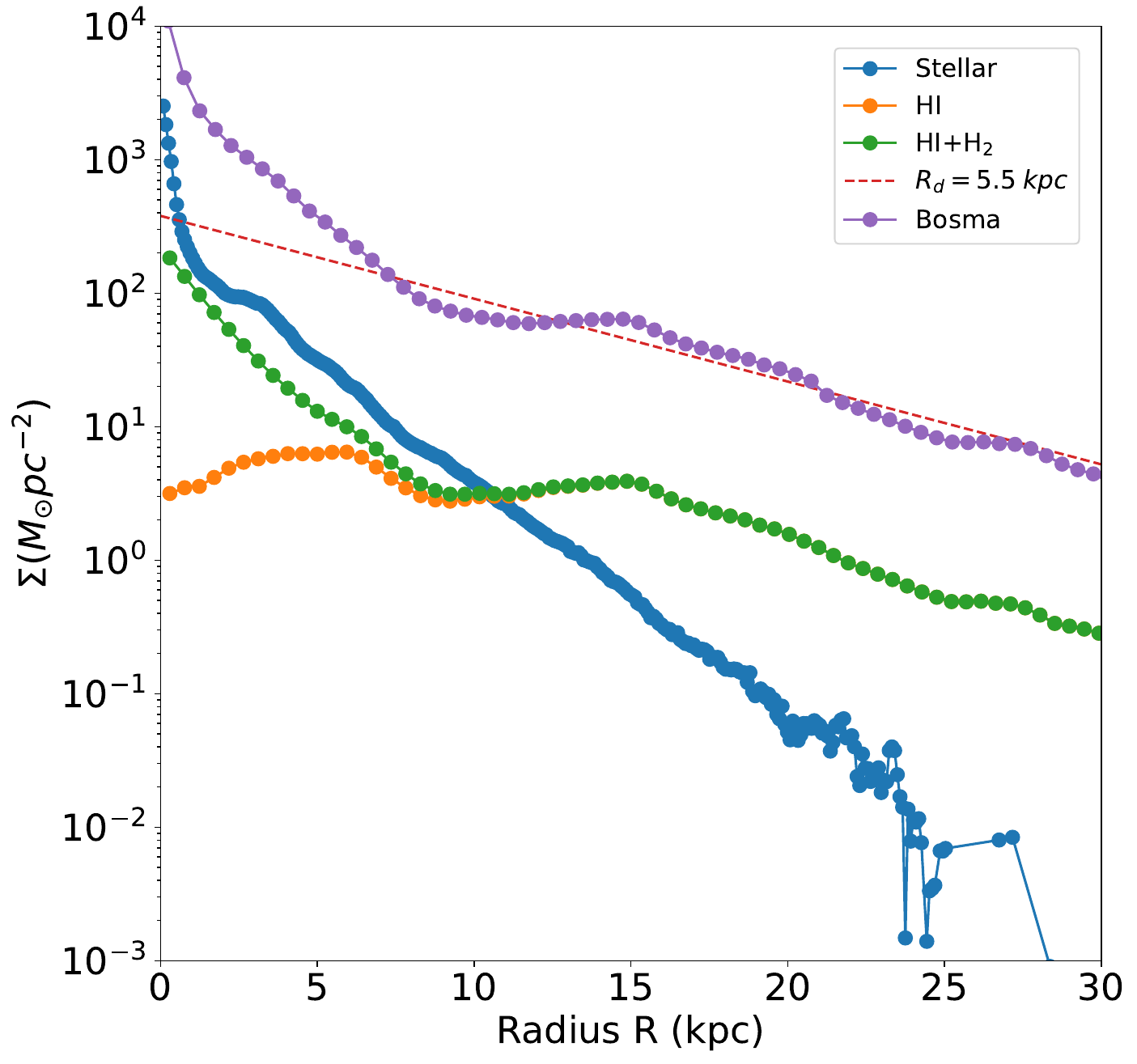}
    \caption{NGC 2903}
    \label{fig:first}
\end{subfigure}
\begin{subfigure}{0.35\textwidth}
    \includegraphics[width=\textwidth]{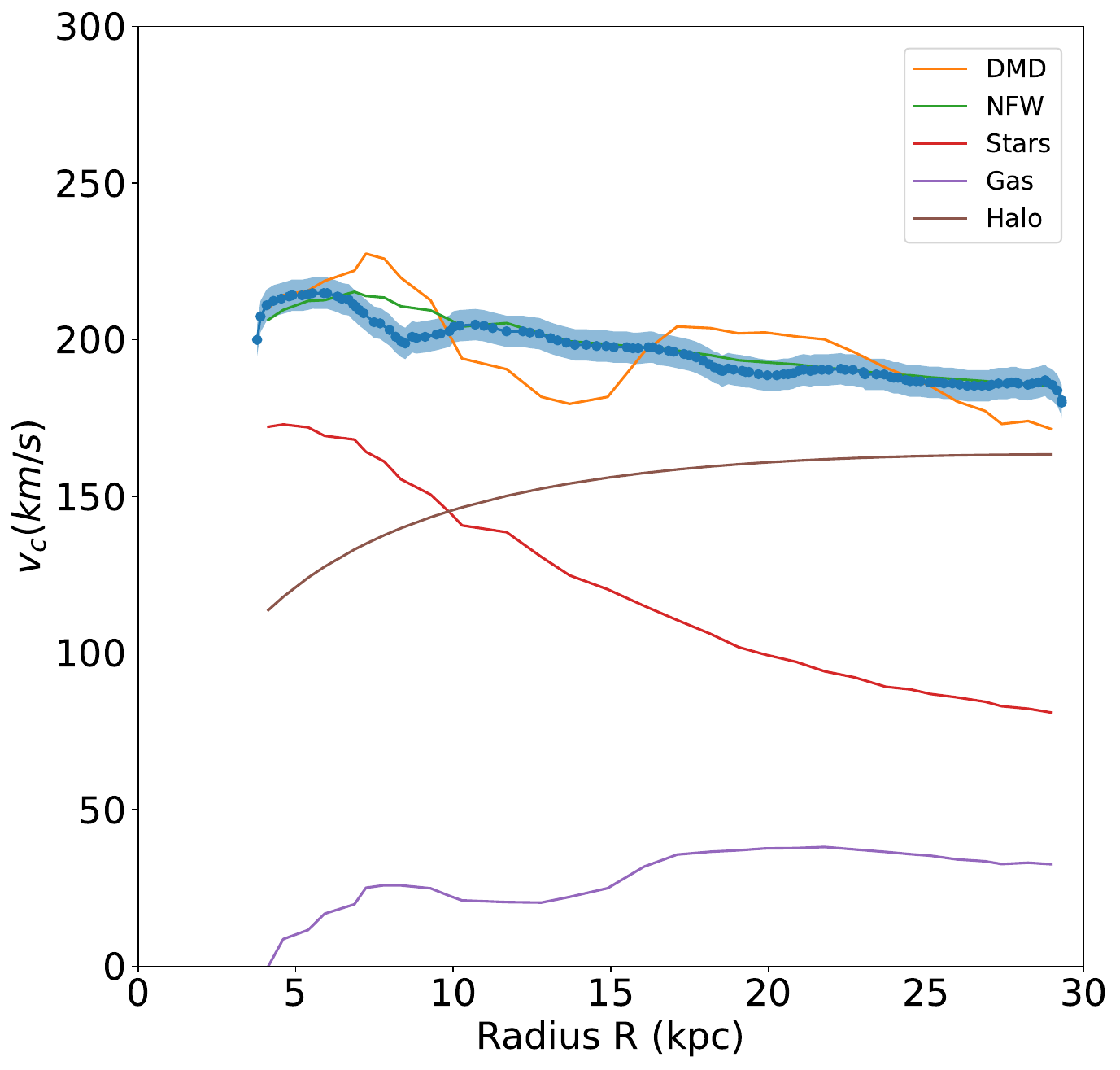}
    \caption{NGC 2903}
    \label{fig:second}
\end{subfigure}
\begin{subfigure}{0.35\textwidth}
    \includegraphics[width=\textwidth]{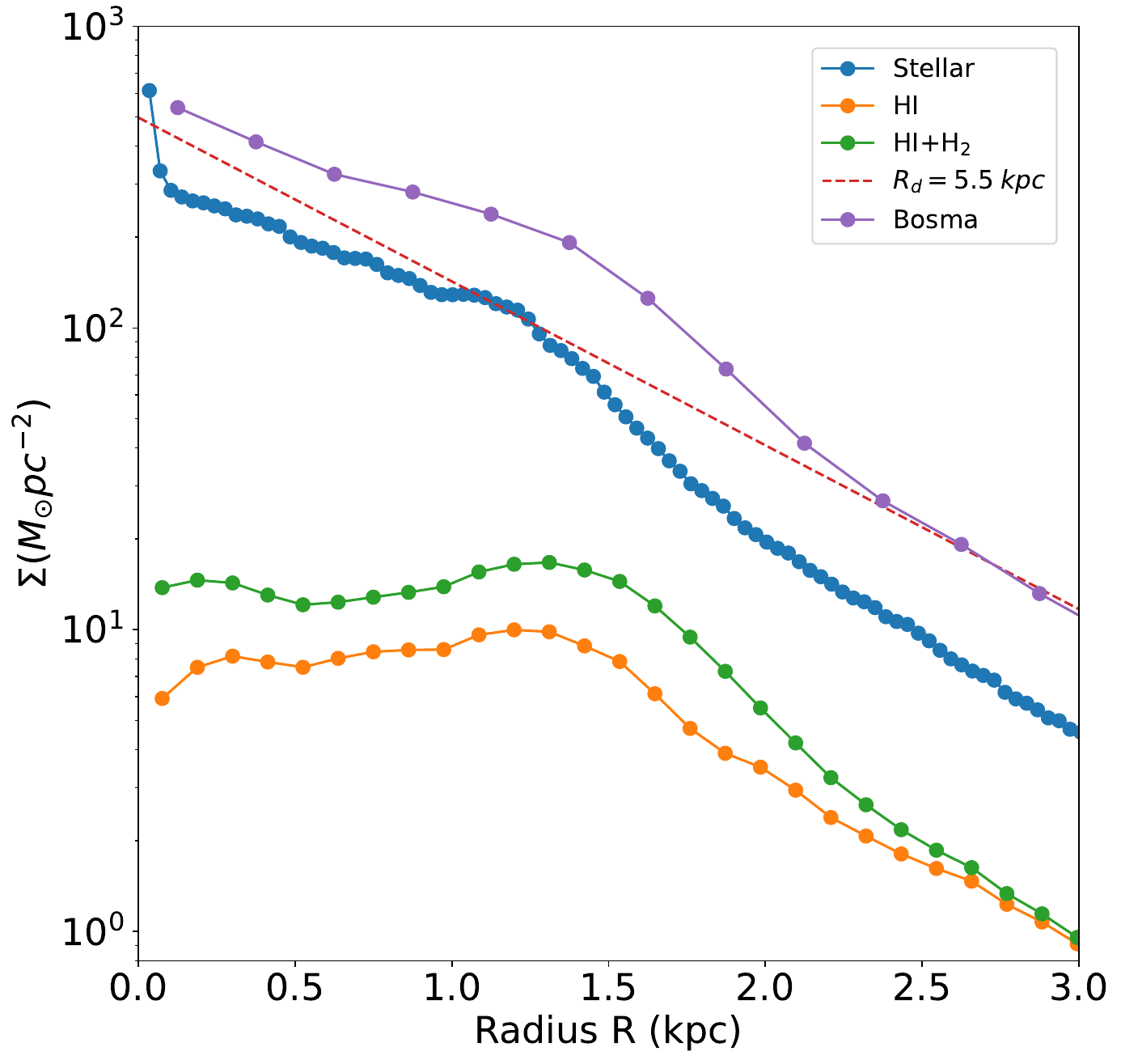}
        \caption{NGC 2976}
    \label{fig:third}
\end{subfigure}
\begin{subfigure}{0.35\textwidth}
    \includegraphics[width=\textwidth]{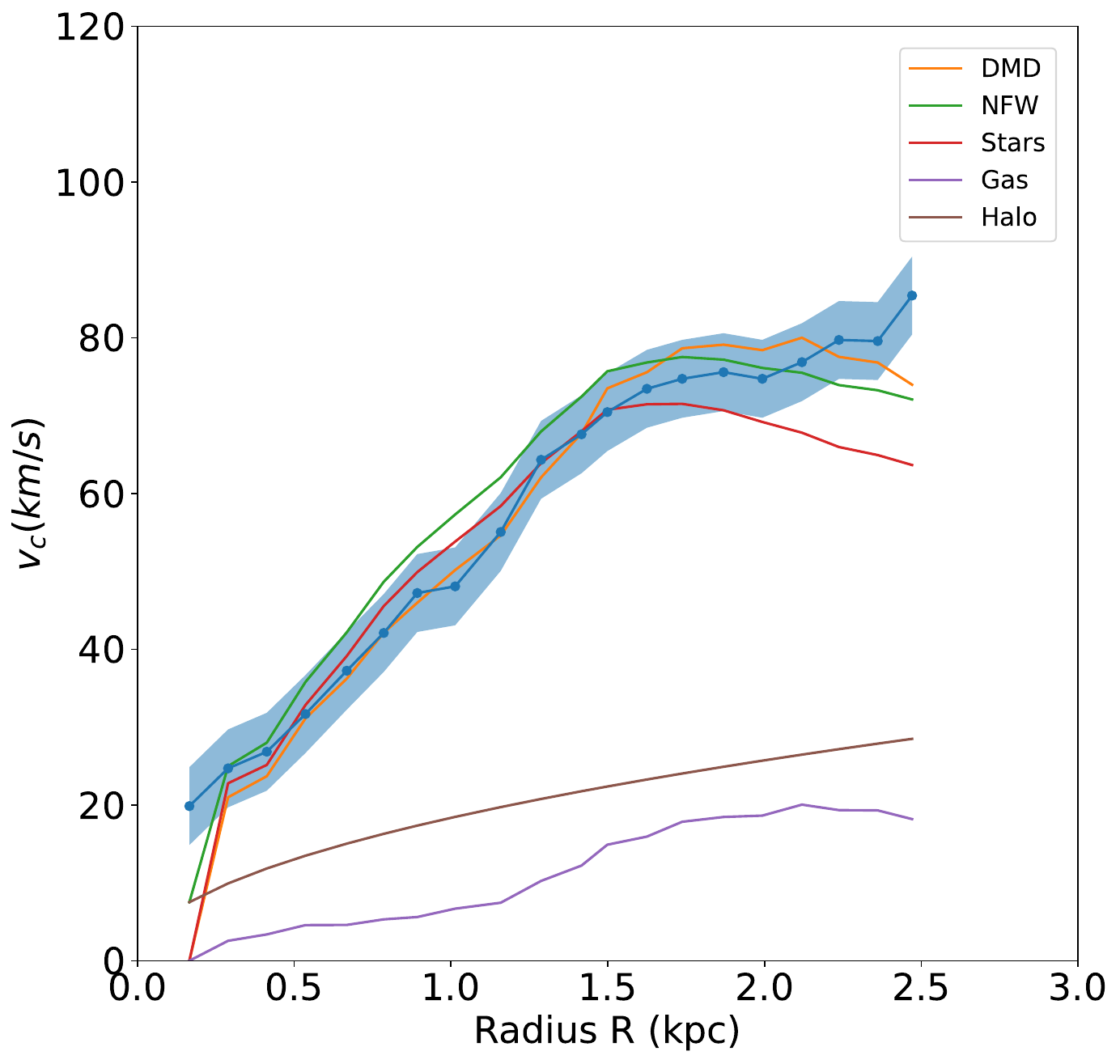}
            \caption{NGC 2976}
    \label{fig:third}
\end{subfigure}
\caption{As Fig.\ref{fig1}}
\label{fig2a}
\end{figure*}

\begin{figure*}
\centering
\begin{subfigure}{0.35\textwidth}
    \includegraphics[width=\textwidth]{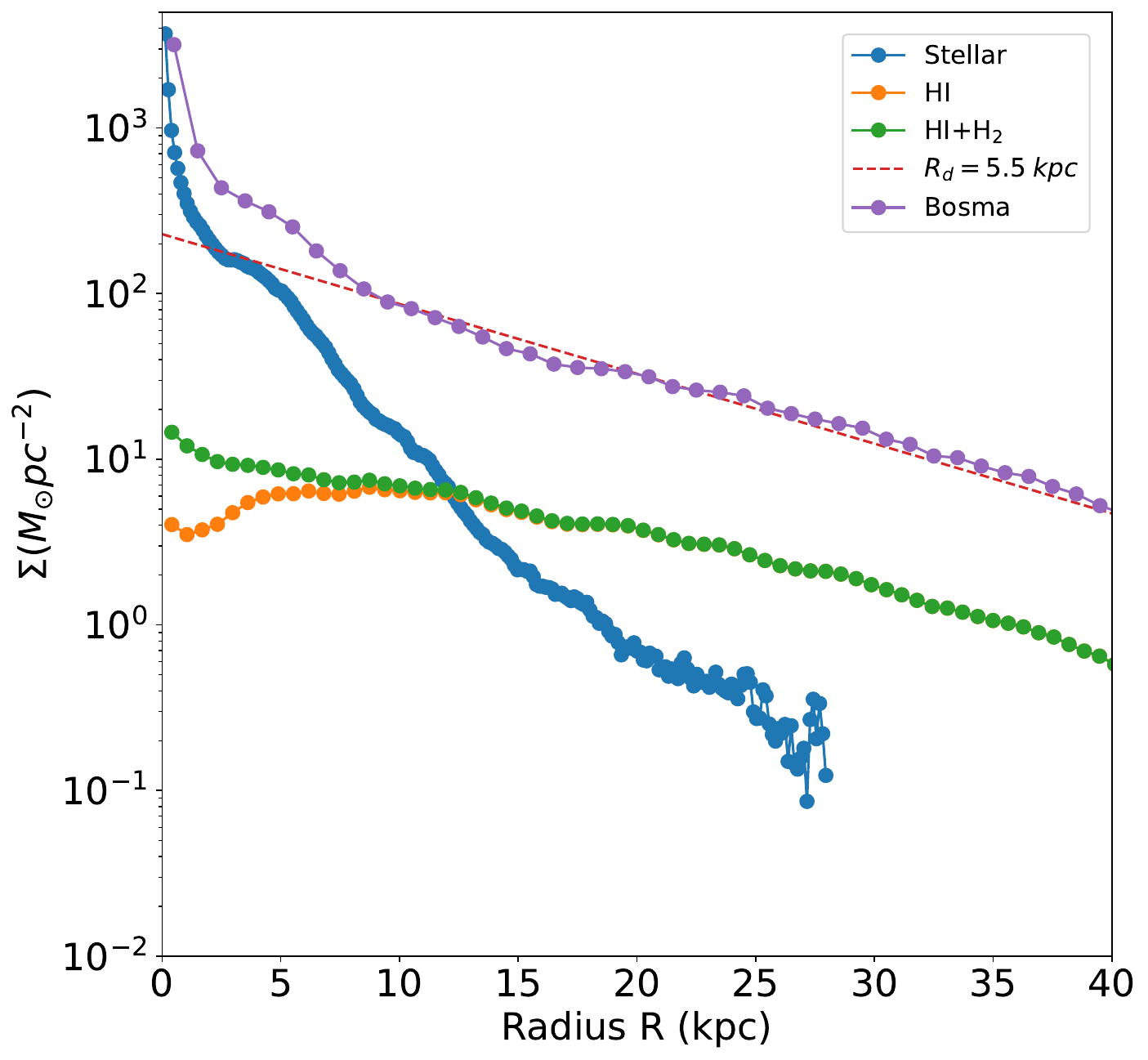}
    \caption{NGC 3198}
    \label{fig:1}
\end{subfigure}
\begin{subfigure}{0.35\textwidth}
    \includegraphics[width=\textwidth]{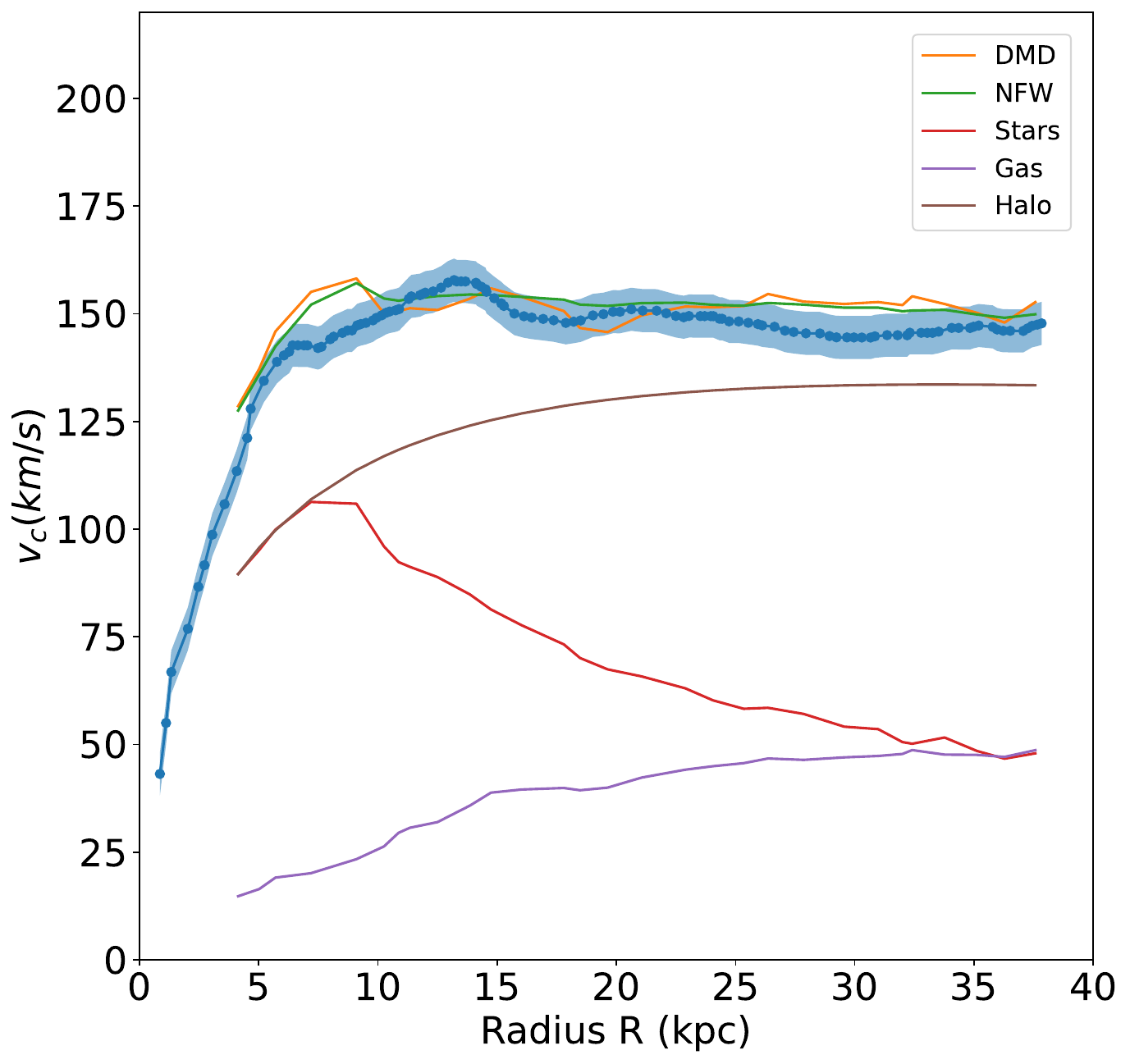}
    \caption{NGC 3198}
    \label{fig:2}
\end{subfigure}
\begin{subfigure}{0.35\textwidth}
    \includegraphics[width=\textwidth]{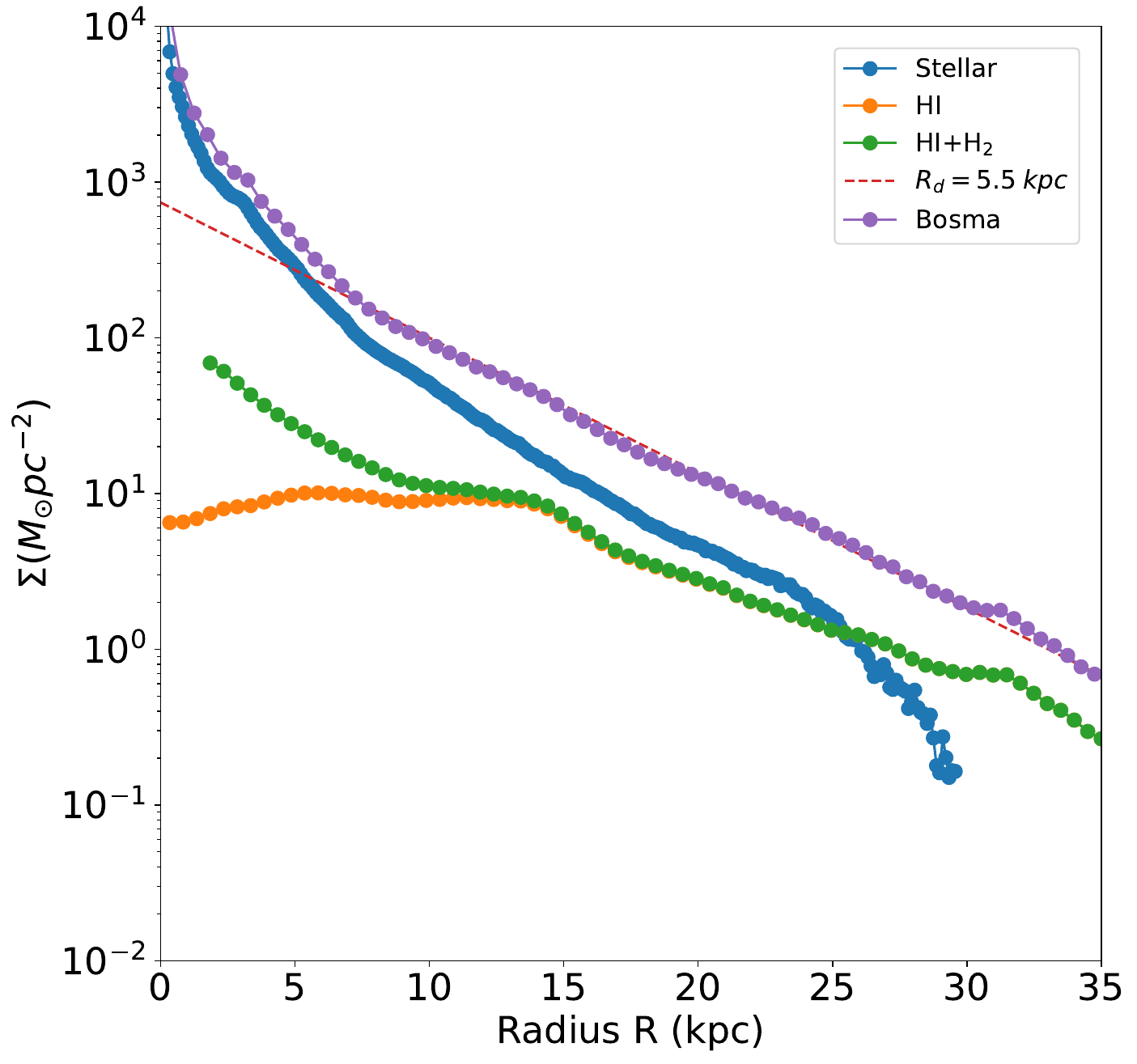}
        \caption{NGC 3521}
    \label{fig:3}
\end{subfigure}
\begin{subfigure}{0.35\textwidth}
    \includegraphics[width=\textwidth]{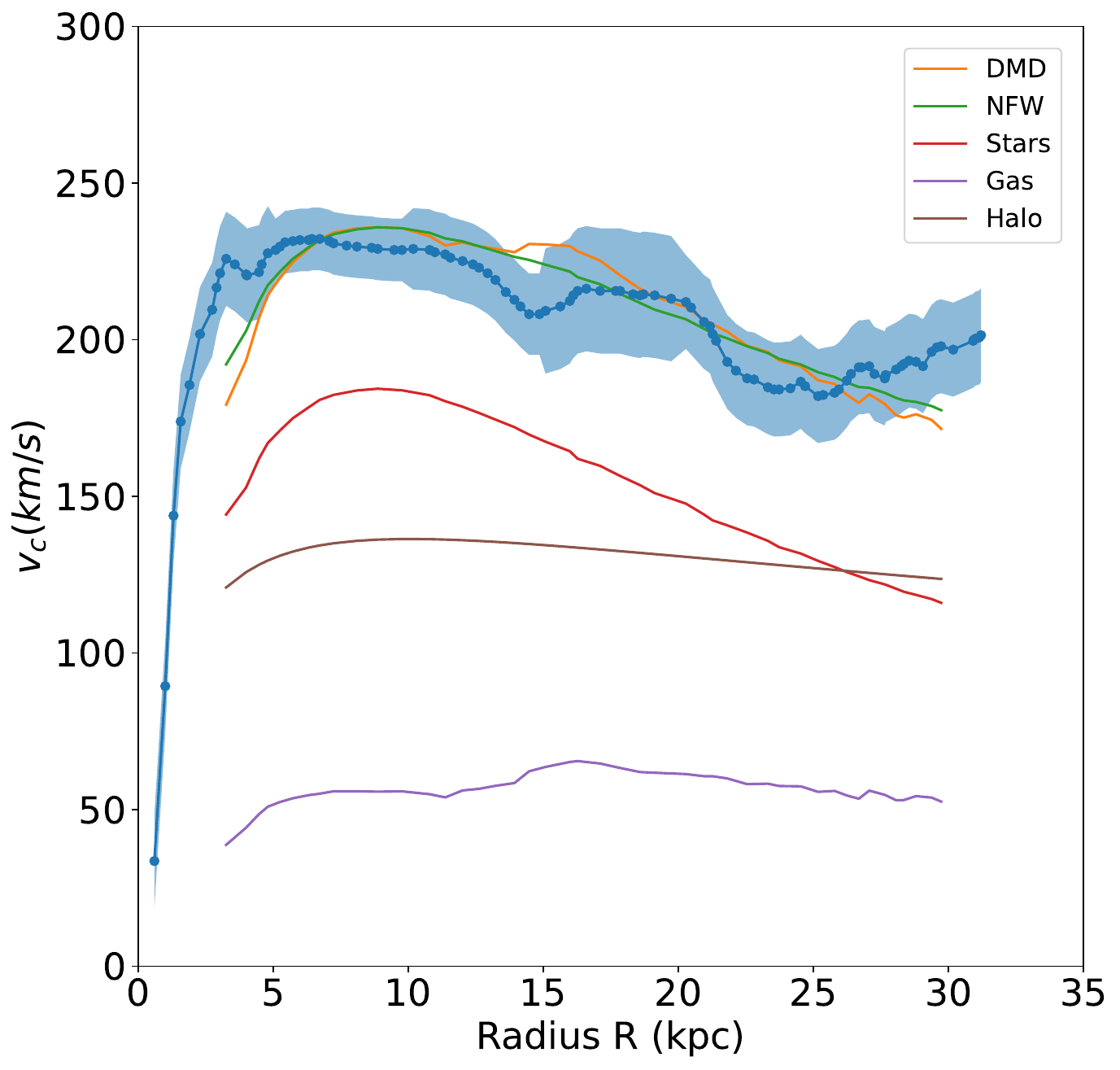}
            \caption{NGC 3521}
    \label{fig:4}
\end{subfigure}
\centering
\begin{subfigure}{0.35\textwidth}
    \includegraphics[width=\textwidth]{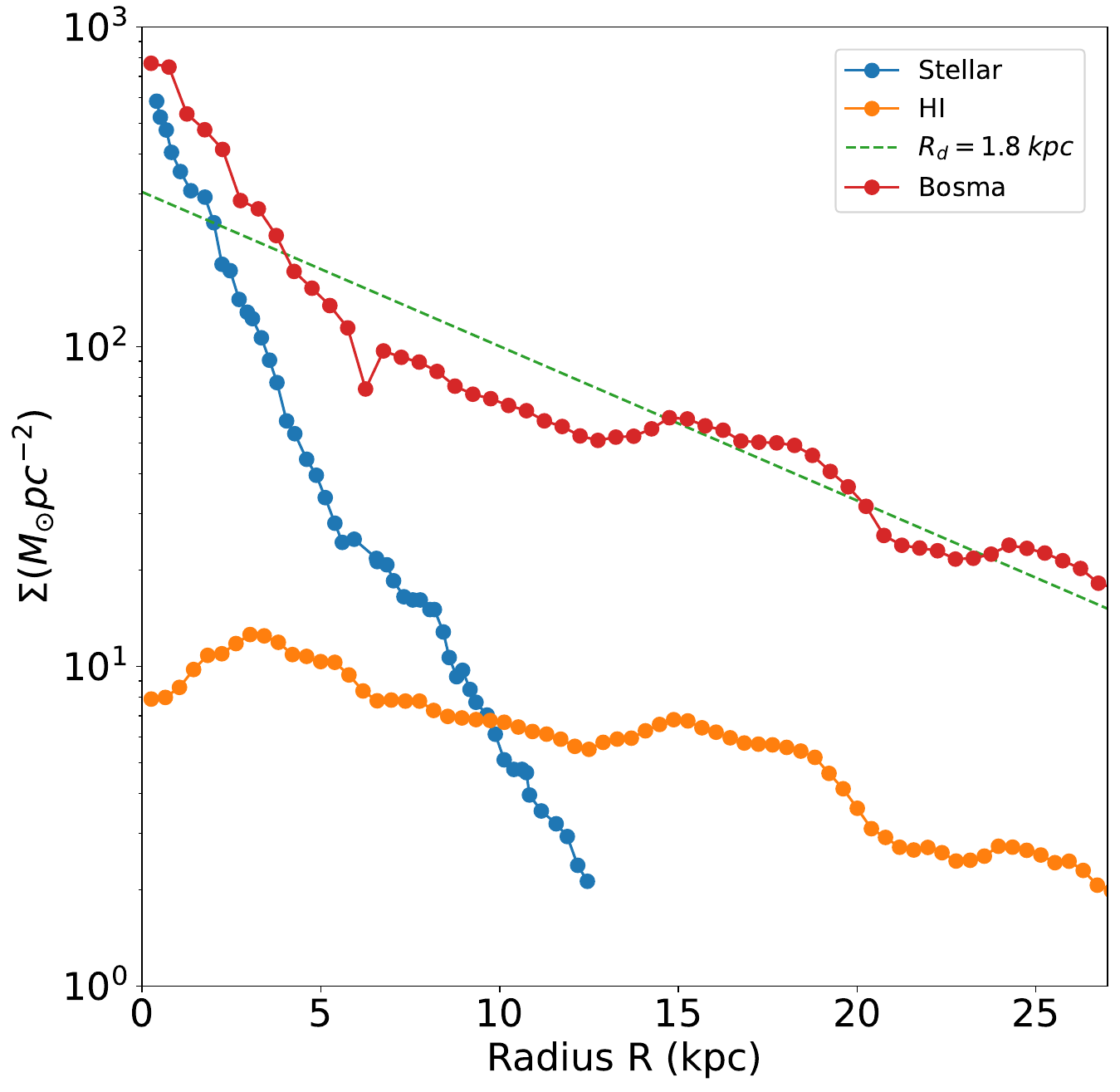}
    \caption{NGC 3621}
    \label{fig:1}
\end{subfigure}
\begin{subfigure}{0.35\textwidth}
    \includegraphics[width=\textwidth]{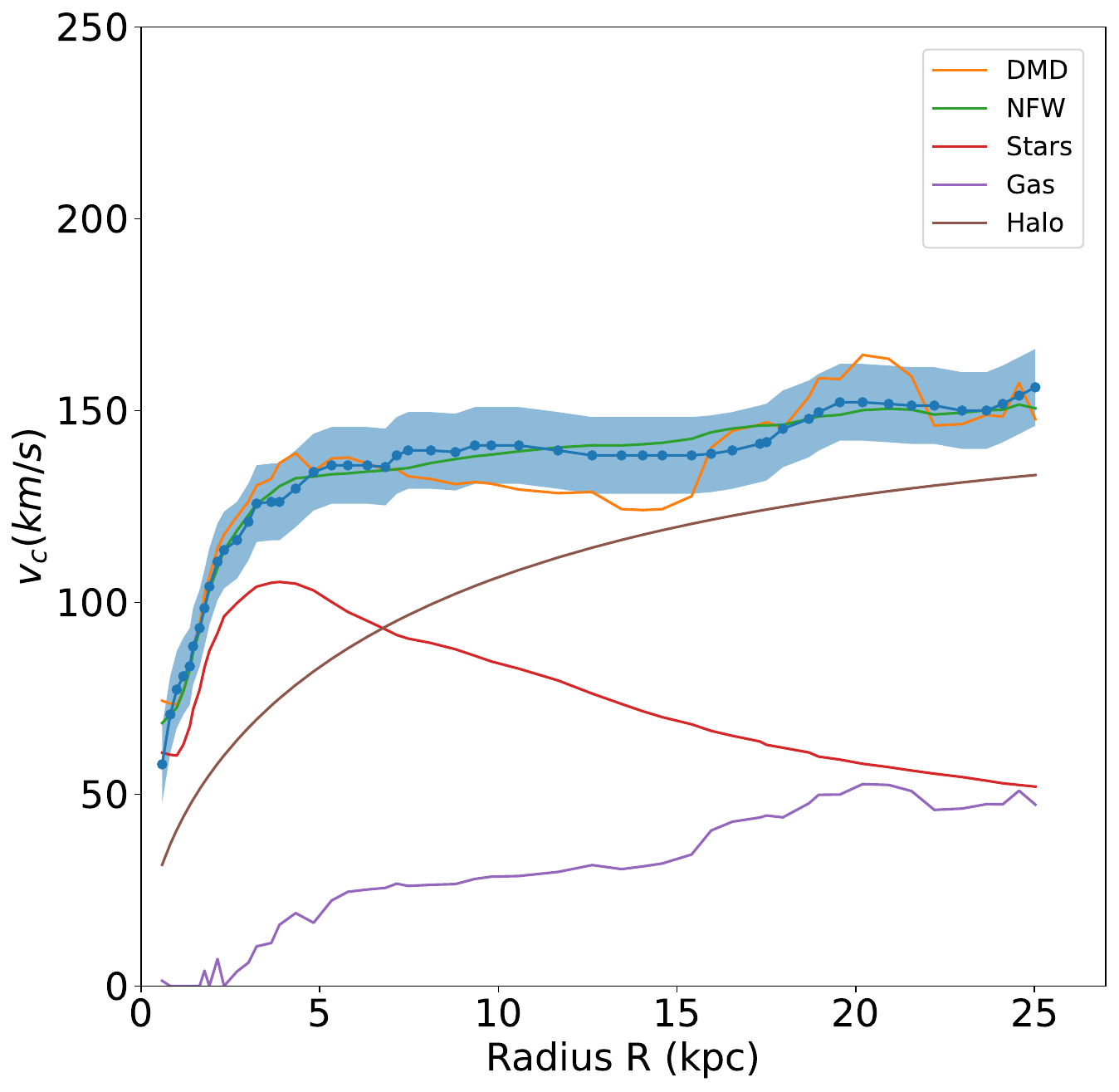}
    \caption{NGC 3621}
    \label{fig:2}
\end{subfigure}
\caption{As Fig.\ref{fig1}.}
\label{fig2b}
\end{figure*}

\begin{figure*}
\begin{subfigure}{0.35\textwidth}
    \includegraphics[width=\textwidth]{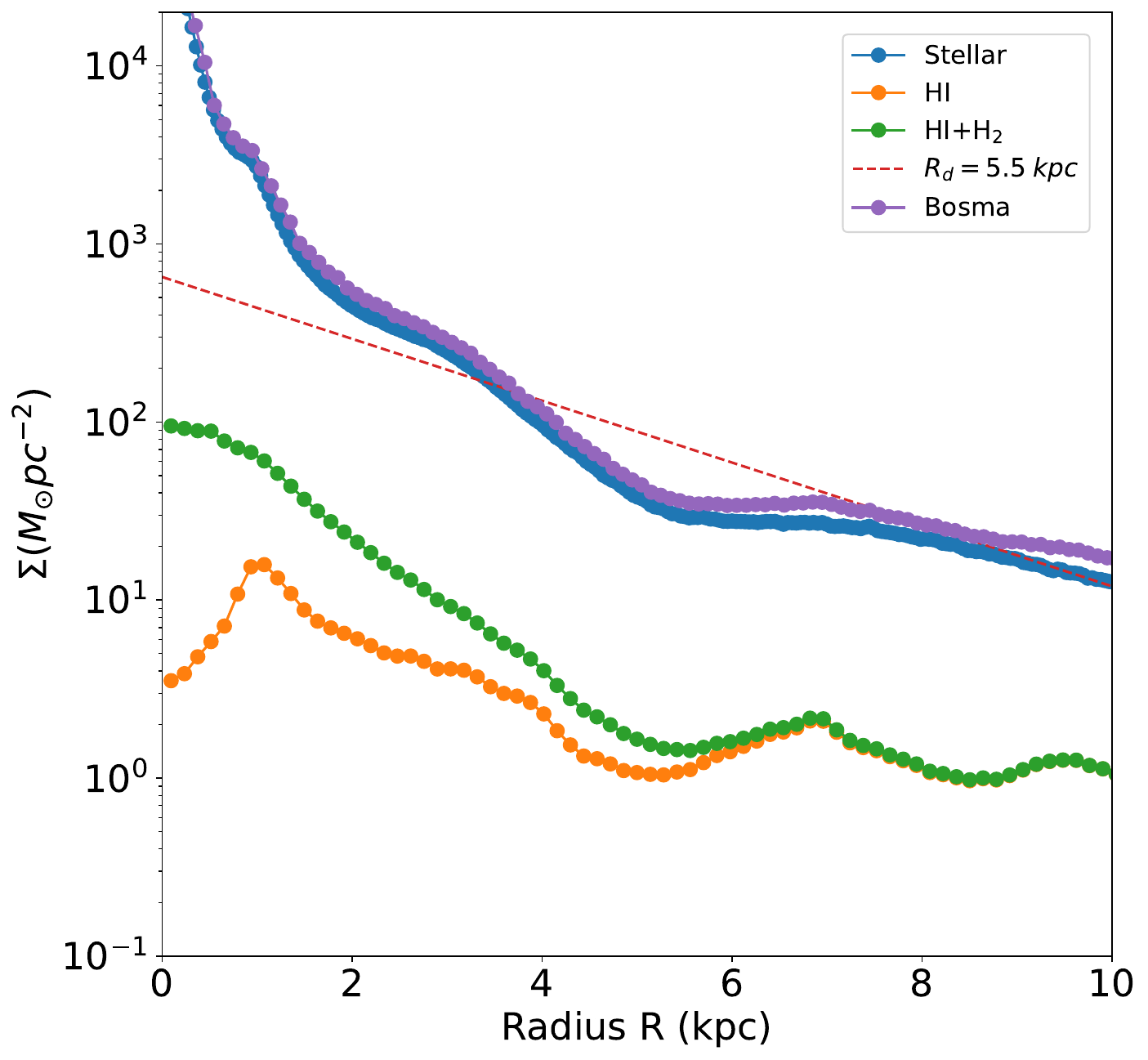}
        \caption{NGC 4736}
    \label{fig:3}
\end{subfigure}
\begin{subfigure}{0.35\textwidth}
    \includegraphics[width=\textwidth]{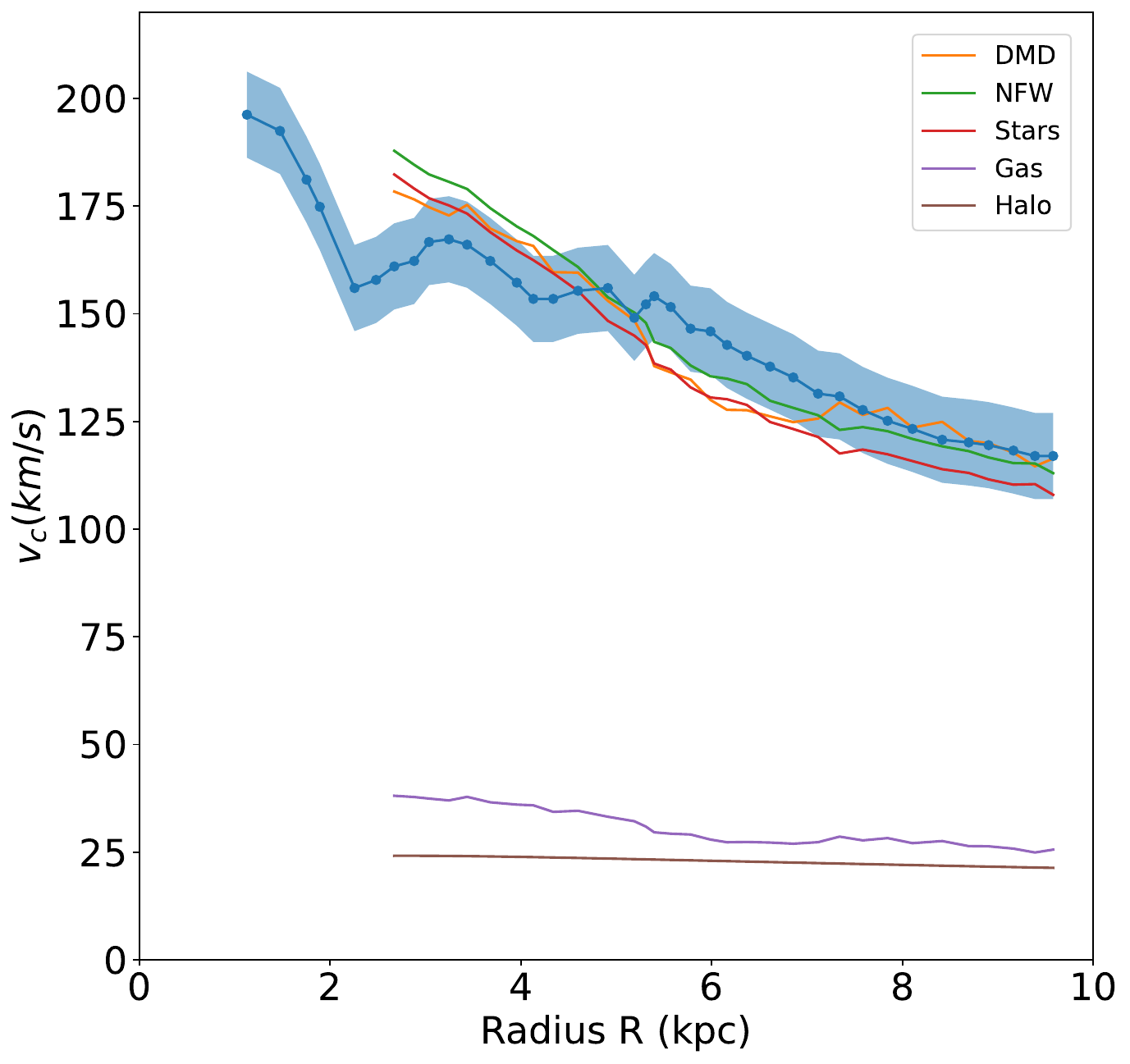}
            \caption{NGC 4736}
    \label{fig:4}
\end{subfigure}
\begin{subfigure}{0.35\textwidth}
    \includegraphics[width=\textwidth]{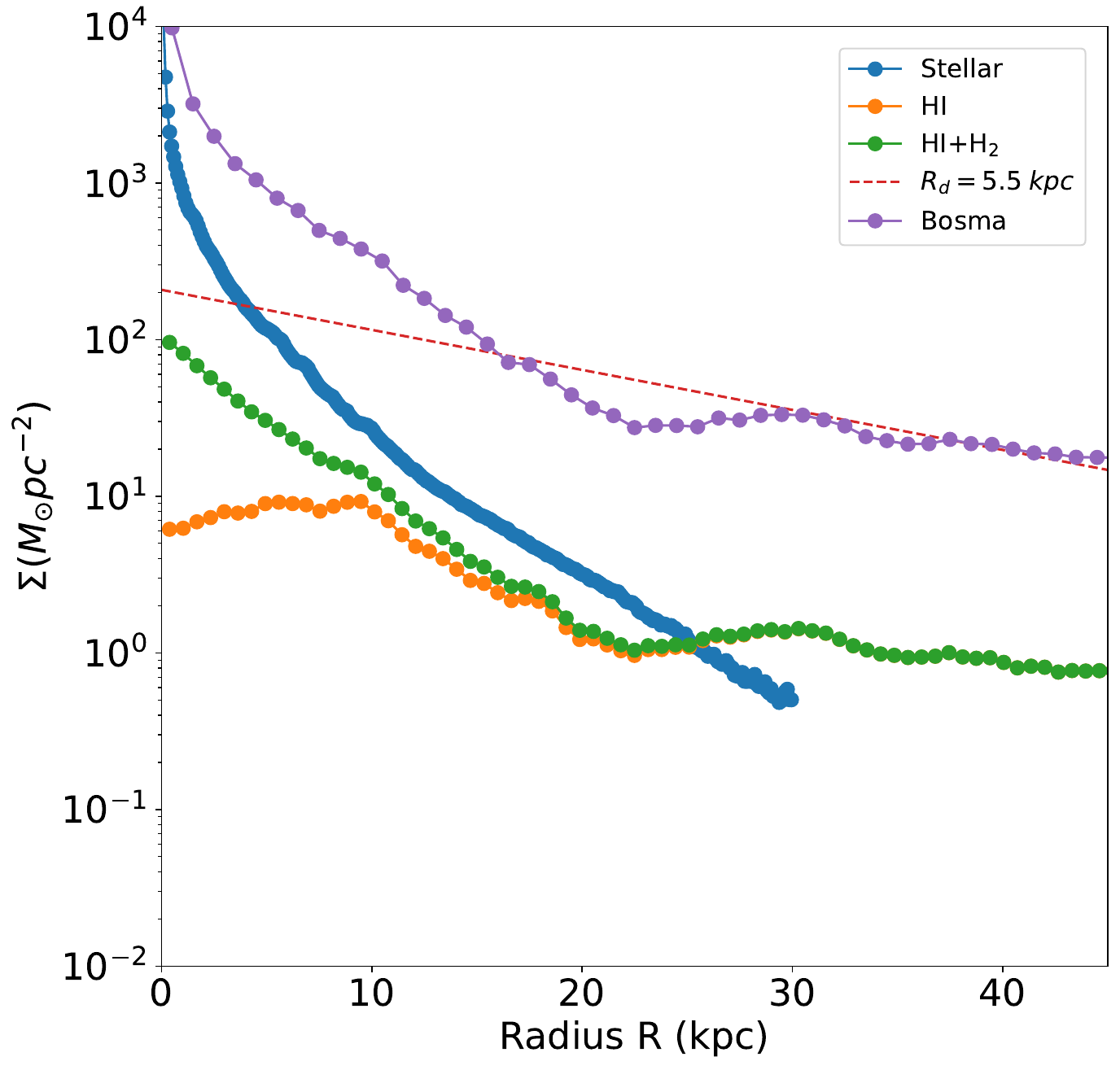}
    \caption{NGC 5055}
    \label{fig:1}
\end{subfigure}
\begin{subfigure}{0.35\textwidth}
    \includegraphics[width=\textwidth]{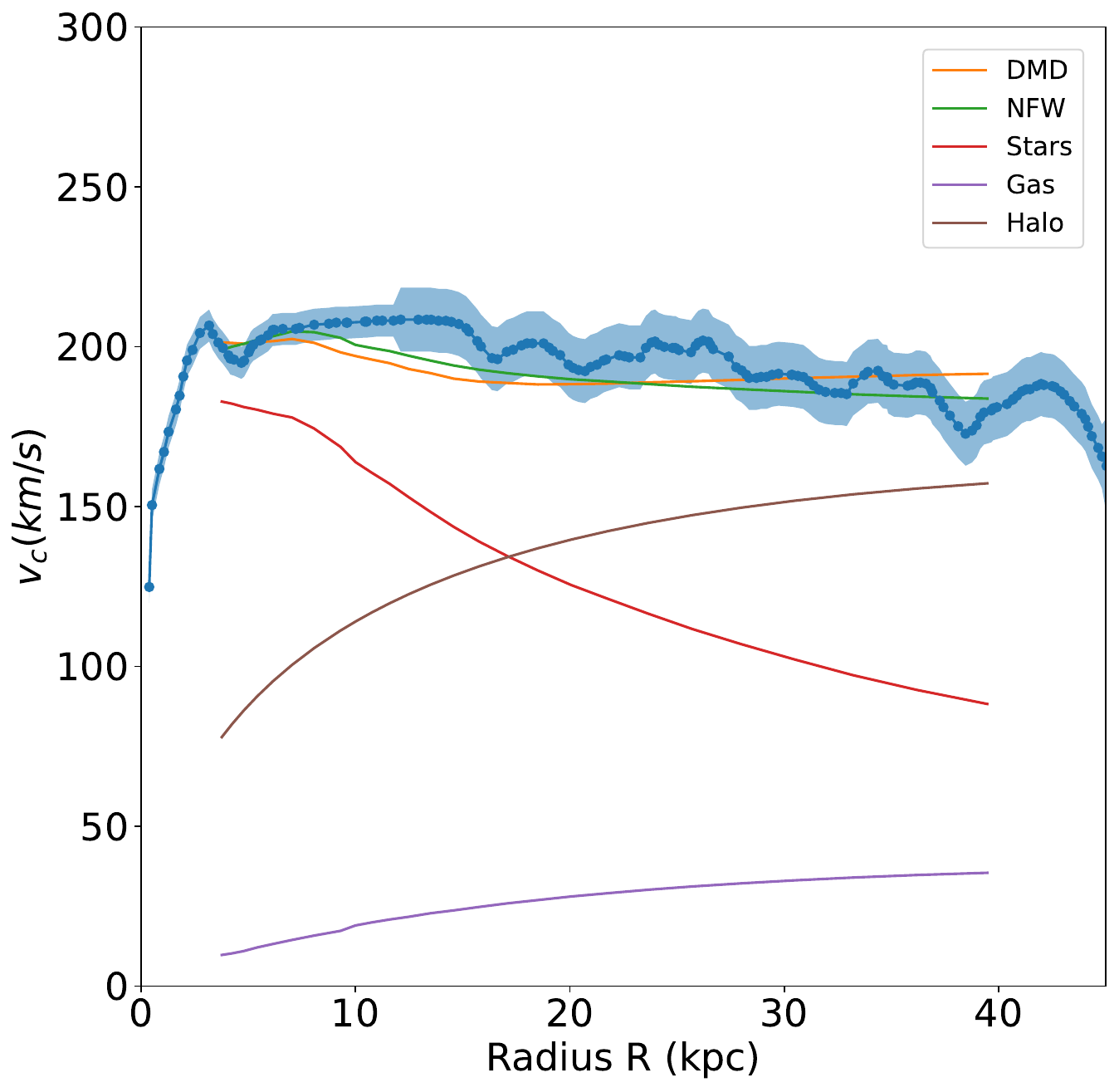}
    \caption{NGC 5055}
    \label{fig:2}
\end{subfigure}
\begin{subfigure}{0.35\textwidth}
    \includegraphics[width=\textwidth]{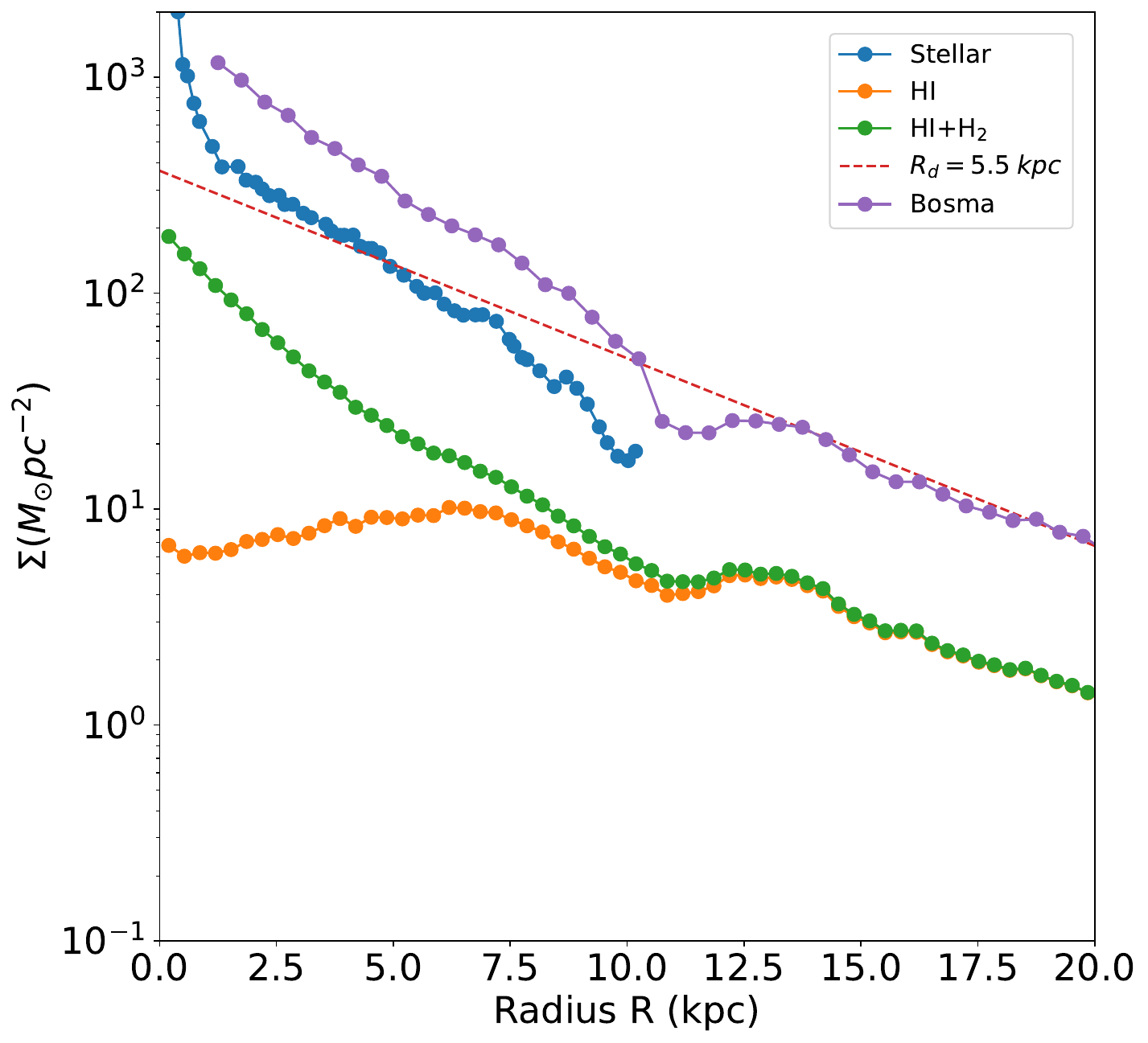}
        \caption{NGC 6946}
    \label{fig:3}
\end{subfigure}
\begin{subfigure}{0.37\textwidth}
    \includegraphics[width=\textwidth]{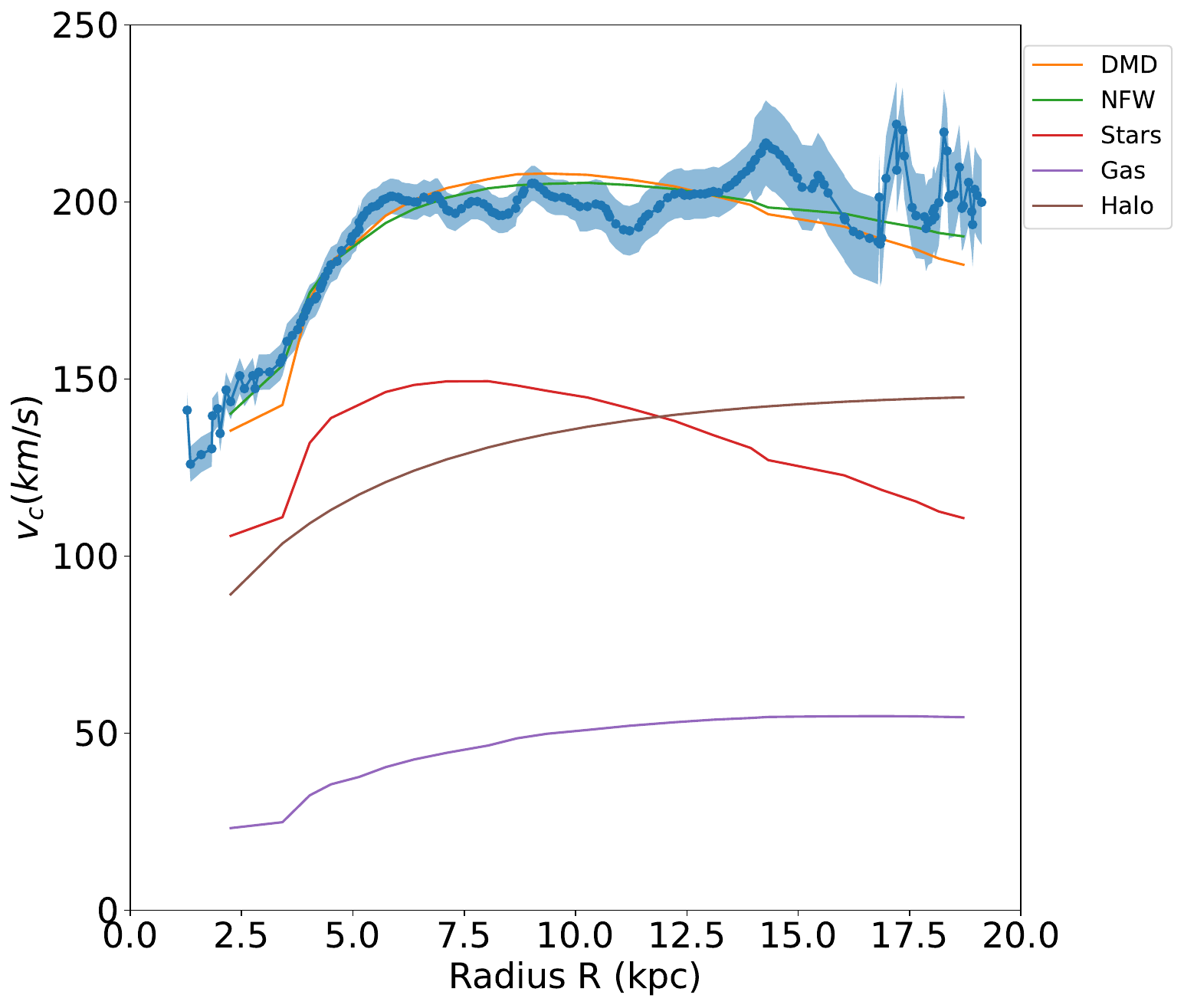}
            \caption{NGC 6946}
    \label{fig:4}
\end{subfigure}
\caption{As Fig.\ref{fig1}.}
\label{fig2c}
\end{figure*}

\begin{figure*}
\begin{subfigure}{0.35\textwidth}
    \includegraphics[width=\textwidth]{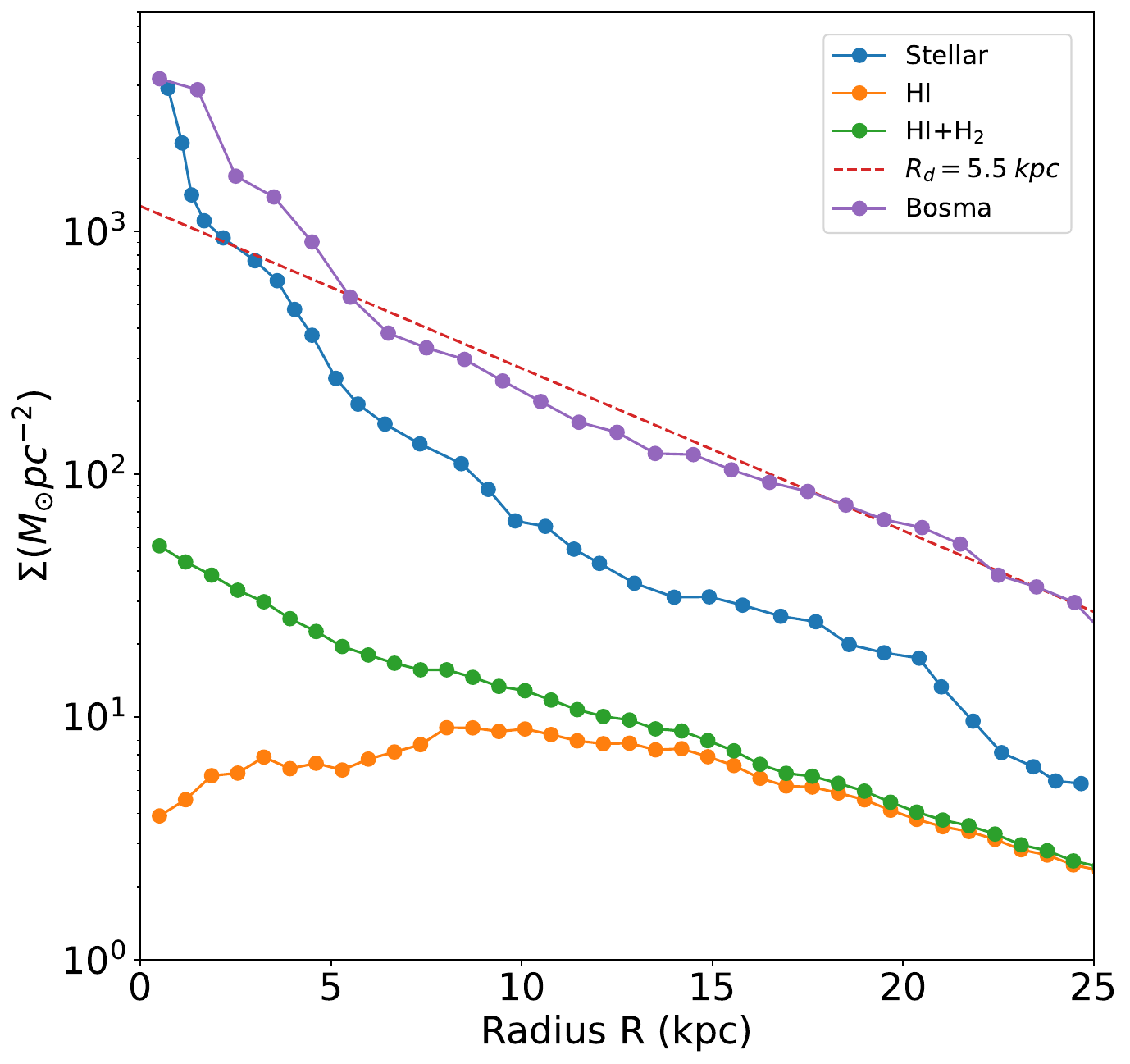}
    \caption{NGC 7331}
    \label{fig:1}
\end{subfigure}
\begin{subfigure}{0.35\textwidth}
    \includegraphics[width=\textwidth]{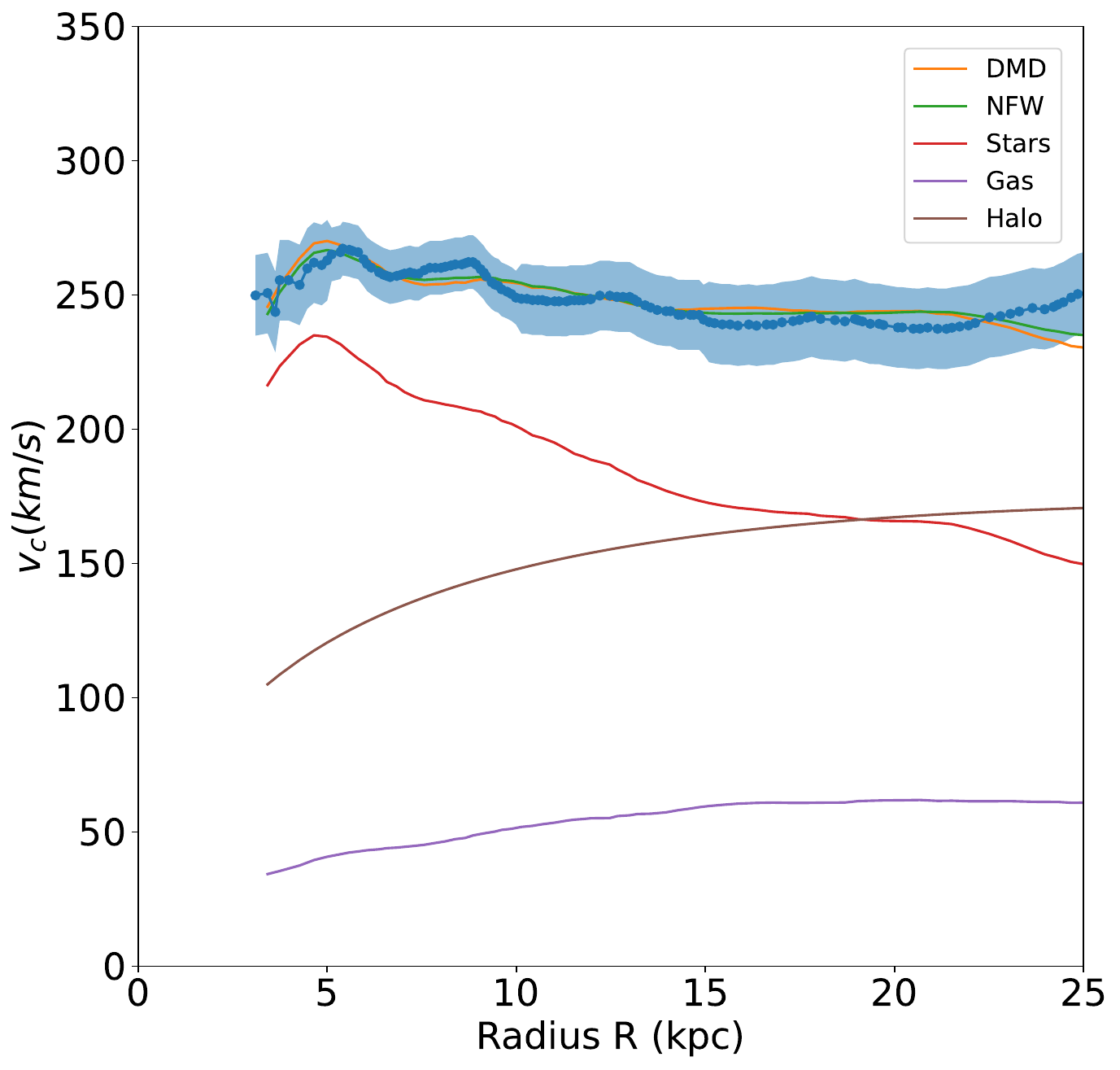}
    \caption{NGC 7331}
    \label{fig:2}
\end{subfigure}
\begin{subfigure}{0.35\textwidth}
    \includegraphics[width=\textwidth]{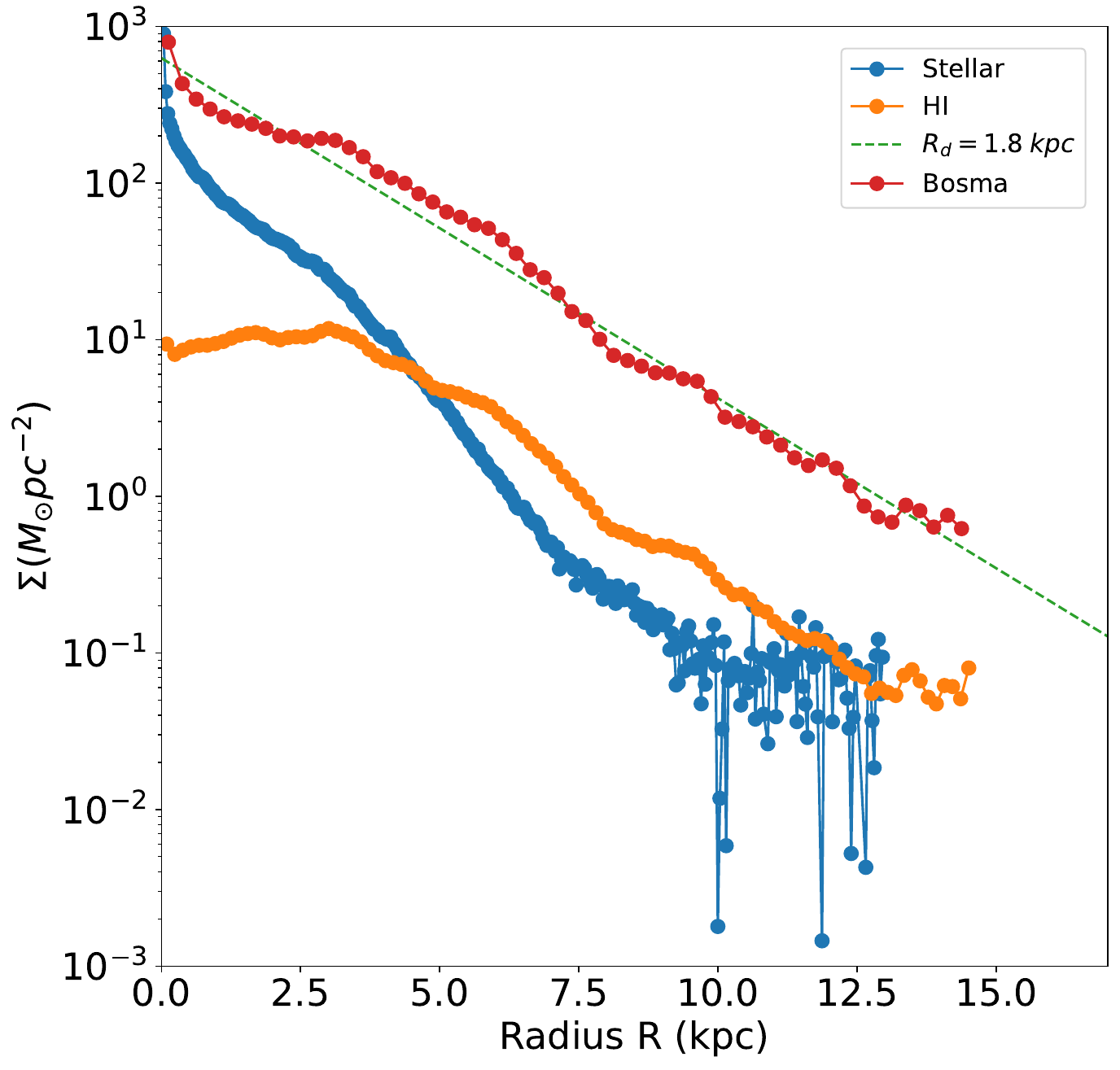}
        \caption{NGC 7793}
    \label{fig:3}
\end{subfigure}
\begin{subfigure}{0.35\textwidth}
    \includegraphics[width=\textwidth]{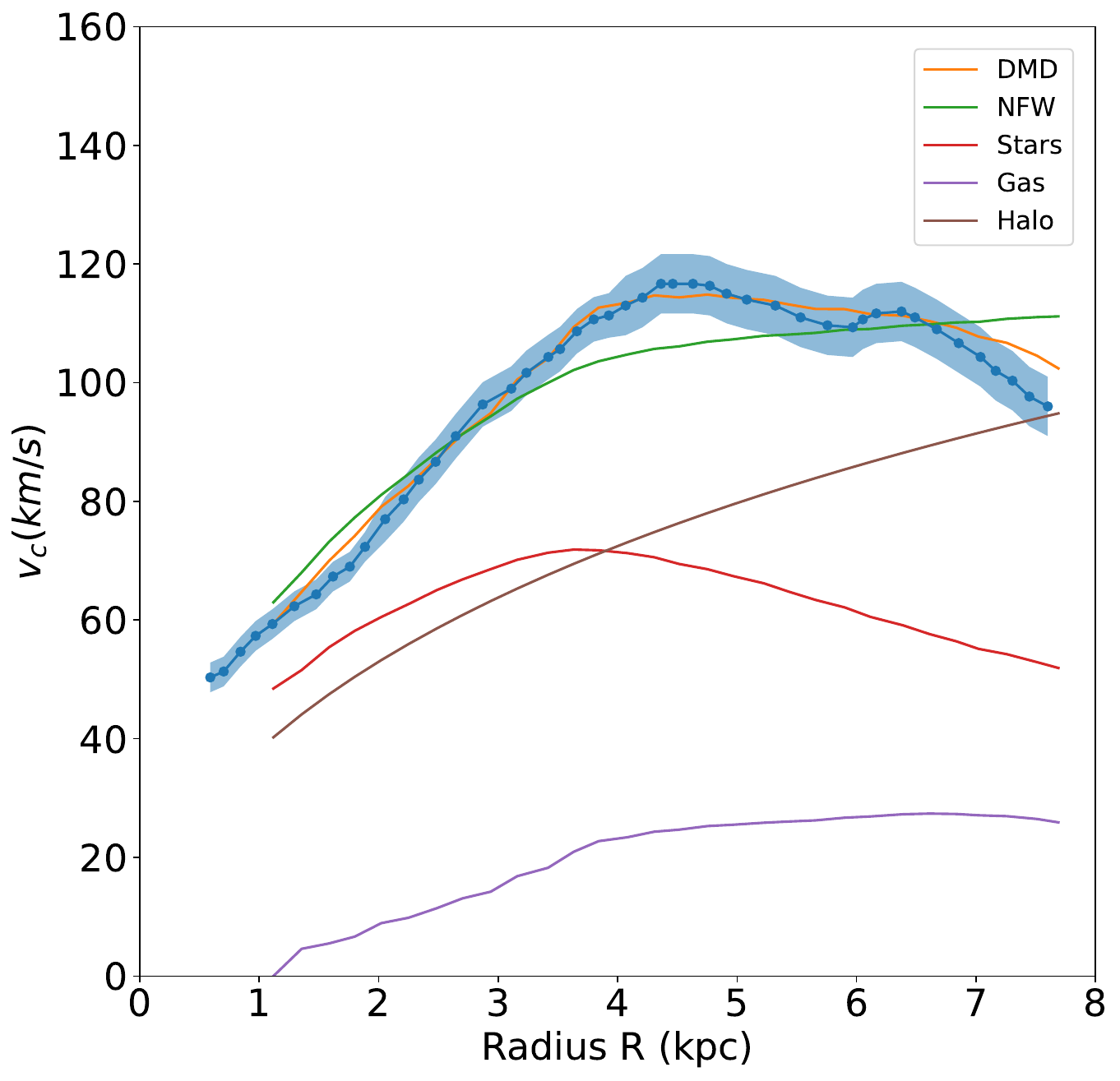}
            \caption{NGC 7793}
    \label{fig:4}
\end{subfigure}
\begin{subfigure}{0.35\textwidth}
    \includegraphics[width=\textwidth]{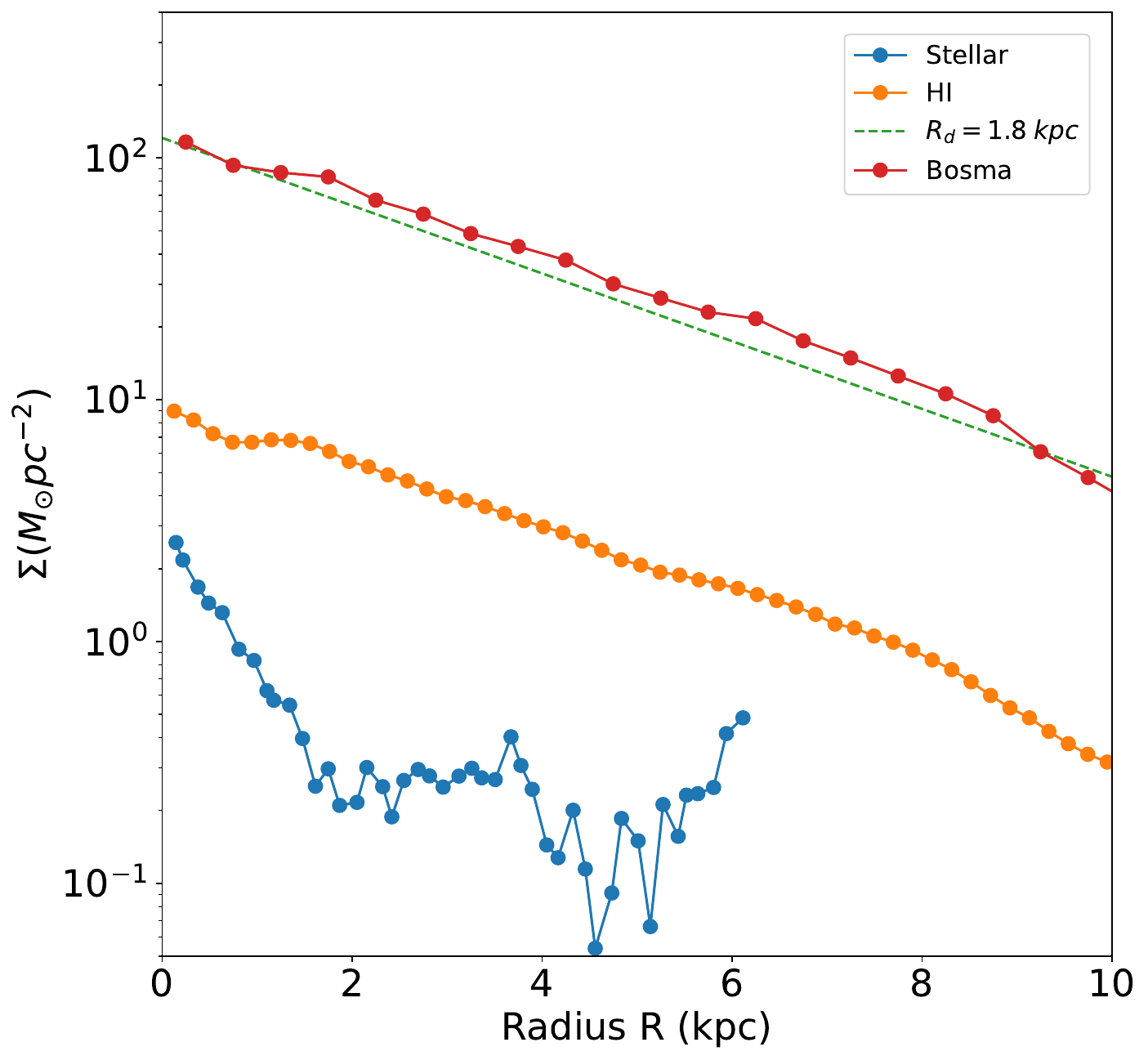}
    \caption{DDO 154}
    \label{fig:1}
\end{subfigure}
\begin{subfigure}{0.35\textwidth}
    \includegraphics[width=\textwidth]{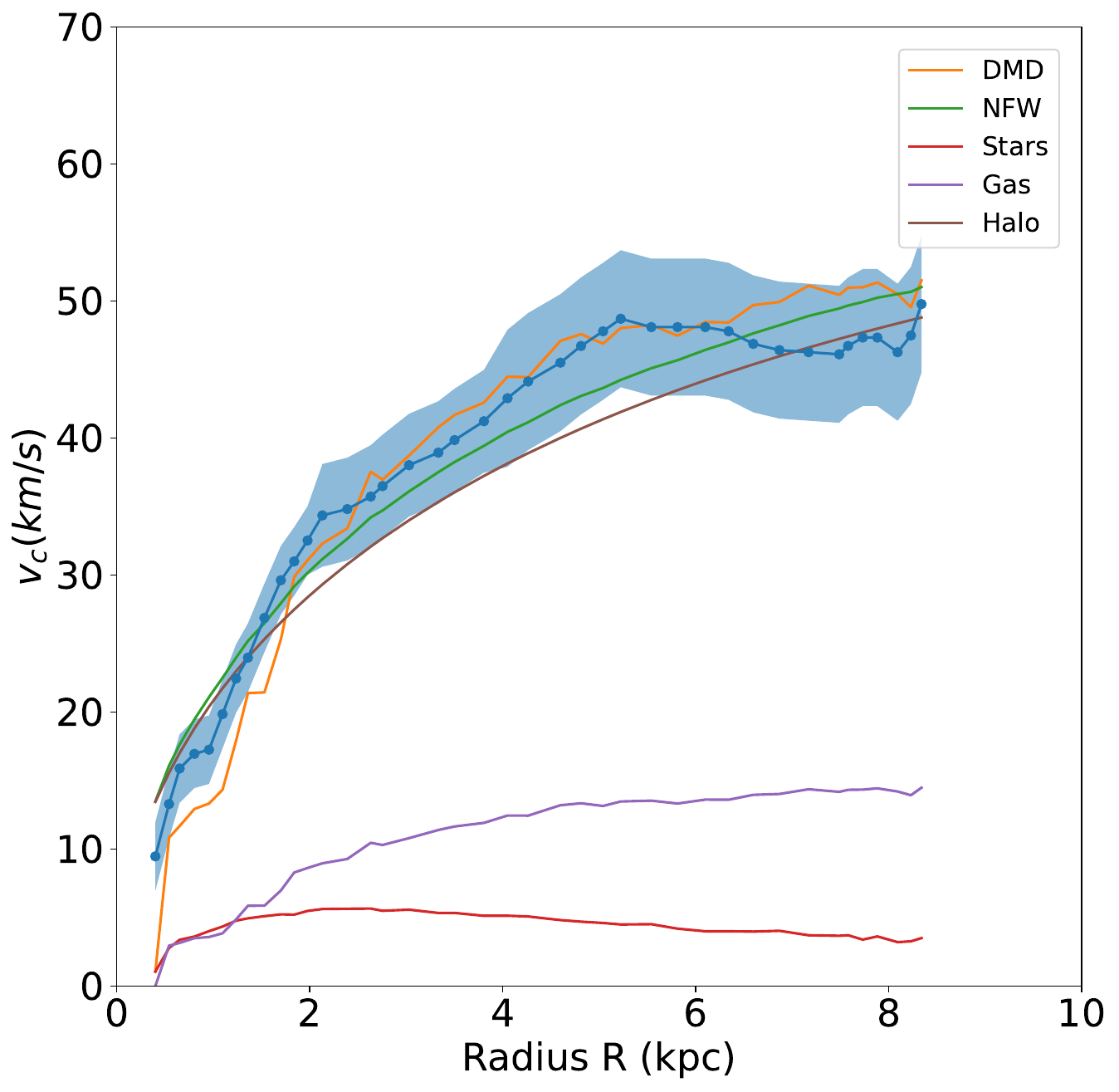}
    \caption{DDO 154}
    \label{fig:2}
\end{subfigure}
\caption{As Fig.\ref{fig1}.}
\label{fig2}
\end{figure*}

\begin{figure*}
\begin{subfigure}{0.35\textwidth}
    \includegraphics[width=\textwidth]{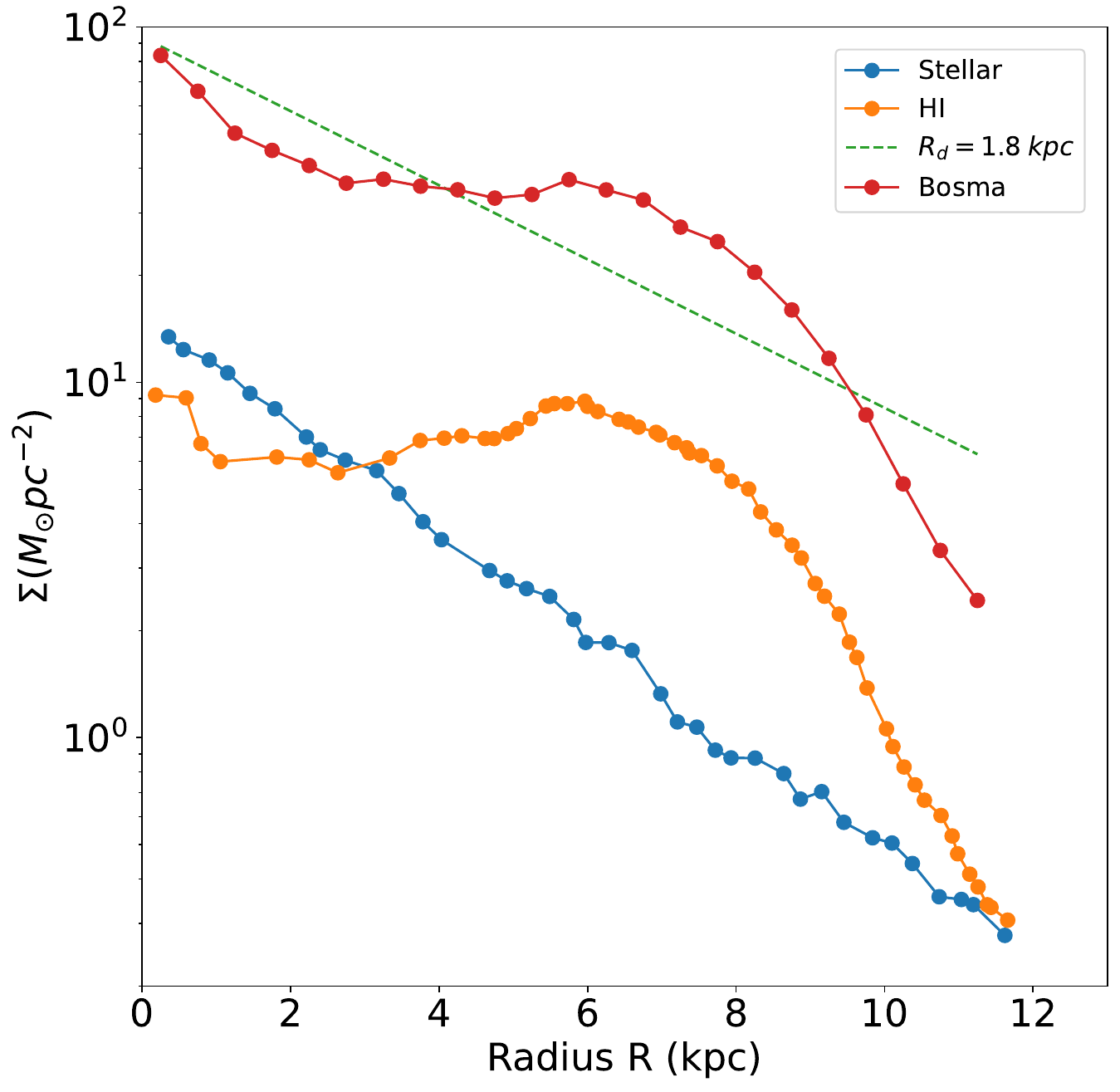}
        \caption{IC 2574}
    \label{fig:3}
\end{subfigure}
\begin{subfigure}{0.35\textwidth}
    \includegraphics[width=\textwidth]{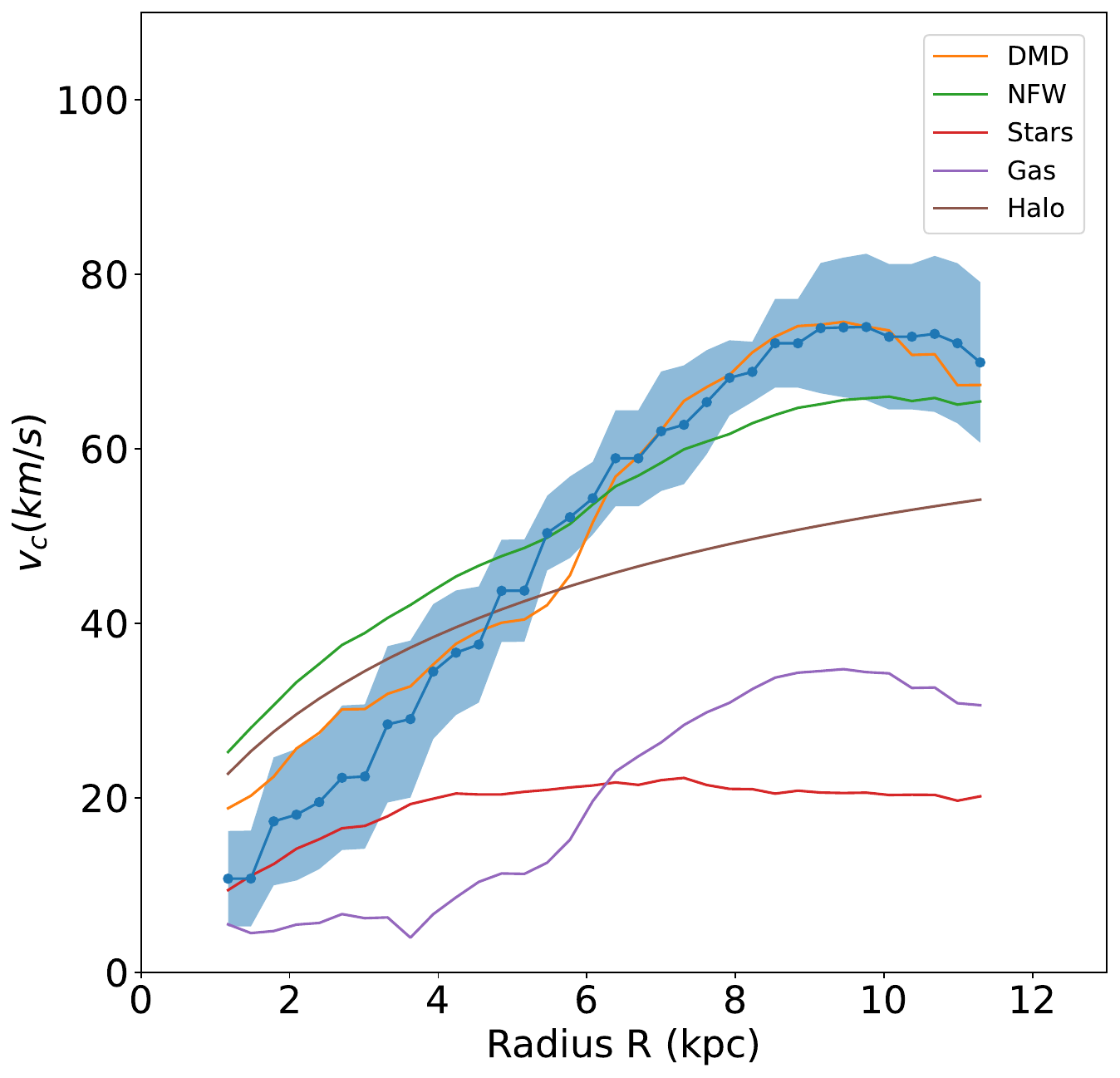}
            \caption{IC 2574}
    \label{fig:4}
\end{subfigure}
\caption{As Fig.\ref{fig1}.}
\label{fig2}
\end{figure*}

\begin{figure*}
\centering
\begin{subfigure}{0.3\textwidth}
    \includegraphics[width=\textwidth]{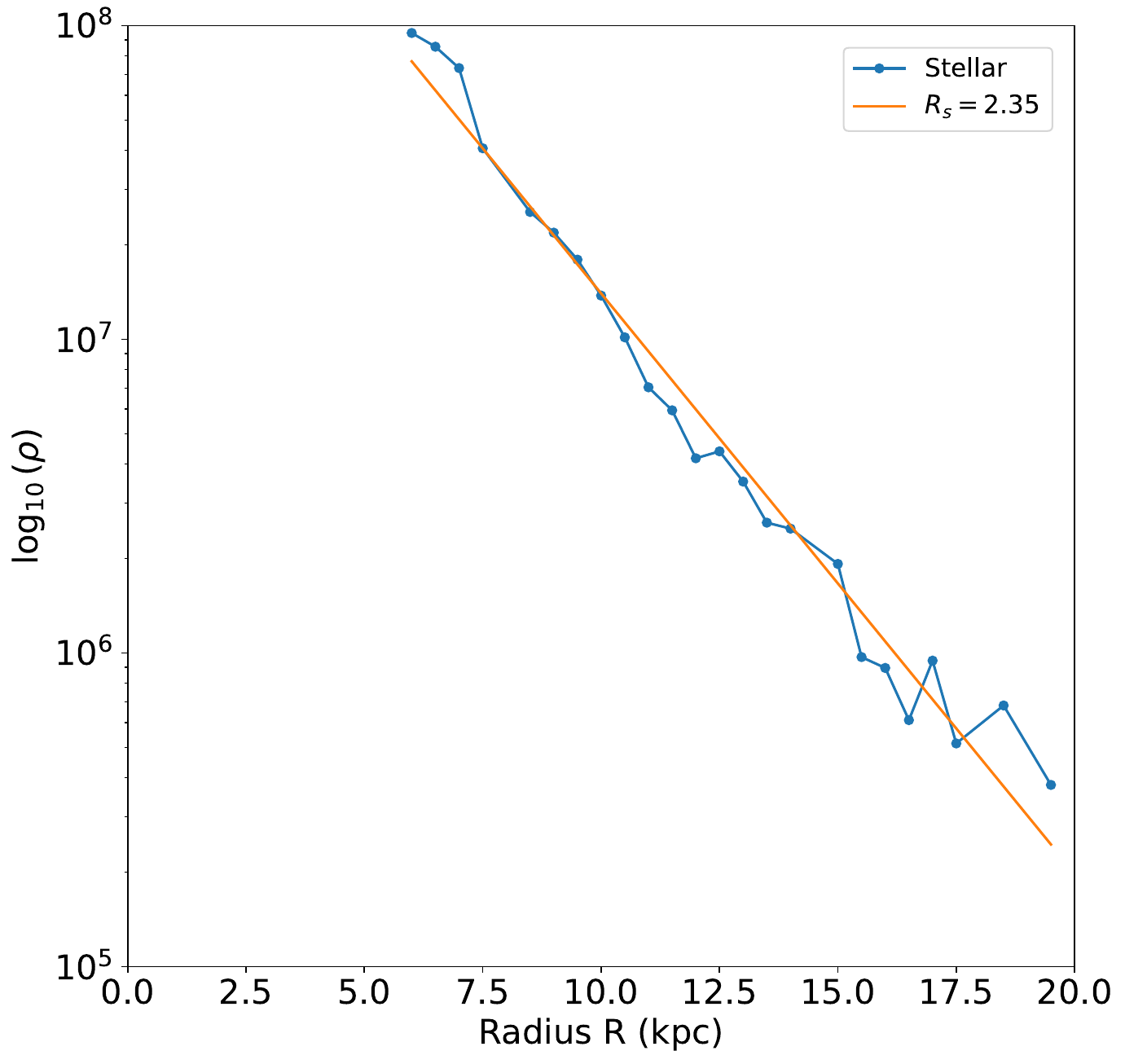}
    \caption{MW}
    \label{fig:first}
\end{subfigure}
\begin{subfigure}{0.3\textwidth}
    \includegraphics[width=\textwidth]{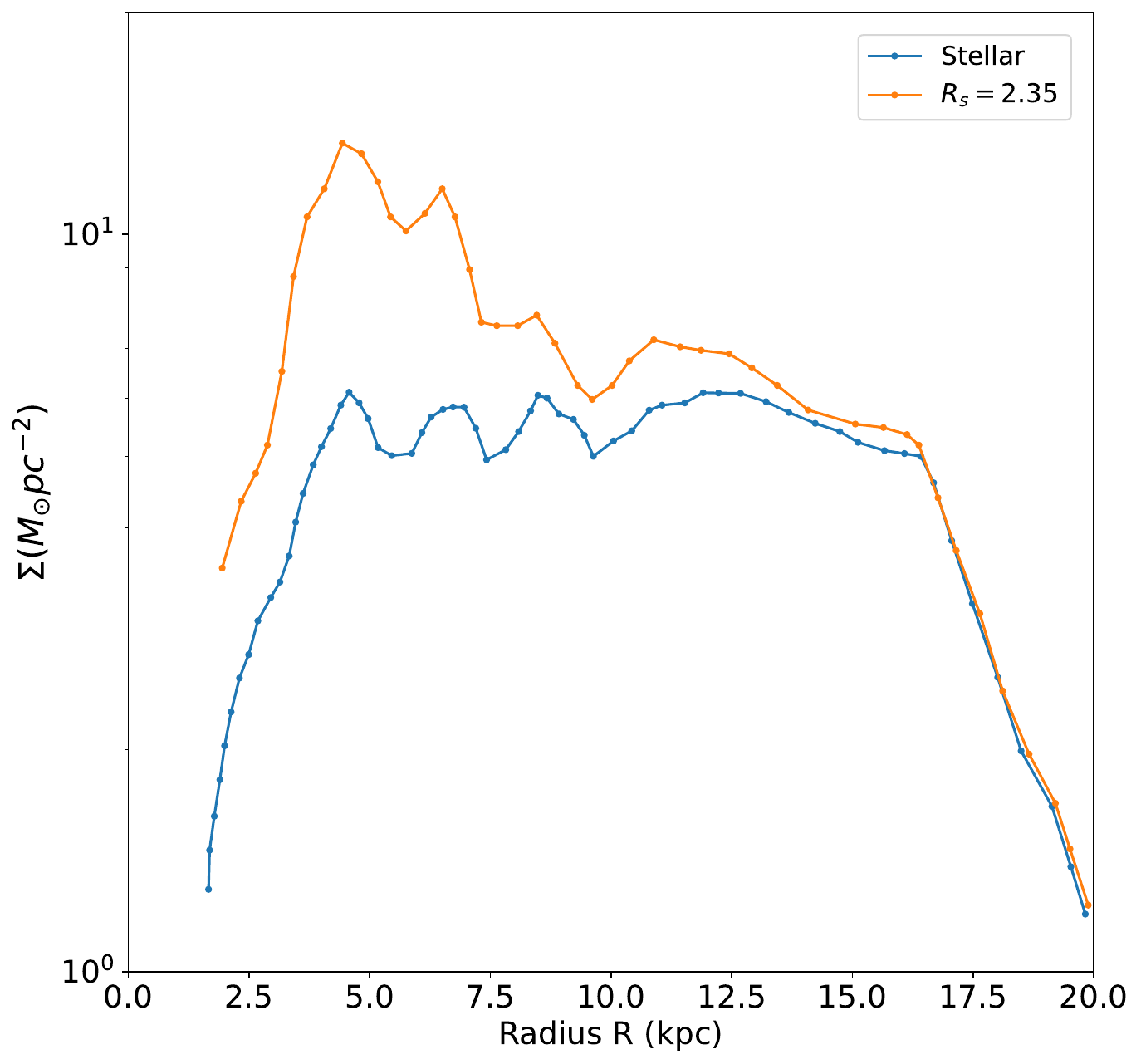}
    \caption{MW}
    \label{fig:second}
\end{subfigure} 
\begin{subfigure}{0.3\textwidth}
    \includegraphics[width=\textwidth]{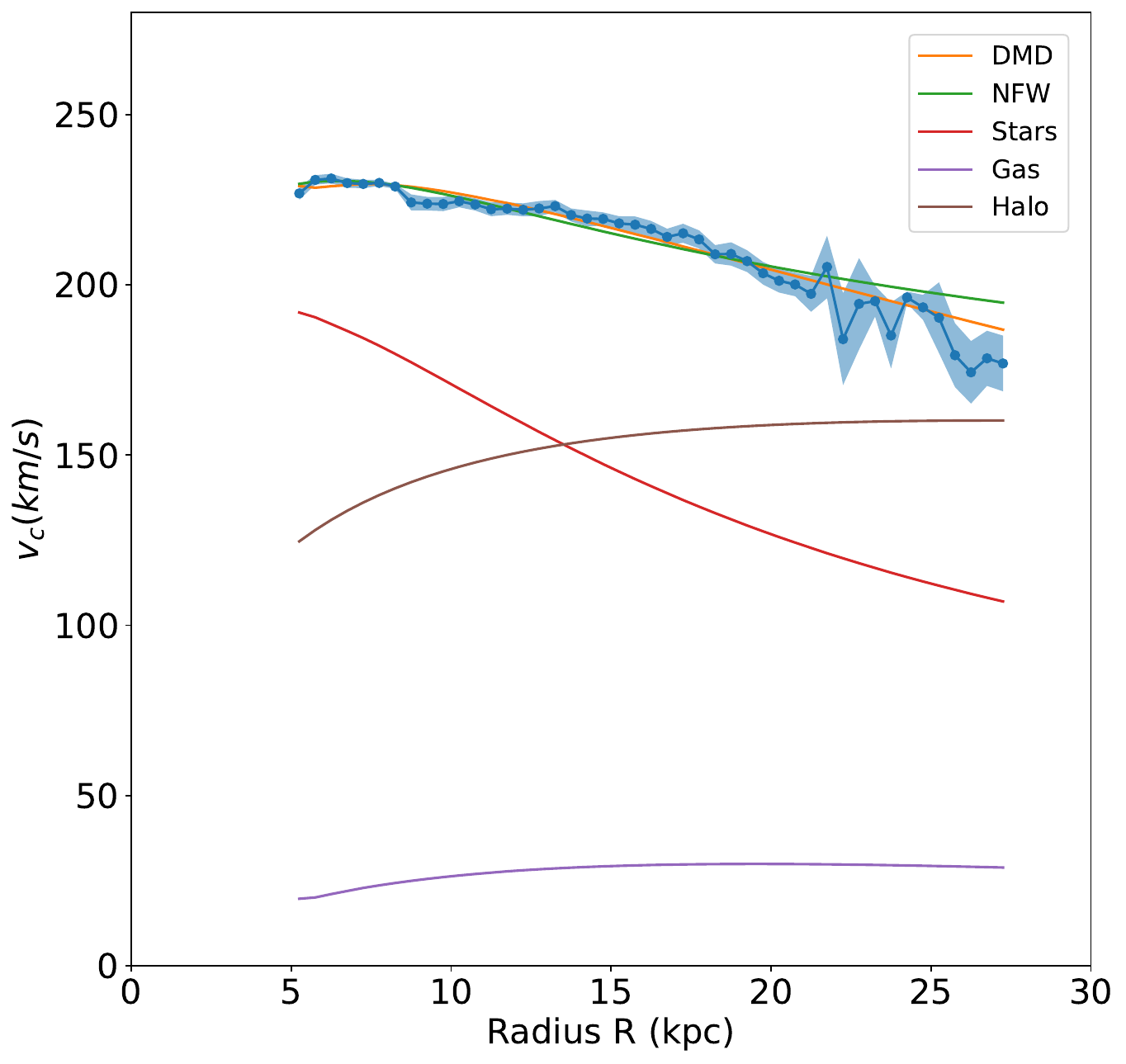}
    \caption{MW}
    \label{fig:second}
\end{subfigure} 
 \caption{
  Data for the Milky Way.
 Left panel: stellar density on the disc (units are number of stars kpc$^{-3}$).
 Center panel: \HI{} and \HI{}+H$_2$ surface density.
 Right panel: rotation curve. The three lines are respectively 
 the best fits with the NFW and DMD mass models and  the best exponential 
 thin disc (ETD) approximation: note that in this later case 
 the small radii behavior is not well fitted because of the it neglects the bulge. } 
\label{fig3} 
\end{figure*}

\subsection{NGC 925}

The surface brightness profile of  the stellar component    (2MASS  data --- see \cite{deBlok_etal_2008}) displays an exponential decay while the surface brightness profiles of both  \HI{}  and  \HI{} +H$_2$ are almost flat  for $R<10$ kpc. The contribution to the gas mass of the H$_2$ component is marginal as  $M_{\text{\HI{}}} = 0.46 \times 10^{10} M_\odot$  \citep{Walter_etal_2008} and $M_{\text{H}_2} = 0.04  \times 10^{10} M_\odot$  \citep{Schruba_etal_2011}.  The Bosma brightness profile shows an approximate exponential decay with $R_{\text{d}} \approx 6 $ kpc from very small scale, where the stellar component dominates at small radii and  the gas component in the outer regions. The maximum radius up to which the rotation curve is measured is  about $2.5 R_{\text{d}}$; at that distance  $v_{\text{c}}(R)$ is still rising.  The very inner   disc rotation curve, i.e., $R<3$ kpc,  can be accounted for by the stellar component alone (where we used the same mass $M_{\text{s}}=0.72 \times 10^{10} M_\odot$ as  \cite{Hessman+Ziebart_2011}). On larger radii, the DMD requires $\approx$ 8 times the  \HI{} +H$_2$ mass to fit the rotation curve. The total DMD mass is about 3.5 times larger than the visibile one.  The NFW model has a virial mass approximately 120 times larger than the baryonic one; correspondingly, the velocity at the virial radius $V_{\text{vir}}\approx 160$ km/s is larger than the observed maximum rotation velocity $\approx 120$ km/s. This problematic situation arises when fitting a linearly rising rotation curve with an NFW model  \citep{McGaugh+deBlok_1998}, leading to a concentration parameter of $c \approx 2.5$ that is larger than the expected value according to Eq. \ref{cnfw} of $c_{\text{vir}} \approx 6.7$.

{  
In our analysis, we have found that both the DMD model and its modified version (mDMD), where the free parameters $\Upsilon_{\text{g}}$ (disc mass-to-light ratio) and $R_{\text{d}}$ (disc scale length) are considered, yield similar reduced $\chi^2$ values. This implies that both models provide reasonable fits to the data, and there is no significant preference for one over the other based on the goodness of fit alone.

We note that the gNFW (generalized NFW) model generally provides a better agreement with the data compared to the NFW model. This is because the gNFW model offers more flexibility at small radii, allowing it to capture the detailed features that characterize the observed rotation curve. 

Although the obtained reduced $\chi^2$ values are relatively high, around $\approx 2$, this is not uncommon in studies of rotation curves. The bumps and wiggles present in the rotation curve data can introduce some level of uncertainty and make achieving a perfect fit challenging. Therefore, the higher reduced $\chi^2$ values are primarily a reflection of the intrinsic complexities  in the rotation curve data rather than a fundamental limitation of the models themselves.
}

\subsection{NGC 2366} 

The stellar surface brightness (2MASS  data --- see \cite{deBlok_etal_2008}), exhibits a single exponential decay for $R>1$ kpc. The  \HI{}  surface brightness displays a constant behavior at small radii, then it shows approximately an exponential decay with the same characteristic scale-length of the stellar component.  No information  of  the H$_2$ surface brightness is available for this galaxy.  The Bosma brightness profile shows an exponential decay with $R_{\text{d}} \approx 1.9$ kpc. The total baryonic mass  is dominated by the   \HI{}  component \citep{Walter_etal_2008} that is about  four times larger than the stellar mass  \citep{Hessman+Ziebart_2011}.  The maximum radius up to which the rotation curve is measured is three time $R_{\text{d}}$. The rotation curve shows a flattening for $R>2$ kpc.  Fitting the rotation curve with the DMD model we find that $\Upsilon_{\text{s}} = 5.8$ and   $\Upsilon_{\text{g}} = 3.2$, so that we obtain that  the total mass in this model is about 3.6 times larger than the baryonic one.  A NFW model requires a halo with a virial mass 34 times larger than the baryonic mass.   The   concentration parameter as calculated from Eq. \ref{cnfw}  is  $c_{\text{vir}} \approx 8.7$ and it is different from the best-fit value $c \approx 5.5$.
{ The DMD and mDMD models give similar reduced $\chi^2\approx 0.2$ value, and the same occurs for the NFW and gNFW models with $\chi^2\approx 0.7$ .} 

\subsection{NGC 2403} 
The stellar surface brightness (S$^4$G data) shows a double exponential decay,  signaling the presence of a bulge at small radii whose mass is estimated to be about 10\% that of the   disc   \citep{Hessman+Ziebart_2011}. The  \HI{}  surface brightness is close to constant in the very inner   disc and then decays exponentially. The  \HI{} +H$2$ surface brightness displays an exponential decay with a characteristic scale-length that is approximately three times  larger than the stellar one $R_{\text{s}}$.  The Bosma brightness profile shows an approximate exponential decay with $R_{\text{d}} \approx 5$ kpc. The maximum radius up to which the rotation curve is measured is about three times $R_{\text{d}}$.  The stellar mass is similar to the  \HI{}  mass \citep{Walter_etal_2008} while the H$2$ contribution \citep{Schruba_etal_2011}  represents only a small fraction of the total gas content. The inner   disc requires three times the mass of the stellar component to fit the rotation curve. The DMD mass is eight times larger than the sum of the stellar and gas mass, while the NFW halo virial mass is almost 45 times larger. In this case, we find from Eq. \ref{cnfw} that $c_{\text{vir}} \approx 7.9$, which is about half the best-fit value of $c \approx 15.0$.
{ In this case, the best fit for the DMD model yields a reduced $\chi^2$ value of 1.1, which is half that of the mDMD model. Both the NFW and gNFW models result in smaller $\chi^2$ values of 0.4..}


\subsection{NGC 2841} 

The stellar surface brightness  (S$^4$G data) shows the typical bulge-  disc transition at small radii, i.e. $\sim 3$ kpc; on large radii  it shows an exponential decay    a characteristic scale-length   $R \approx 3.4$ kpc, which is about three times smaller than the  \HI{}   exponential scale-length.  As for other galaxies, we neglect the contribution of the bulge (whose mass is about 20\% that of the   disc \cite{Hessman+Ziebart_2011}) and consider the range for $R > 4$ kpc for a fitting with  the mass models. The H$_2$ mass represents 20\% of the total gas content  \citep{Schruba_etal_2011}, and gives  a non-negligible contribution to the gas surface brightness only at small radii. The  \HI{} +H$_2$ surface brightness  shows an approximate exponential decay with  characteristic scale  $R_{\text{g}} = 10.6$ kpc.  The Bosma brightness profile is dominated by the rescaled gas component at all radii but the very inner ones where the bulge dominates and thus it has the same  characteristic scale of the  \HI{}  component. The maximum radius up to which the rotation curve is measured is 4.5 times $R_{\text{d}}$: this rotation curve is one of those that extends most in terms of $R/R_{\text{d}}$.  The peak of the rotation curve is at $\approx 10$ kpc and the velocity is $\approx 320$ km/s while in  the outermost regions, $R>40$ kpc the rotation curve  decreases  of about 15\%, reaching $\approx 270$ km/s: this is thus an example of a slowly decaying rotation curve.  To explain the behavior of the rotation  at large radii a substantial amount of  DM  is required.  Indeed, we find that $\Upsilon_{\text{s}} \approx 1.3$ and $\Upsilon_{\text{g}} \approx 47$ so that the main mass contribution in this model is given by the rescaled gas component.  The total mass for the DMD case is 5.4 times larger than the baryonic mass, while the virial mass of the NFW halo is about 18 times larger.  From Eq. \ref{cnfw}, we find that $c_{\text{vir}} \approx 6.5$, while the best-fit value of $c \approx 16.7$.

{ The DMD and mDMD models yield similar reduced $\chi^2$ values of approximately 2, while the NFW and gNFW models produce smaller values, around $\chi^2\approx 0.4$.}


\subsection{NGC 2903} 

Because this galaxy has a bulge, the stellar surface brightness (S$^4$G data)  shows a double exponential decay, with $R_{\text{d}} = 0.25$ kpc for $R < 1$ kpc and $R_{\text{d}} = 2.5$ kpc for $R > 4$ kpc. The mass of the bulge is about 10\% of that of the   disc  \citep{Hessman+Ziebart_2011} and we exclude it from the fit. The the  \HI{}  surface brightness profile is close to flat for $R<10$ kpc, whereas the  \HI{} +H$_2$ surface brightness profile shows a double exponential decay, with H$2$ being dominant for $R < 7$ kpc, having a characteristic scale of $R_{\text{g}} \approx 2.5$ kpc, while $R_{\text{g}} \approx 8$ kpc for $R > 10$ kpc. We performed the fit with the  \HI{}  profile alone as the H$_2$ gives a very high contribution at small radii where the bulge dominates.   The Bosma brightness profile shows an approximate exponential decay with $R_{\text{d}} \approx 7$ kpc for $R>10$ kpc. As for the case of NGC 2841, also this galaxy shows a slowly decaying rotation curve in the observed range of distances, passing from 215 km/s at 5 kpc to 185 km/s at 30 kpc, i.e. a decrease of about 15\%. The maximum radius up to which the rotation curve is measured is three times $R_{\text{d}}$. The sum of the stellar and gas components is sufficient to explain the inner   disc rotation curve while in the outer parts, the best fit with a DMD model gives $\Upsilon_{\text{g}} \approx 18$. The DMD mass is less than 4 times larger than the baryonic mass, while the NFW virial mass is about 13 times larger. From Eq. \ref{cnfw}, we find $c_{\text{vir}} \approx 7.4$, which is different from the best-fit value of $c \approx 13.2$.

{ 
The mDMD model obtains a best-fit value for $R_{\text{g}}$ that is twice the one measured in the gas surface density profile. This higher value of $R_{\text{g}}$ enables a superior fit for the mDMD model compared to the DMD model, resulting in a  $\chi^2$ that is reduced by one-third. In contrast, both the NFW and gNFW models yield smaller values, approximately $\chi^2\approx 0.3$.
}


\subsection{NGC 2976} 
The stellar surface brightness profile  (S$^4$G data)  shows a double exponential decay with $R_{\text{d}} = 1.2$ kpc for $R<1$ kpc and $R_{\text{d}} = 0.7$ kpc for $R>1$ kpc. This is not indicative of a typical bulge, as the characteristic scale-length  in the inner part is only slightly larger than in the outer part. The  \HI{} +H$_2$ surface brightness profile shows some distinct features, including a flat behavior at small radii ($R<1.3$ kpc), an exponential decay with $R_{\text{d}} = 0.6$ kpc for $1<R<4$ kpc, and $R_{\text{d}} = 3$ kpc for $R>4$ kpc. However, the rotation curve is determined only up to $R=3$ kpc, so the last regime is not relevant.  The Bosma brightness profile shows an approximate exponential decay with $R_{\text{d}} \approx 1$ kpc and the maximum radius up to which the rotation curve is measured is three times $R_{\text{d}}$.  This galaxy shows signs of strong tidal interactions in its  \HI{}  distribution \citep{Bigiel+Blitz_2012}.  The velocity profile is almost entirely accounted for by the stars, and the DMD is only 1.5 larger than the baryonic matter component. However, the NFW halo virial mass is 80 times  larger than the baryonic matter component. The value of $c_{\text{vir}} \approx 8.3$ is twice the best-fit value of $c \approx 2.7$, suggesting that Eq. \ref{cnfw} is not satisfied.

{ 
The four models produce similar reduced $\chi^2$ values, approximately around 1.7. These elevated values can be attributed to the prominent bumps present in the rotation curve of this galaxy.
}

\subsection{NGC 3198} 
Even this galaxy presents a stellar surface brightness profile  (S$^4$G data)  with a bulge-  disc structure, i.e. a double exponential behavior: at small radii ($R < 1.5$ kpc), it decays exponentially with a scale-length of $R_{\text{s}} = 0.7$ kpc, and then at larger radii it shows a slope with $R_{\text{s}} = 3.4$ kpc. The mass of the bulge is 20\% that of   disc: as for the other galaxies we have excluded the very inner   disc from the fit. The  \HI{}  surface brightness is close to constant at small radii and then shows an exponential decay with a scale-length of $R_{\text{g}} \approx 12$ kpc. The contribution of H$_2$ \citep{Schruba_etal_2011} is small and limited to very small radii. The  \HI{} +H$2$ surface brightness has an exponential decay from the very inner   disc with the same characteristic scale-length as the  \HI{} .  The Bosma brightness profile shows an approximate exponential decay with $R_{\text{d}} \approx 10$ kpc, i.e. that of the gas component as the stellar component dominates only in the very inner   disc. As the maximum radius up to which the rotation curve is measured is about 40 kpc, this is four times $R_{\text{d}}$. In the DMD case, the stellar component accounts for about half of the observed rotation curve up to 10 kpc as for the best fit model  $\Upsilon_{\text{s}}=1.8$.  At larger radii the rescaled gas component, with $\Upsilon_{\text{g}}=8.1$ dominates the mass distribution and the total DMD mass is about five times larger than the baryonic component.  In the case of the NFW halo the best fit virial mass is about 13 times larger than the baryonic component.  The best fit concentration parameter is again different from the phenomenological one: Eq. \ref{cnfw} gives $c_{\text{vir}} \approx 10.3$, while  $c \approx 7.6$.

{ 
The DMD and mDMD models yield nearly identical reduced $\chi^2$ values, while the same is true for the NFW and gNFW models.
} 


\subsection{NGC 3521} 
The stellar surface brightness profile  (S$^4$G data)  exhibits a double exponential behavior. At small radii (for $R<1$ kpc), it decays with a scale-length of $R_{\text{s}}=1.4$ kpc, and at larger radii  has a scale-length of $R_{\text{s}}=4$ kpc. The  \HI{}  surface brightness is nearly constant at small radii, and for $R>10$ kpc, it exhibits an exponential decay with a scale-length of $R_{\text{g}} \approx 7 $ kpc. The contribution of H$_2$ \citep{Schruba_etal_2011} is limited to within 10 kpc. The Bosma brightness profile shows an approximate exponential decay with $R_{\text{d}} \approx 5$ kpc so that the maximum radius up to which the rotation curve is measured is five times $R_{\text{d}}$. The stellar component is almost enough to account for the observed rotation curve up to 15 kpc. The DMD mass model requires  $\Upsilon_{\text{g}}=3.1$, while $\Upsilon_{\text{s}}=1.5$: in this case the total mass only 1.8 times larger than the baryonic one.  For the NFW mass model, the halo virial mass is about two times larger than  the baryonic mass.  
The  best fit  concentration parameter,  $c \approx 35$, is smaller than the valued predicted by   Eq. \ref{cnfw} gives  $c_{\text{vir}} \approx  8$.

{
The DMD, NFW, and gNFW models all result in similar reduced $\chi^2$ values. However, in this instance, the mDMD model's best fit yields a high value of $\chi^2=3.2$. This is attributed to the mDMD model's inability to accurately fit the innermost portion of the rotation curve.
}


\subsection{NGC 3621} 
The stellar surface brightness profile (2MASS  data --- see \cite{deBlok_etal_2008}) has a single exponential behavior with a scale-length of $R_{\text{s}} =$ 2.3 kpc. The  \HI{}  surface brightness shows a depletion at small radii, i.e. $R< 3$ kpc and then an exponential decay with a scale-length of $R_{\text{g}} \approx $ 10 kpc. No data are available  for the  H$_2$ surface density profile. The Bosma brightness profile shows an approximate exponential decay with $R_{\text{d}} \approx 10$ kpc and the maximum radius up to which the rotation curve is measured is twice  $R_{\text{d}}$.  The stellar and gas masses are very similar, i.e. $M_{\text{s}} \approx M_{\text{g}} \approx 1.4 \times 10^{10} M_\odot$  \citep{Hessman+Ziebart_2011}. The stellar component is sufficient to account for the observed rotation curve at very small radii, i.e., $R<5$  kpc.  The best fit  of the DMD mass model  requires $\Upsilon_{\text{s}} \approx 1.5$ whereas the gas component is about 9 times larger.  The total DMD mass is about 5 times larger than the baryon mass. In the the NFW mass model, the halo virial mass is estimated to be 26 times the visible component.  In this case the best-fit value of the concentration parameter, $c \approx 6.8$, is very close to the value predicted by Eq. \ref{cnfw}, i.e. $c_{\text{vir}} \approx 7.3$.

{
All models yield reduced $\chi^2$ values below 1, indicating excellent fits. }

\subsection{NGC 4736} 
The stellar component surface brightness  (S$^4$G data) exhibits a complex behavior that can be modeled as three exponential decays. The first is characterized by $R_{\text{s}} = 0.2$ for radii less than 1 kpc, followed by $R_{\text{s}} = 1.0$ for 1 kpc $< R < 6$ kpc, and $R_{\text{s}} = 5.0$ for $R > 5$ kpc. We fit the models only for $R > 1$ kpc, ignoring the bulge. Despite this, the profile of the stellar surface brightness still displays a unique behavior at small radii, and the rotation curve also exhibits a very clear decline. The  \HI{}  surface brightness is nearly constant, while the  \HI{} +H$2$ (data from \citep{Schruba_etal_2011}) surface brightness shows approximately a  double exponential behavior with $R_{\text{d}} = 1.0$ for $R < 5$ kpc and $R_{\text{d}} = 10.0$ for $R > 5$ kpc. The Bosma brightness profile does not follow a simple exponential decay in the same range of radii where the rotation curve was measured and can fitted by a rough exponential decay with $R_{\text{d}} \approx 2.2$ kpc but with a large modulation about this behavior both at small and a large radii. The maximum radius up to which the rotation curve is measured is 4  times $R_{\text{d}}$. This is one of the most extended rotation curves in the sample.  The peak of the rotation curve is at $\approx 1$ kpc and the velocity is $\approx 195$ km/s while in  the outermost regions, $R\approx 9$ kpc the rotation curve is decreased  of about 40\%, reaching $\approx 120$ km/s. The stellar component alone can take into account the observed rotation curve at small radii: this is the reason why the best fit with te Bosma model gives $\Upsilon_{\text{s}} \approx 3$. In this respect we note that the stellar mass we sued was $M_{\text{s}}=4 \times 10^{10} M_\odot$ which is three times larger the value reported by \cite{Hessman+Ziebart_2011} (excluding the bulge's mass). The DMD mass is only  10\% larger than the baryonic mass. In the NFW mass model the halo virial  represents a small fraction of the baryonic mass, showing that the stellar component is sufficient to take into account almost the whole mass budget.  Finally  the best-fit value concentration parameter $c \approx 13$, larger than $c_{\text{vir}} \approx 13$. predicted from  Eq. \ref{cnfw}.

{
The DMD, mDMD, NFW, and gNFW models all yield similar reduced $\chi^2$ values, approximately around 1. This relatively high value is attributed to the presence of multiple bumps and wiggles in the rotation curve.
}


\subsection{NGC 5055} 

The stellar surface brightness  profile  (S$^4$G data) has a double exponential behavior, typical signature of the bulge, with  $R_{\text{s}}=0.5$  for $R<3$ kpc and $R_{\text{s}}=5.0$ for $R_{\text{d}}>3$ kpc. The  \HI{}   surface brightness  has a plateau at small radii. The  \HI{} +H$2$ (data from \citep{Schruba_etal_2011}) surface brightness  shows  a double exponential behavior with $R_{\text{g}} =5.0$ for $R<20$ kpc and $R_{\text{g}}\approx 20 $ for $R>22$ kpc.  We performed the fit with the  \HI{}  profile alone as the H$_2$ gives a very high contribution at small radii where the bulge dominates.  The Bosma brightness profile shows an approximate exponential decay with $R_{\text{d}} \approx 20$ kpc and the maximum radius up to which the rotation curve is measured is twice  $R_{\text{d}}$. The peak of the rotation curve is at $\approx 10$ kpc and the velocity is $\approx 210$ km/s while in  the outermost regions, $R>40$ kpc the rotation curve is decreased of about 20\%, reaching $\approx 165$ km/s.  The sum of the stellar and gas components is almost sufficient to take into account the inner   disc behavior (we avoid to fit the bulge as for the other cases whose mass is about 10\% of that of the   disc  \cite{Hessman+Ziebart_2011}).  The total DMD mass is about four times larger than the visibile mass while the NFW virial mass is  16 larger than the visibile mass.  The best-fit concentration parameter is $c \approx 2.7$ that is close to $c_{\text{vir}} \approx 6.6 $ from  Eq. \ref{cnfw}.

{
The DMD, NFW, and gNFW models all result in similar reduced $\chi^2$ values, each smaller than 1. In contrast, the mDMD model obtains a value of $\chi^2=5.3$ due to the gas component's inability to accurately fit the inner region of the rotation curve.}

\subsection{NGC 6946} 
The stellar surface brightness  profile (2MASS  data --- see \cite{deBlok_etal_2008})has a double exponential behavior with   $R_{\text{s}} =0.6$  for $R<1.5$ kpc and $R_{\text{s}}=2.8$ for $R_{\text{d}}>2$ kpc: this is an indication for the presence of a bulge whose mass is estimated to be 0.27 $\times 10^{10} M_\odot$ while the mass of the   disc is 4.2 $\times 10^{10} M_\odot$ \citep{Hessman+Ziebart_2011}. We avoid to include the bulge in the fit and we consider only $R>2$ kpc.  The  \HI{}  +H$_2$   surface brightness  profile (data from \citep{Schruba_etal_2011}) shows  approximately  an exponential behavior with $R_{\text{g}} \approx 4$, while the  \HI{}   surface brightness has a plateau at small radii.   The Bosma brightness profile shows an approximate exponential decay with $R_{\text{d}} \approx 4$ kpc and the maximum radius up to which the rotation curve is measured is five times  $R_{\text{d}}$.   We performed the fit with the  \HI{}  profile alone as the H$_2$ gives a very high contribution at small radii where the bulge dominates.   The rotation curve shows and increase at small radii and the a flattening. The stellar plus gas component are sufficient to take into account the rotation curve in the inner   disc. At larger radii the gas mass must be scaled by a factor    $\Upsilon_{\text{g}} \approx 7$.  The total DMD mass is  about 1.8 times  larger than the baryonic mass whereas the NFW virial mass is  respectively  6  times  larger than the baryonic mass. The best-fit concentration parameter is $c \approx 12$ and is larger than  $c_{\text{vir}} \approx 7 $  predicted by Eq. \ref{cnfw}. 

{ 
Both the DMD and mDMD  yield small values of  reduced $\chi^2$ values, approximately around 2. Conversely, the NFW and gNFW models produce reduced $\chi^2$ values of around 1.
}

\subsection{NGC 7331} 

The stellar surface brightness    profile  (2MASS  data --- see \cite{deBlok_etal_2008}) shows a double   exponential decay signaling the presence of a bulge with    $R_{\text{s}}=2$ kpc at small radii and $R_{\text{s}} \approx 6$ kpc at large ones. The  \HI{}    surface brightness profile is flat at small radii and then it shows   an exponential behavior with $R_{\text{g}} \approx 7 $; the  \HI{} +H$_2$  surface brightness exponentially decays  with the same characteristic scale from small scales.  The Bosma brightness profile shows an approximate exponential decay with $R_{\text{d}} \approx 7$ kpc and the maximum radius up to which the rotation curve is measured is three times  $R_{\text{d}}$.  We avoid to fit the bulge as for the other cases whose mass is about 10\% of that of the   disc  \citep{Hessman+Ziebart_2011}. The sum of the stellar and gas components is sufficient  takes into account the observed rotation curve at small radii, i.e. $R<5$ kpc.   The total DMD mass is  about 2  times  larger than the visibile mass while the NFW virial mass is about 7  times  larger than the visibile mass. The best-fit concentration parameter is $c \approx 11.9$ that is close to $c_{\text{vir}} \approx 7.1 $ from  Eq. \ref{cnfw}.

{ 
All models produce reduced $\chi^2$ values that are small, around 0.2.
}

\subsection{NGC 7793} 
The stellar surface brightness  (S$^4$G data) shows a simple exponential decay with   $R_{\text{s}} =1.2$ whereas the  \HI{}    surface brightness  profile is flat at small radii and then  decays exponentially with a  characteristic scale-length close to the stellar one  $R_{\text{s}} \approx 2$.  No data are available for the H$_2$ distribution. The Bosma brightness profile shows an approximate exponential decay with $R_{\text{d}} \approx 2$ kpc   and the maximum radius up to which the rotation curve is measured is four times  $R_{\text{d}}$. The peak of the rotation curve is at $\approx 4.5$ kpc and the velocity is $\approx 116$ km/s while in  the outermost regions, $R>7$ kpc the rotation curve is decreased of about 17\%, reaching $\approx 95$ km/s. The inner   disc rotation curve, i.e., for $R<3$ kpc,  can be explained by the stellar component solely. The total DMD mass is  about 3 times  larger than the baryonic mass. The halo  virial mass is 200  times  larger than the baryonic mass: this problematic situation occurs when one fits  a  linearly rising rotation curve with a  NFW model  \citep{McGaugh+deBlok_1998}. The concentration parameter value: $c_{\text{vir}} \approx  7.0$  from  Eq. \ref{cnfw} is larger than  the  best-fit  value  $c \approx 5.6$.

{ 
In this particular case, the best-fit model is the DMD, which results in a $\chi^2$ value of 0.3, whereas the other models yield $\chi^2$ values greater than 1.
}

\subsection{DDO 154} 
The stellar surface brightness  profile (2MASS  data --- see \cite{deBlok_etal_2008})  shows decay at small radii and the it flattens although being very noisy:  the stellar component is however a negligible component of the whole mass budget as the mass of the gas component is almost 20 times larger. Instead, the  \HI{}    surface brightness  shows  an exponential decay with a scale-length $R_{\text{s}} = 3.1$.  No data are available for the H$_2$ distribution.  The Bosma brightness profile shows an approximate exponential decay with $R_{\text{d}} \approx 3.1$ kpc  and the maximum radius up to which the rotation curve is measured is almost three times  $R_{\text{d}}$.  In the DMD mass model the total mass is  about 12 times  larger than the baryonic mass. In the NFW mass model the halo virial mass is instead about 190 times  larger than the baryonic mass:  even in this case the best-fit concentration parameter is smaller than the  values given by from  Eq. \ref{cnfw},  i.e., $c \approx 4.5 $ instead of $c_{\text{vir}}=9.1$.

{ 
The mDMD model produces a $\chi^2$ value of approximately 0.1, similar to the NFW and gNFW models, while the DMD model yields a $\chi^2$ of 1.2.
}

\subsection{IC 2574} 
The stellar surface brightness  profile  (2MASS  data --- see \cite{deBlok_etal_2008})  has an exponential behavior with $R_{\text{s}} =2.9$ while the  \HI{}  surface brightness  is constant at small radii, i.e. $R<7$ kpc, and then it  shows a sharp decay.  No data are available for the H$_2$ distribution. The Bosma brightness profile shows an approximate exponential decay with $R_{\text{d}} \approx 4$ kpc: however, especially at large radii, the exponential approximation works badly because in that range of radii the surface brightness is dominated by the rescaled   \HI{}  one. The maximum radius up to which the rotation curve  is measured is twice times  $R_{\text{d}}$.  Both fits with the DMD and the NFW work well.    The observed  \HI{}  mass is twice the stellar mass and  the total DMD mass is  about 3 times  larger than the baryonic one. The NFW virial mass is  about190  times  larger than the baryonic mass.  Finally the best-fit concentration factor  $c \approx 4.5 $, is smaller than the value obtained from  Eq. \ref{cnfw}, i.e., $c_{\text{vir}}=9.0$.

{ 
Both the DMD and mDMD models result in small values of reduced $\chi^2$, approximately around 0.6. In contrast, the NFW model yields $\chi^2=2$, and the gNFW model yields 0.8.
}

\subsection{Milky Way} 
\cite{SylosLabini_etal_2023a} made a DMD fit to the rotation curve of the MW measured by the latest data releases of the Gaia mission by \cite{Eilers_etal_2019,Wang_etal_2023}. They have obtained that, by fixing $\Upsilon_{\text{s}}=1$ the gas amplification is about $\Upsilon_{\text{g}}=16$. The sum of the stellar and gas components have a more complicated behavior than a simple exponential function. However, a exponential decay with $R_{\text{d}}=5.7$ kpc well fits with rotation curve if the mass is $M_{\text{d}}=17 \times 10^{10} M_\odot$. Note that  the best fit values of \cite{SylosLabini_etal_2023a}  give a  local density of $\sim 0.3$ Gev cm$^{-3}$ and for the Bosma case 13,2 Gev$^{-3}$.

{ 
All models produce reduced $\chi^2$ values that are between 1 and 2.}

\end{document}